\documentclass[usenatbib]{tex/mn2e}

\usepackage[]{graphicx}
\usepackage{caption}
\usepackage{amssymb,amsmath}
\usepackage{natbib}
\usepackage{longtable}
\newcommand{\hii}{H\,{\sc ii}~} 
\newcommand{\ha}{\mbox{H$\alpha$}} 
\newcommand{\hb}{\mbox{H$\beta$}}

\newcommand{\nfields}{14~}
\newcommand{\ms}{``Main Sequence''}
\newcommand{\msp}{``Main Sequence.''~}
\newcommand{\oiii}{[OIII]$\lambda4959,\lambda5007$}

\begin{document}

\title[CLU I: Galaxy Catalogs from Preliminary Fields]{Census of the Local Universe (CLU) Narrow-Band Survey I: Galaxy Catalogs from Preliminary Fields}

\author[D. Cook et al.]
{David O. Cook,$^{1,2}$
Mansi M. Kasliwal,$^{1}$
Angela Van Sistine,$^{3}$
David L. Kaplan,$^{3}$
\newauthor
Jessica S. Sutter,$^{4}$
Thomas Kupfer,$^{1}$
David L. Shupe,$^{2}$
Russ R. Laher,$^{2}$
\newauthor
Frank J. Masci,$^{2}$
Daniel A. Dale,$^{4}$
Branimir Sesar$^{5}$
Patrick R. Brady,$^{2}$
Lin Yan,$^{4}$
\newauthor
Eran O. Ofek,$^{6}$
David H. Reitze,$^{1}$
Shrinivas R. Kulkarni$^{1}$
\\
$^1$California Institute of Technology, 1200 East California Blvd, Pasadena, CA 91125, USA; dcook$@$ipac.caltech.edu\\
$^2$ IPAC/Caltech , 1200 E California Blvd, Pasadena, CA 91125, USA\\
$^3$University of Wisconsin-Milwaukee, P.O. Box 413, Milwaukee, WI 53201, USA\\
$^4$University of Wyoming, 1000 University Ave, Laramie, WY 82071, USA\\
$^5$Max Planck Institute for Astronomy, Königstuhl 17, D-69117 Heidelberg, Germany\\
$^6$Weizmann Institute of Science, Rehovot, Israel\\}
\maketitle

\begin{abstract}

We present the Census of the Local Universe (CLU) narrow-band survey to search for emission-line (\ha) galaxies. CLU-\ha~has imaged $\approx$3$\pi$ of the sky (26,470~deg$^2$) with 4 narrow-band filters that probe a distance out to 200~Mpc. We have obtained spectroscopic follow-up for galaxy candidates in 14 preliminary fields (101.6~deg$^2$) to characterize the limits and completeness of the survey. In these preliminary fields, CLU can identify emission lines down to an \ha~flux limit of $10^{-14}$~$\rm{erg~s^{-1}~cm^{-2}}$ at 90\% completeness, and recovers 83\% (67\%) of the \ha~flux from catalogued galaxies in our search volume at the $\Sigma$=2.5 ($\Sigma$=5) color excess levels. The contamination from galaxies with no emission lines is 61\% (12\%) for $\Sigma$=2.5 ($\Sigma$=5). Also, in the regions of overlap between our preliminary fields and previous emission-line surveys, we recover the majority of the galaxies found in previous surveys and identify an additional $\approx$300 galaxies. In total, we find 90 galaxies with no previous distance information, several of which are interesting objects: 7 blue compact dwarfs, 1 green pea, and a Seyfert galaxy; we also identified a known planetary nebula. These objects show that the CLU-\ha~survey can be a discovery machine for objects in our own Galaxy and extreme galaxies out to intermediate redshifts. However, the majority of the CLU-\ha~galaxies identified in this work show properties consistent with normal star-forming galaxies. CLU-\ha~galaxies with new redshifts will be added to existing galaxy catalogs to focus the search for the electromagnetic counterpart to gravitational wave events.  


\end{abstract}


\section{Introduction}
Large-area, blind surveys of emission-line or UV-excess galaxies have led to important extra-galactic discoveries over the past few decades. The first systematic search was the Markarian Survey \citep{markarian81} which in total scanned 15,200 deg$^2$ from 1965-1980 down to an characteristic magnitude of 15.5~mag in the V-band. Just over 1500 galaxies were discovered in this search including starburst galaxies, AGN, QSOs, Sefert galaxies, and blue compact dwarfs (BCDs). Many of these galaxies exhibit some of the most extreme properties still known in the local Universe \citep[low metallicity, high star formation rates, optical line ratios, etc.;][]{mickaelian14}, and have facilitated a better understanding of star formation and galaxy evolution.

Several studies have since searched smaller areas of the sky to greater depths, such as the second Markarian Survey \citep{markarian83}, University of Michigan (UM) survey \citep{UM1}, Case survey \citep{case1}, Kitt Peak International Spectral Survey – KISS \citep{kiss00}, the Hamburg/SAO Survey \citep[HSS;][]{hss99}. Many of these studies focused on finding samples of BCDs to investigate their relationship to primordial environments due to their low metallicities \citep[i.e., as low as 1/40 $Z_{\odot}$ or 12+log(O/H)=7.1 for DDO68, SBS 0335–052, and I Zw 18;][]{thuan05,izotov05,izotov07} and their high gas content (see Thuan 1991 for a review). In the decades after these surveys, a handful of extremely low metallicity BCDs have been found \citep{izotov06,izotov07,hirschauer16,izotov18}, but none with 12+log(O/H)$<$6.9 suggesting a metallicity floor in the local Universe. However, larger area surveys that can select BCDs could fill out the extremely low metallicity end of the distribution and provide constraints on star formation and galaxy evolution models. 

The next advancement in large area galaxy surveys was the Sloan Digital Sky Survey \citep[SDSS;][]{york00}. The SDSS spectroscopic galaxy survey obtained spectra of half a million extragalactic sources in $\sim$9400~deg$^2$ using fiber-fed plates \citep{sdss12}. These data directly led to the discovery of the \ms~of star-forming galaxies, where the current star-formation rate (SFR) of a galaxy forms a tight relationship with its total stellar mass \citep{brinchmann04,daddi07,salim07,peng10}. This, now fundamental, relationship led to further discoveries which found that the \ms~shows an invariant scatter but an increased normalization at higher redshifts \citep{elbaz07,noeske07,heinis14}. Thus, these studies have provided the means to quantify the evolution of galaxy properties over time, and can provide insights into star formation via galaxies whose properties fall off the \msp

The SDSS survey also gave rise to the discovery of extreme emission-line galaxies at intermediate redshifts ($0.11\leq \rm{z} \leq0.36$) known as ``green peas.'' These objects were identified based on their point source-like appearance where the strong \oiii~doublet (EW$\geq$100~\AA) is located in the green-coded SDSS $r$-band filter giving them their green colors \citep{lintott08}. In-depth studies of green peas have shown that these objects have compact morphologies, high SFRs, and low metallicities \citep{cardamone09,amorin10,izotov11}. Thus, these galaxies bare a striking resemblance to high-redshift galaxies (Lyman-break galaxies and Ly$\alpha$ emitters) that are thought to contribute a significant fraction of ionizing photons during the epoch of reionization \citep{finkelstein12}. In addition, the similar metallicities and abundance ratios of green peas and BCDs suggest that these two are related objects \citep{izotov11}, but this connection is currently under debate. Large area surveys that can select ``green peas" and BCDs will provide better statistics to help determine if a connection exists.

The discovery of the galaxy \ms~has allowed astronomers to quantify galaxy properties over time and the low scatter in this relationship has provided a reference point from which to put extreme galaxies into context. However, a better understanding of the physical processes responsible for the \ms~scatter and the extreme deviations will largely come from high sensitivity and spatial resolution examinations of galaxies located in the local Universe. As such, an all-sky, or nearly all-sky, survey designed to find a more complete sample of extreme and normal star-forming galaxies in the local Universe would provide better statistics of galaxy evolutionary trends and would sample more extreme environments from which to test star formation theories.

In this paper we introduce such a survey: the Census of the Local Universe (CLU) emission-line (\ha) galaxy survey. CLU-\ha~is a narrow-band survey of the entire northern sky above a declination of $-20\deg$ with the aim of constraining galaxy distances out to 200~Mpc. We have imaged $\approx$3$\pi$ sr of the sky as part of the Intermediate Palomar Transient Factory \citep[iPTF;][]{law09} with four contiguous narrow-band filters. Using the \ha~emission line, CLU is designed to find emission-line galaxies from redshift 0 to 0.047 ($\approx$200~Mpc). As a consequence, CLU will provide distance constraints and SFRs to normal star-forming galaxies with moderate EW emission lines and provide a more complete census of galaxies in the local Universe. In addition, CLU will also provide lists of extreme galaxies such as: BCDs in the local Universe, and green peas at intermediate redshifts whose redshifted [OIII] lines will also be detectable in our survey. A more complete list of extreme galaxies in the local Universe across 3/4ths the sky will probe more extreme environments and put these galaxies in context of the galaxy \msp

Another astrophysical impact of the CLU survey is the search for electromagnetic (EM) counterparts to gravitational waves (GW) now detectable from the Laser Interferometer Gravitational Wave Observatory \citep[LIGO;][]{abbott16}. Linking the new discovery tool medium of gravitational waves to our understanding of electromagnetic radiation observable from our current telescopes will have significant impacts on our understanding of the Universe. However, the sky localization of LIGO GW events can range from 30--1000 square degrees on the sky making this search a challenge. Targeted follow-up observations of likely host galaxies can narrow down the search area by a factor of 100 \citep{nissanke13,gehrels16}. As such, the CLU narrow-band filters were designed to probe out to the projected sensitivity distance of merging objects thought to produce EM counterparts \citep[i.e., neutron stars at 200~Mpc;][]{aligo}.

In this paper, we introduce the CLU-\ha~survey, provide the survey parameters (e.g., sky coverage, narrow-band filter properties, and survey limits), detail the galaxy candidate selection criteria, show examples of extreme objects found in preliminary fields, and discuss science applications (e.g., extreme galaxies, star-forming galaxies, and EMGW counterpart searches) for this unique survey.

\section{CLU-\ha~Survey Description}
The CLU-\ha~survey endeavors to discover as many \ha-emitting galaxies as possible out to 200~Mpc (z=0.047) and provide a relatively well constrained redshift range for each emission-line galaxy. In addition, the \ha~fluxes measured for all galaxies (both previously known and newly discovered in the local Universe) will provide a recent SFR for each galaxy \citep[i.e., the integrated SFR over the past $\sim$10~Myr; e.g.,][]{kennicutt98,murphy11,kennicutt12}. Thus CLU-\ha~will not only yield the positions and well constrained distances for new galaxies in the local Universe, but will also provide nearly uniform \ha~fluxes and SFRs for both newly discovered and previously known galaxies. 


\subsection{Narrow-band Filters}

CLU-\ha~searches for emission-line galaxies using 4 wavelength-adjacent, narrow-band filters with a combined wavelength range of $6525-6878$~\AA. Each emission-line galaxy can be identified via a flux excess in one filter (the ``On" filter) signifying the presence of an emission line compared to a filter that covers only the adjacent continuum (the ``Off" filter). Thus, CLU-\ha~provides a distance constrained by the width of our narrow-band filters.

\begin{figure}
  \includegraphics[scale=0.5]{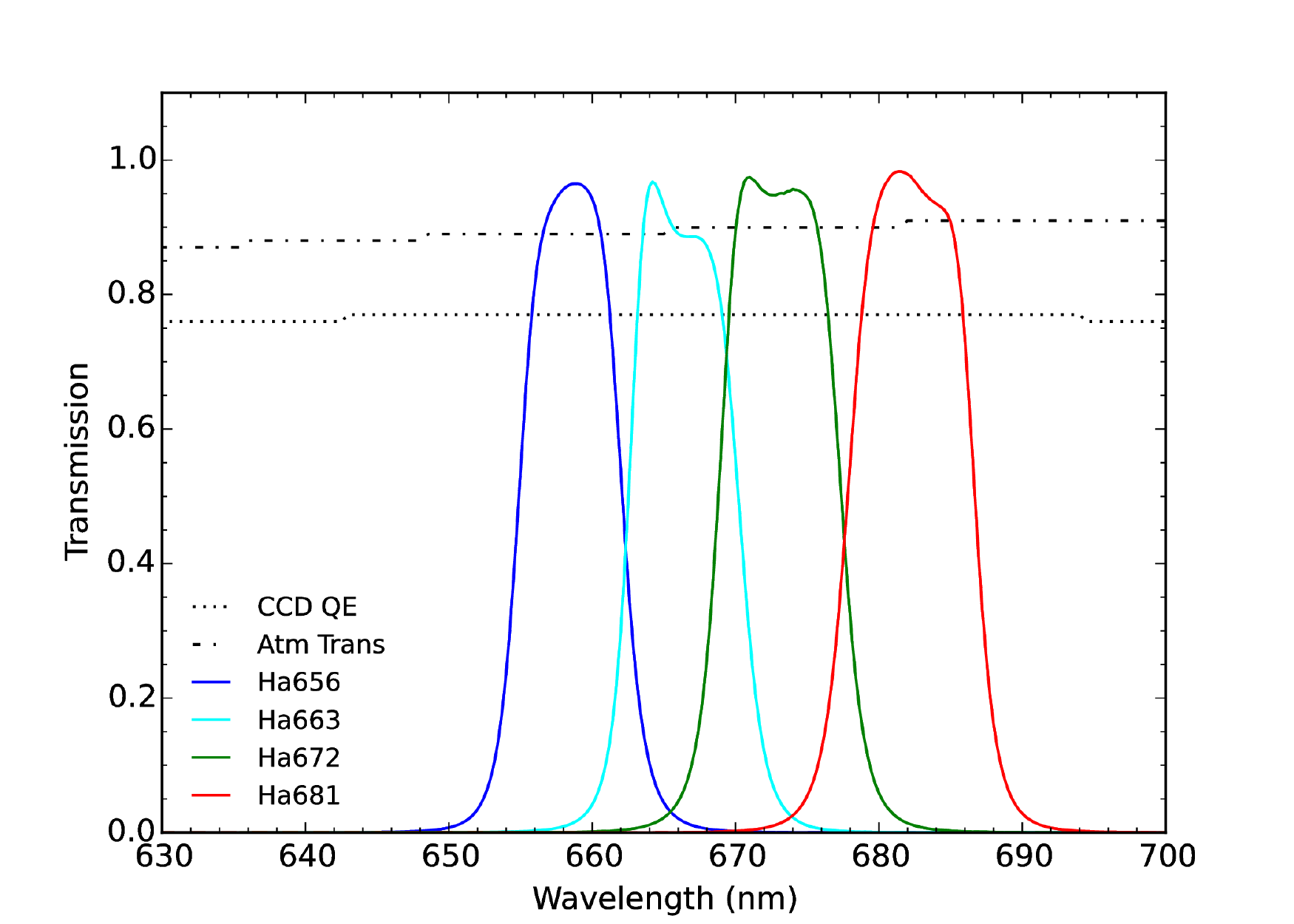}
  \caption{The measured filter transmission profiles of the 4 CLU-\ha~narrow-band filters where the blue, cyan, green, and red curves represent \ha1, \ha2, \ha3, and \ha4 filters, respectively. The horizontal dotted and dashed lines represent the CCD quantum efficiency and the atmospheric transmission at Palomar Observatory, respectively. }
   \label{fig:filttrans}
\end{figure}

The transmission curves for our narrow-band filters are shown in Figure~\ref{fig:filttrans}, where the solid lines represent the transmission from the manufacturer, the dotted line represents the CCD quantum efficiency of the camera and the dash-dotted line represent the atmospheric transmission at Palomar \citep{ofek12}. 

The filter widths and central wavelengths are calculated via:

\begin{equation}
    \Delta\lambda \equiv \int{T(\lambda)d\lambda},
\end{equation}

\begin{equation}
    \bar{\lambda} \equiv \frac{\int{\lambda T(\lambda)d\lambda}}{\int{T(\lambda)d\lambda}},
\end{equation}

\noindent where T is final transmittance of the filter curves with the peak normalized to unity. Table~\ref{tab:filtprop} presents the properties of our 4 narrow-band filters. 

The first filter is centered near the z=0 \ha~emission line ($\lambda=6563$~\AA) and the wavelength range of the last filter extends to z=0.047 (i.e., $\sim$200~Mpc) \ha~emission line. The first two filters (\ha1 and \ha2) make up the first filter pair and the 3rd and 4th filters (\ha3 and \ha4) make up the second filter pair. We note that the small overlap between the filter pairs can result in decreased efficiency of emission line detection for emission lines whose wavelengths fall in this gap; however, the overlap is only 10~\AA~and is not likely to affect a large number of sources.

\begin{table}
\begin{center}
{Narrowband \ha~Filter Properties}
\begin{tabular}{ccccc}
\hline
\hline

Filter	& Filter	& Filter	     & Redshift	\\
name	& $\lambda$	& $\Delta\lambda$     & range	\\
	& (\AA)		& (\AA)		      & (\#)	\\
\hline
  \ha1  & \textbf{6584.2}	& \textbf{76.1} 	 & \textbf{-0.0026 $<$ z $<$ 0.0090}  \\  
  \ha2  & \textbf{6663.7}	& \textbf{77.9} 	 & \textbf{~0.0094 $<$ z $<$ 0.0213}	\\
  \ha3  & \textbf{6730.9}	& \textbf{90.1} 	 & \textbf{~0.0187 $<$ z $<$ 0.0324}	\\
  \ha4  & \textbf{6822.1}	& \textbf{92.1} 	 & \textbf{~0.0325 $<$ z $<$ 0.0471}	\\
\hline
\end{tabular}\\
\end{center}
\caption{The properties of the CLU narrowband filters, where the columns present the filter names, central wavelength, FWHM, and redshift range, from left-to-right. The first filter (\ha1) is centered on rest-frame \ha~while the last filter’s FWHM extends to 200 Mpc.}
\label{tab:filtprop}
\end{table}

\subsection{Observational Strategy}
Observations were taken on the Oschin 48 inch telescope at the Palomar Observatory with a 7.92~deg$^2$ mosaic imager composed of 12 CCD detectors (or chips) and a 1$\arcsec$ per pixel image scale. However, one of the chips (chip \#3) is non-functional resulting in an 11-chip imager with a field of view 7.26~deg$^2$ \citep[for details see,][]{law09,rau09}. The survey uses an observational strategy of 3 spatially-staggered, overlapping grids that are limited to a declination of greater than $-20^{\circ}$, where each grid contains N$=3626$ iPTF fields ($26,470$ deg$^2$ of the sky). Observations using the \ha1 and \ha2 filters cover the entire sky above $-20^{\circ}$ declination.  Observations using the \ha3 and \ha4 filters also cover the sky above $-20^{\circ}$ declination, but avoid the Galactic plane ($|b|\gtrsim 3^{\circ}$). A single 60 second exposure is taken for each field in each grid resulting in three images for the majority of positions on the sky. The multiple images facilitate cosmic ray rejection, filling in chip gaps and the non-functional chip, and deeper final co-added images.

\begin{figure*}
  \includegraphics[scale=0.6]{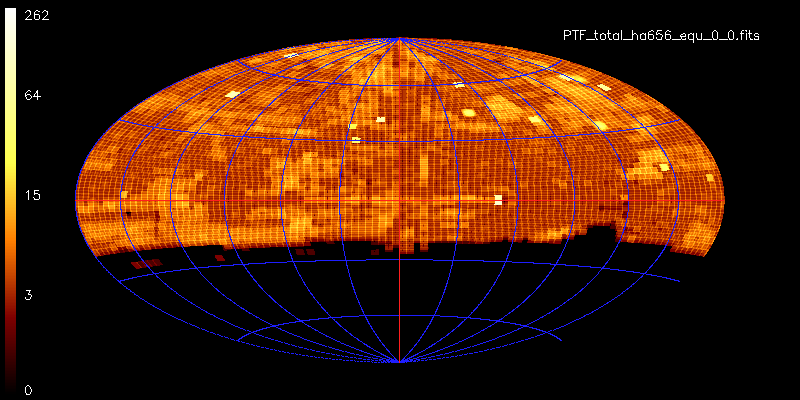}
  \caption{The sky coverage map in equatorial coordinates for the first narrow-band pair of filters (\ha1~and \ha2) in our survey, where the red vertical line represents a RA=0${\fdg}$ The Individual colored boxes represent one pointing with a 7.26~deg$^2$ field of view. The color bar indicates how many observations have been taken for each pointing}
   \label{fig:skymap1}
\end{figure*}

\begin{figure*}
  \includegraphics[scale=0.6]{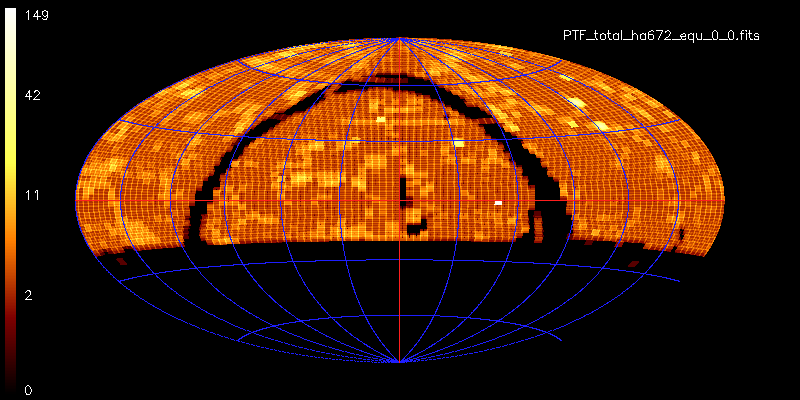}
  \caption{The sky coverage map in equatorial coordinates similar to Figure~\ref{fig:skymap1} but for the second narrow-band pair of filters (\ha3~and \ha4) in our survey. The dark contiguous sections represents the Galactic plane which is avoided in the second filter pair. }
   \label{fig:skymap2}
\end{figure*}

Data acquisition ended in March 2017 (due to decommissioning of the iPTF camera) with 98.3\% of fields observed in \ha1 and \ha2 and 91.3\% observed in \ha3 and \ha4. However, the second filter pair is 95\% completed at a Galactic latitude limit of three degrees (95\% at $|b|\gtrsim 3^{\circ}$) and 99\% completed at latitudes above 11$^{\circ}$ (99\% at $|b|\gtrsim 11^{\circ}$).  Figures~\ref{fig:skymap1} and \ref{fig:skymap2} present the sky coverage maps in equatorial coordinates (where 0$^{\circ}$ Right Ascension is represented as the center, vertical red line) of the CLU-\ha~survey for the first filter pair (\ha1~and \ha2) and the second filter pair (\ha3~and \ha4), respectively. The colors for each iPTF pointing box in these figures represent the number of observations, where lighter/brighter colors indicate more observations. The Galactic plane is easily identified in Figure~\ref{fig:skymap2} as a dark strip of fields with little to no observations in the second filter pair.  The majority of the fields within our targeted sky coverage have three observations. In addition to our survey, there were multiple iPTF projects that observed specific regions of the sky using the CLU narrow-band filters and can provide even deeper \ha~imaging for these fields.  The locations that these programs targeted are apparent in Figures~\ref{fig:skymap1} and \ref{fig:skymap2} as the brightest regions with $\sim$100-200 observations.

Although the ultimate goal of the CLU-\ha~survey is to coadd the images from the 3 staggered, overlapping grid patterns, the preliminary analysis here uses the single exposures from only one of these grids in \nfields fields (see \S\ref{sec:prefields}).

\subsection{Data Reduction and Source Catalogs}
Data reduction and source extraction are carried out in an automated pipeline built by the Infrared Processing and Analysis Center (IPAC)\footnote{http://www.ipac.caltech.edu/} specifically for the iPTF survey. The full description of this pipeline can be found in \cite{laher14}, but we provide a brief overview here. 

The IPAC reduction pipeline consists of both ``off-the-shelf" and custom software which have been extensively tested on millions of images from iPTF broadband images. After each night of data is acquired, the IPAC pipeline performs a bias subtraction and applies a flat-field correction.  The astrometric solution for each processed image is computed via \texttt{SCAMP} \citep{scamp} on one of three stellar catalogs: SDSS-DR7 \citep{sdss7}, UCAC3 \citep{ucac3}, or USNO-B1 \citep{usnob1}. 

After data reduction is completed, \texttt{Source Extractor} \citep{sex96} is used to generate a source catalog for every chip and filter image in each field. The fluxes are reported using many aperture definitions; however, we utilize the fluxes from aperture photometry at 5 pixels ($\sim5\arcsec$) in diameter for galaxy candidate selection (see \S\ref{sec:candsel}) and the point spread function (PSF) fitted fluxes of stars for calibration purposes (see below). The 5 pixel aperture is used to select narrow-band excess sources since the \ha~colors showed smaller scatter in this aperture size compared to larger 8 and 10 pixel diameter apertures. The median and standard deviation 5$\sigma$ detection limits of a point source with a 5$\arcsec$ diameter aperture for each chip image in all 14 preliminary fields is $18.6 \pm 0.30$, $18.7 \pm 0.28$, $18.8 \pm 0.28$, $18.8 \pm 0.25$ AB mag for filters \ha1, \ha2, \ha3, and \ha4, respectively.


\subsection{Calibration} \label{sec:cal}
Calibration is carried out for each chip and filter image combination, where Pan-STARRS DR1 \citep[PS1;][]{ps1} stars (N$>100$) are matched with CLU-\ha~sources. We match all CLU-\ha~sources with the following criteria: 1) are relatively isolated (i.e., no source within $10\arcsec$), 2) have a photometric error less than 0.1~mag, and 3) have no \texttt{Source Extractor} photometry flags to PS1 stars with the following criteria: 1) a $g-i$ color between 0 and 1.7~mag, 2) a photometric error less than 0.1~mag, 3) a PSF minus Kron magnitude in the $i$-band of less than 0.05~mag to select stars.\footnote{https://confluence.stsci.edu/display/PANSTARRS/How+to+\\separate+stars+and+galaxies} The photometric zeropoints and color correction terms are determined by fitting a linear relationship between $\ha-r$ (CLU$-$PS1) and $g-i$ (PS1-PS1) colors, where both CLU-\ha~and PS1 magnitudes are PSF fitted magnitudes. 

\begin{figure}
  \includegraphics[scale=0.5]{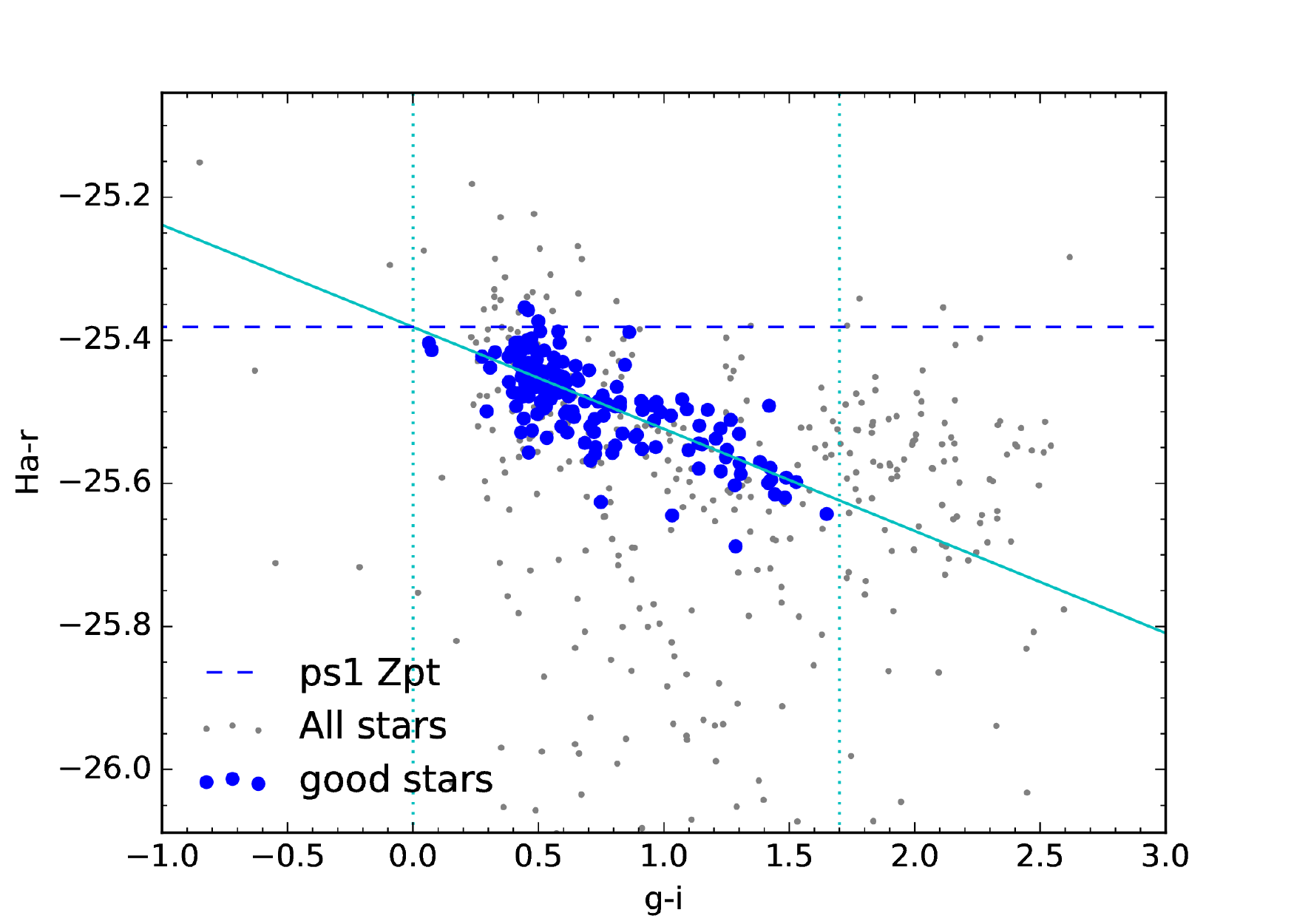}
  \caption{The \ha-r versus $g-i$ color plot which is used to define our photometric calibration (i.e., zeropoint and color correction). The blue filled points are all stars selected for calibration, the horizontal blue line represents the zeropoint, and the solid cyan line is the fit to calibrations stars and is used for color corrections.}
   \label{fig:sdsscal}
\end{figure}  

Figure~\ref{fig:sdsscal} graphically presents our calibration method, where the grey dots represent all CLU-\ha~sources matched to PS1 sources, the blue dots represent matched ``good" stars that meet our criteria, the dashed line represents the zeropoint, and the solid-cyan line represents the fit to ``good" stars. The error in the fit is used as the zeropoint error and is added in quadrature to the photometric error of each source. Typical errors in the zeropoints are $\sim0.05$~mag. The fitted line in Figure~\ref{fig:sdsscal} is used to perform a color correction to the magnitdues of the \ha~sources.

Due to the large field of view of the iPTF instrument, variations in the sensitivity across the imaging array exist after the images are processed with the IPAC image reduction pipeline and need to be taken into account when deriving the final calibrated magnitudes of extracted sources. This variation can be quantified in the zero-point of stars located across the entire field of view. The procedure for deriving a sensitivity variation map is described in detail by \cite{ofek12}, but we provide a brief overview here. Using a minimum of 200 images for each chip and filter image we measure the zero-points for all stars that meet our calibration star criteria (see paragraph above), then combine them into bins of 32~pixels on a side. The median zero-point residual for each bin, normalized by the median zero-point across the entire image, are used as the zero-point variation correction. This correction is applied to the final calibrated magnitude for each source given the position on each chip and the filter used in the observation. The zero-point variation across a single chip has a typical range of 0.01-0.05~mag and the standard deviation of the zeropoint variation is $\sim$0.02~mag for all filters.





\subsection{Ancillary Data} \label{sec:ancdata}
We supplement our CLU-\ha~fluxes with information from the SDSS DR12 \citep{sdss12}, GALEX all sky \citep{galex}, and WISE all sky surveys \citep{wise} in addition to PS1 PSF and Kron magnitudes. We utilize the model $ugriz$ magnitudes from SDSS DR12. In addition, we cross-match the CLU-\ha~sources against entries in the SDSS `galSpecLine' table and extract the redshift, \ha~line flux (`h\_alpha\_flux'), and equivalent width (EW; `h\_alpha\_eqw'). We also extract FUV and NUV kron fluxes from the GALEX all-sky imaging survey \citep[AIS;][]{galexgr67}, and the instrumental profile-fit photometry of the first and fourth WISE bands (`w1mpro' and `w4mpro'). These ancillary data are used to cull contaminants and measure several physical properties of galaxies.

\section{Candidate Selection} \label{sec:candsel}
In this section, we describe our galaxy candidate selection methods. We have chosen to test these methods in \nfields preliminary fields where we have performed a spectroscopic follow-up campaign. In addition, we quantify the success rates of our selection method and the limits of the survey.


\subsection{Preliminary Fields} \label{sec:prefields}
The preliminary fields were chosen based on the following criteria: 1) the CLU-\ha~images must contain \ha~filter pairs taken on the same night and have observations in all four filters to ensure a complete analysis, 2) the fields must have SDSS coverage to provide a list of galaxies with known redshifts, and 3) a declination close to $30^{\circ}$ to facilitate spectroscopic follow-up from Palomar Observatory. The basic properties of the \nfields preliminary fields are listed in Table~\ref{tab:prelim}.

In addition, we have chosen to include one field which contains a galaxy cluster. The field labeled ``p3967" in Table~\ref{tab:prelim} is spatially coincident with the Coma cluster whose redshift falls in the wavelength range of our third narrow-band filter (\ha3). We include this field to robustly test the limits of our selection methods as this cluster contains hundreds of galaxies whose \ha~EWs span a large range from ellipticals with no emission lines to star bursts with strong emission lines \citep{mahajan10}. However, the majority of the preliminary fields occupy relatively sparse sections of the sky (i.e., outside the Galactic plane and not coincident with any galaxy clusters). Each pointing in our \ha~survey covers 7.26~deg$^2$ on the sky; thus our \nfields preliminary fields cover a total of 101.6~deg$^2$.


\subsection{Candidate Selection Criteria}
The identification of emission-line candidates follows the general methods of previous emission-line galaxy surveys using narrow-band filters \citep[e.g.,][]{bunker95,pascual01,fujita03,geach08,sobral09,ly10,jlee12,sobral12,stroe15}. The photometry used for the selection process are taken from the Source Extractor catalogs. We use the 5 pixel diameter aperture photometry for all selection criteria. In addition, we require a candidate to have a 5$\sigma$ detection in the ``On" band and set non detections to the 5$\sigma$ upper limit (see Table~\ref{tab:prelim}). 


\begin{figure*}
  \begin{center}
  \includegraphics[scale=0.8]{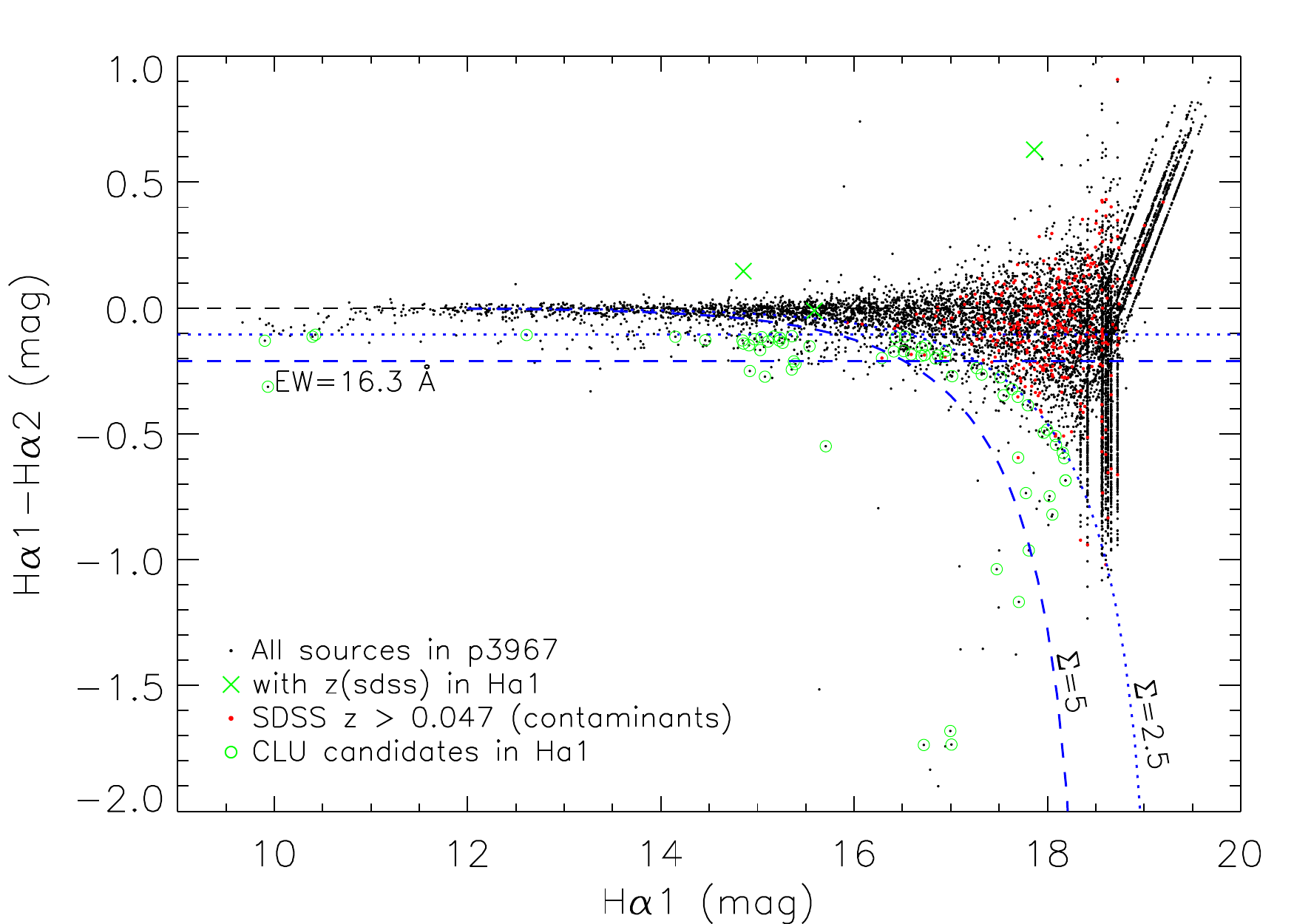}
  \caption{The \ha~color-magnitude diagram of objects in the field labeled ``p3967"}, where the y-axis represents the ``On-Off'' color and the x-axis is the ``On'' magnitude. The black points represent all CLU-\ha~sources, the red dots represent contaminant galaxies, the green X's represent known galaxies with a redshift that fall in the wavelength range of the \ha1 filter, and the green circles represent the galaxy candidates selected. The dotted and dashed lines represent the $\Sigma$ cuts of 2.5 and 5, respectively. The ``On-Off" color defined by the standard deviation of bright stars is converted into an \ha~EW limit and is labeled for the $\Sigma=5$ cut (horizontal dashed line) with a value of 16.3~\AA. The curved lines are derived from a signal-to-noise criteria where the ``On-Off" color scatter increases towards fainter magnitudes. Sources with colors that exceed the $\Sigma$ thresholds and are not selected as candidates were identified as contaminants (e.g., stars; see \S\ref{sec:cont}).
   \label{fig:ha1cand}
   \end{center}
\end{figure*}  

\begin{figure}
  \begin{center}
  \includegraphics[scale=0.5]{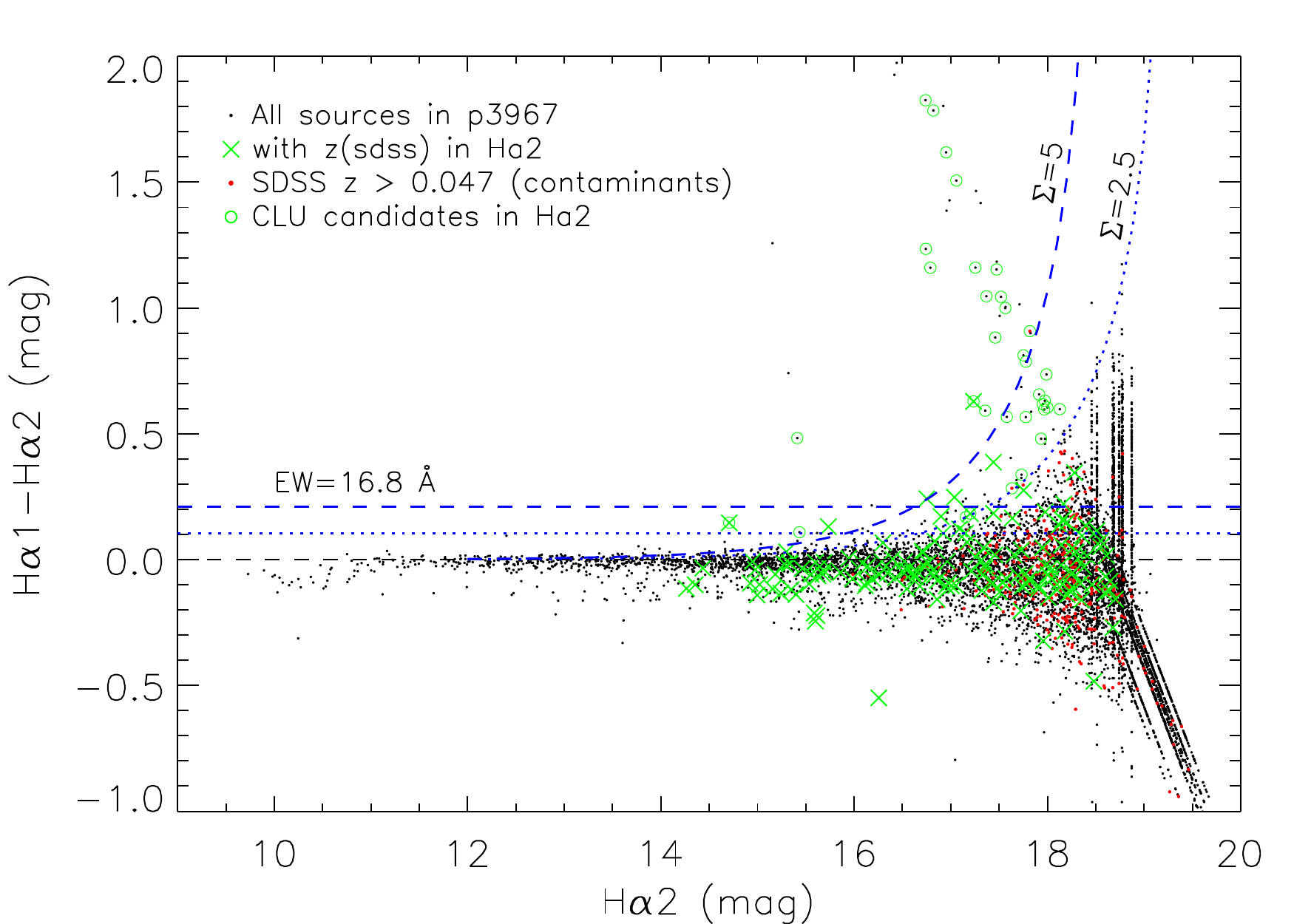}
  \caption{The \ha~color-magnitude diagram used for emission-line sources in the \ha2 filter similar to that in Figure~\ref{fig:ha1cand}. The ``On-Off" color defined by the standard deviation of bright stars is converted into an \ha~EW limit and is labeled for the $\Sigma=5$ cut (horizontal dashed line) with a value of 16.8~\AA.}
   \label{fig:ha2cand}
   \end{center}
\end{figure}

Emission-line candidates are initially selected based on the significance of the excess in ``On-Off" color quantified by the parameter $\Sigma$, where the flux is greater in one filter (the ``On" filter) due to the presence of an emission line compared to the corresponding continuum filter (the ``Off" filter). Each filter pair (\ha1/\ha2 and \ha3/\ha4) provides both the ``On" and ``Off" photometry. For example, \ha2 is used as the ``Off" filter for \ha1 selection, while \ha1 is used as the ``Off" filter for \ha2 selection.

The significance of the narrow-band colors ($\Sigma$) is defined as the number of standard deviations between the color excess (``On-Off") in counts and the random scatter of counts expected for a source with zero color (Bunker et al. 1995). This can be expressed as:
 \begin{equation}
     \Sigma = \frac{c_{\rm on} - c_{\rm off}}{ \delta},
 \end{equation}
 
\noindent where $c_{\rm on,off}$ are the counts for the ``On,Off" filters and $\delta$ is the sky fluctuations in both images combined in quadrature. The combined photometric uncertainty is expressed as:

\begin{equation}
  \delta = \sqrt{\pi r^{2} (\sigma_{\rm{on}}^2 + \sigma_{\rm{off}}^2)};
\end{equation}

\noindent where $\sigma_{\rm{on}/\rm{off}}$ represent the median sky count fluctuations in 100 sky regions for the ``On" and ``Off" images and $r$ is the radius of the photometric aperture. This criteria is similar to a standard signal-to-noise selection and can be expressed in terms of measured colors and magnitudes via:

\begin{equation}
\rm{m}_{off}-m_{on} = -2.5 log(1 - \Sigma\delta10^{-0.4(ZP-\rm{m}_{on})}) 
\end{equation}

\noindent where $\rm{m}_{on/off}$ are the calibrated magnitudes in the ``On" and ``Off" filters and $ZP$ is the photometric zero-point for the ``On" filter. 

\begin{figure}
  \begin{center}
  \includegraphics[scale=0.5]{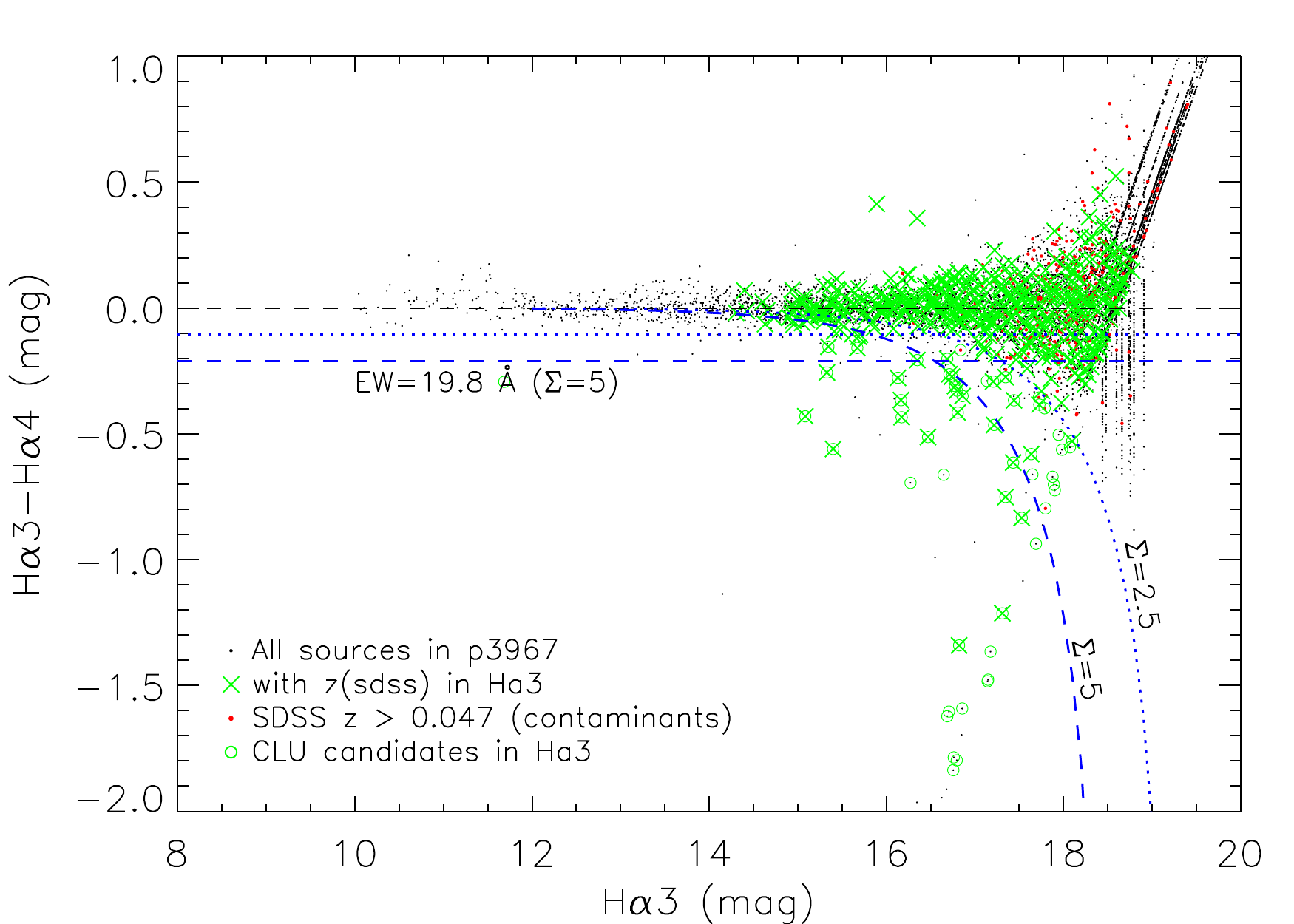}
  \caption{The \ha~color-magnitude diagram used for emission-line sources in the \ha3 filter similar to that in Figure~\ref{fig:ha1cand}. The overabundance of galaxies for this filter is due to the Coma cluster which is dominated by early-type galaxies with little on-going star formation. The lack of star formation in these galaxies is evident from the lack of significant \ha~``On-Off'' colors. The ``On-Off" color defined by the standard deviation of bright stars is converted into an \ha~EW limit and is labeled for the $\Sigma=5$ cut (horizontal dashed line) with a value of 19.8~\AA.}
   \label{fig:ha3cand}
   \end{center}
\end{figure}

\begin{figure}
  \begin{center}
  \includegraphics[scale=0.5]{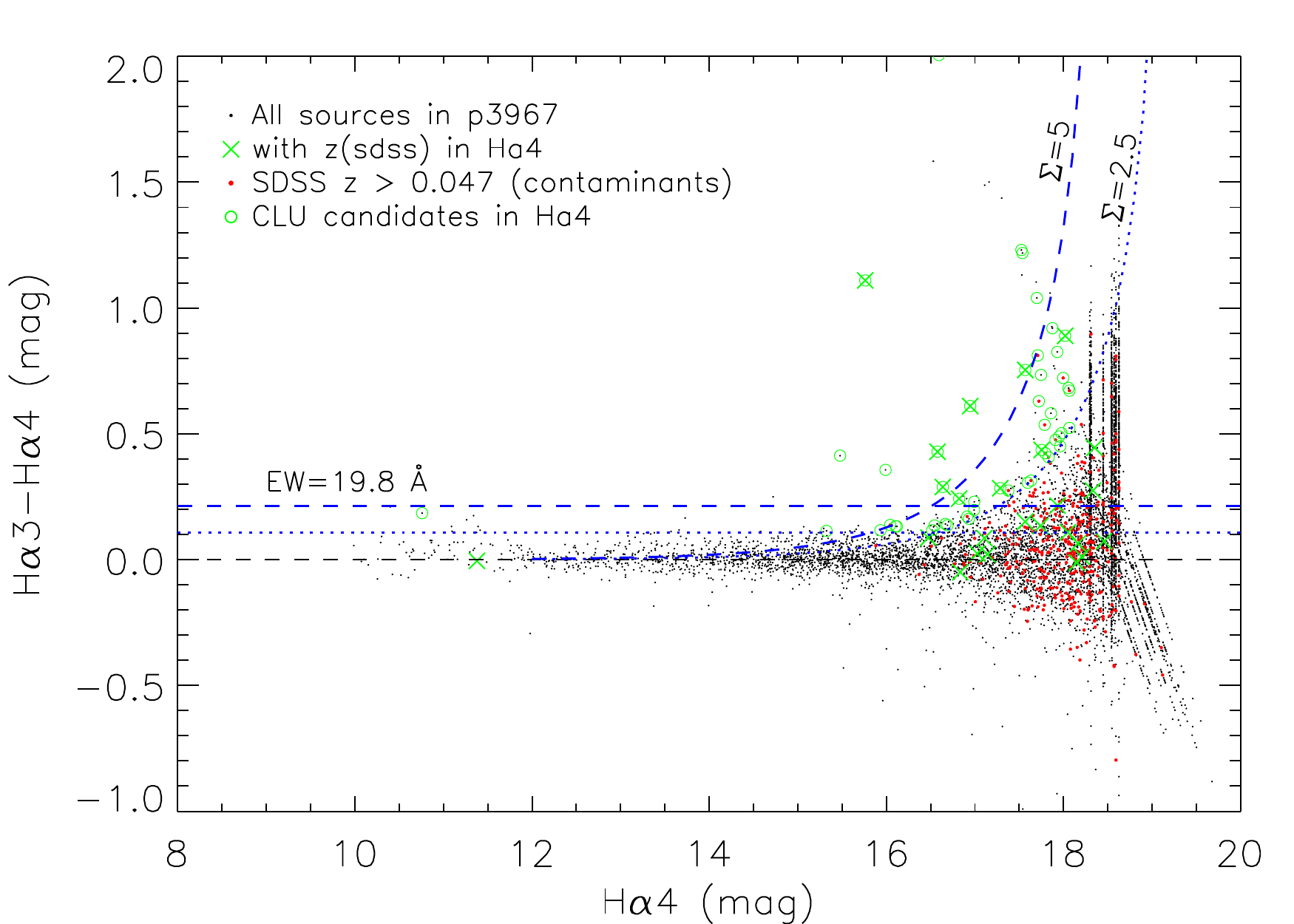}
  \caption{The \ha~color-magnitude diagram used for emission-line sources in the \ha4 filter similar to that in Figure~\ref{fig:ha1cand}. The ``On-Off" color defined by the standard deviation of bright stars is converted into an \ha~EW limit and is labeled for the $\Sigma=5$ cut (horizontal dashed line) with a value of 19.8~\AA.}
   \label{fig:ha4cand}
   \end{center}
\end{figure}

The relationship between $\Sigma$, color, and magnitude is represented as curved lines in a color-magnitude diagram and are graphically presented in Figures~\ref{fig:ha1cand}$-$\ref{fig:ha4cand} for each \ha~filter in one field (``p3967"). While all four of these figures are similar in form, we enlarge the first figure for clarity. In each of these figures the black points represent all sources in the field, the red dots represent galaxies with a spectroscopic redshift (from SDSS) greater than the redshift coverage of our filter set ($z>0.047$), the green X's represent galaxies with a spectroscopic redshift (from SDSS) that fall in the wavelength range of the appropriate filter, and the green circles represent the galaxy candidates selected in each filter.

Inspection of Figures~\ref{fig:ha1cand}$-$\ref{fig:ha4cand} also shows a non-zero color scatter even at brighter magnitudes. Thus, we impose a second condition based on the standard deviation of colors for bright continuum sources (i.e., stars) with magnitudes between 12 and 15~mag. The second criteria is represented as horizontal lines in Figures~\ref{fig:ha1cand}$-$\ref{fig:ha4cand}, and represent the minimum color below which we cannot accurately infer the presence of an emission line. Furthermore, since the EW of an emission line is simply the ratio of the line flux and the continuum flux density, we can calculate an EW limit based on the ``On-Off" color scatter in bright continuum sources. Following the prescription of \cite[][see Equation 7]{stroe15}, the EW can be calculated via:

\begin{equation}
    \rm{EW} = \Delta\lambda_{on}\Big(10^{-0.4\Delta m} - 1\Big),
\end{equation}

\noindent where $\Delta\lambda_{on}$ is the FWHM of the ``On" filter and $\Delta m$ is the ``On-Off" color. The EW cuts based on the ``On-Off" colors for each filter in Figures~\ref{fig:ha1cand}$-$\ref{fig:ha4cand} are labeled on the horizontal lines. 

Both of the selection criteria adopted here \citep[see,][]{bunker95,fukugita07,ly10,jlee12,sobral09,sobral12,stroe15} represent the number of standard deviations away from the expected random scatter and can be combined into a single $\Sigma$ value. For each source, the number of standard deviations is computed above each criteria (bright stars and the noise in the images), and the final $\Sigma$ is defined as the smaller of the two. For example, the $\Sigma$ of a bright source (e.g., $m_{on}\lesssim$~17~mag) will be calculated relative to the scatter in continuum sources, and the $\Sigma$ of a faint source (e.g., $m_{on}\gtrsim$~17~mag) will be calculated relative the signal-to-noise curve.

We utilize two $\Sigma$ cuts to define the extremes of our candidate selection methods. The first is a cut of $\Sigma=2.5$ which will contain the majority of all star-forming galaxies but will contain a relatively higher contamination from galaxies with no emission lines in our narrow-band filters. The second is a cut of $\Sigma=5$ which will contain a reduced fraction of all star-forming galaxies but will contain a low contamination fraction (see \S\ref{sec:galcomp}). The median EW cut and standard deviation for all filters in the \nfields preliminary fields is $7.3\pm1.1$ \AA~and $15.2\pm2.4$ \AA~for $\Sigma=2.5$ and $\Sigma=5$, respectively.

We provide an electronic version of the CLU candidates with $\Sigma$ values above 2.5. In the future, we will generate source catalogs in larger areas of the sky where $\Sigma$ values will be calculated for each source and users can apply their own cuts based upon their science requirements. For a better understanding of the success and contamination rates at different $\Sigma$ values see \S\ref{sec:galcomp}.



\subsection{Contaminant Removal} \label{sec:cont}
The two color excess cuts introduced in the last section will effectively select any source that has a significant \ha~``On-Off'' color. However, the resulting galaxy candidates can still be contaminated by continuum sources with steep blue or red continuum slopes, high-redshift galaxies with an emission or absorption line whose wavelength has been redshifted into the wavelength range spanned by one of our filters, and by cosmic rays or chip defects (e.g., hot pixels, column defects, etc.). 

Point sources with a steep blue or red continuum can be reduced by requiring that our candidates be spatially extended. As our survey is only sensitive to a distance of 200~Mpc and our angular resolution is limited to a FWHM$\sim2\arcsec$, we estimate that galaxies larger than 2~kpc at 200~Mpc will be extended in our data. Comparison to a statistically complete sample of star-forming galaxies in the local Universe where the sample is dominated by dwarf galaxies \citep[LVL;][]{kennicutt08,dale09,lee11,cook14a} reveals that the semi-major axis histogram peaks between $2-4$~kpc suggesting that we will likely detect the majority of galaxies even at our furthest distance by requiring our galaxy candidates to be extended. Similar results are found for the size of \ha~disks in galaxies out to z=0.1 \citep{dale99}. We exclude point sources based upon the recommended PS1 star/galaxy separation using PSF minus Kron $i$-band magnitudes. 

Unfortunately, the PS1 star-galaxy classifications often mislabel saturated stars as galaxies at $r$-band magnitudes brighter than 12-14~mag\footnote{https://panstarrs.stsci.edu/} leaving $\sim$300 candidates that are clearly stars. We can remove $\sim$60\% of these contaminants via a bright \ha~magnitude cut of 12~mag given the brightest quoted saturation magnitude in PS1. We note that the brightest confirmed galaxy in the preliminary fields is $\sim$13.4~mag. The remaining contaminating stars are removed via visual classification. 

The removal of cosmic rays and chip defects is achieved by requiring a source to be spatially coincident with a PS1 source. Since the PS1 detection limits (e.g., $r\sim23.2$~mag) are deeper than the CLU-\ha~single-image exposures, any cosmic ray or chip defect with no spatial overlap with a PS1 source will be easy to flag and remove. Visual inspection of sources removed from our analysis via this method show that no real sources have been removed. Furthermore, visual inspection of our galaxy candidates show no contamination from cosmic rays, but do show a small percentage of contamination from chip-column defects that randomly coincide with a PS1 source; these contaminants are easily removed via visual inspection. We note that the removal of cosmic rays and chip defects will be greatly reduced or unnecessary for future analyses using stacked images.




The final sources of contamination are high-redshift galaxies whose emission or absorption lines have been shifted into the wavelength range of our filters. It is not possible to distinguish between strong emission lines redshifted into the wavelength range of our filters and \ha~emission at lower redshift. We note that there is a dearth of strong emission lines blue-ward of our filters, where the closest lines are the \oiii~doublet near $\lambda=$5000~\AA~(z$\sim$0.3). However, galaxies with extreme \oiii~emission at z$\sim$0.3 are likely to be interesting objects (i.e., green peas) and can be used for studies of galaxy evolution and star formation. In fact, we find a newly discovered z$\sim$0.3 green pea contaminant in the preliminary fields (see \S\ref{sec:intcanddisc}) and future CLU efforts will be aimed at finding and studying more of these objects.

\begin{figure}
  \begin{center}
  \includegraphics[scale=0.5]{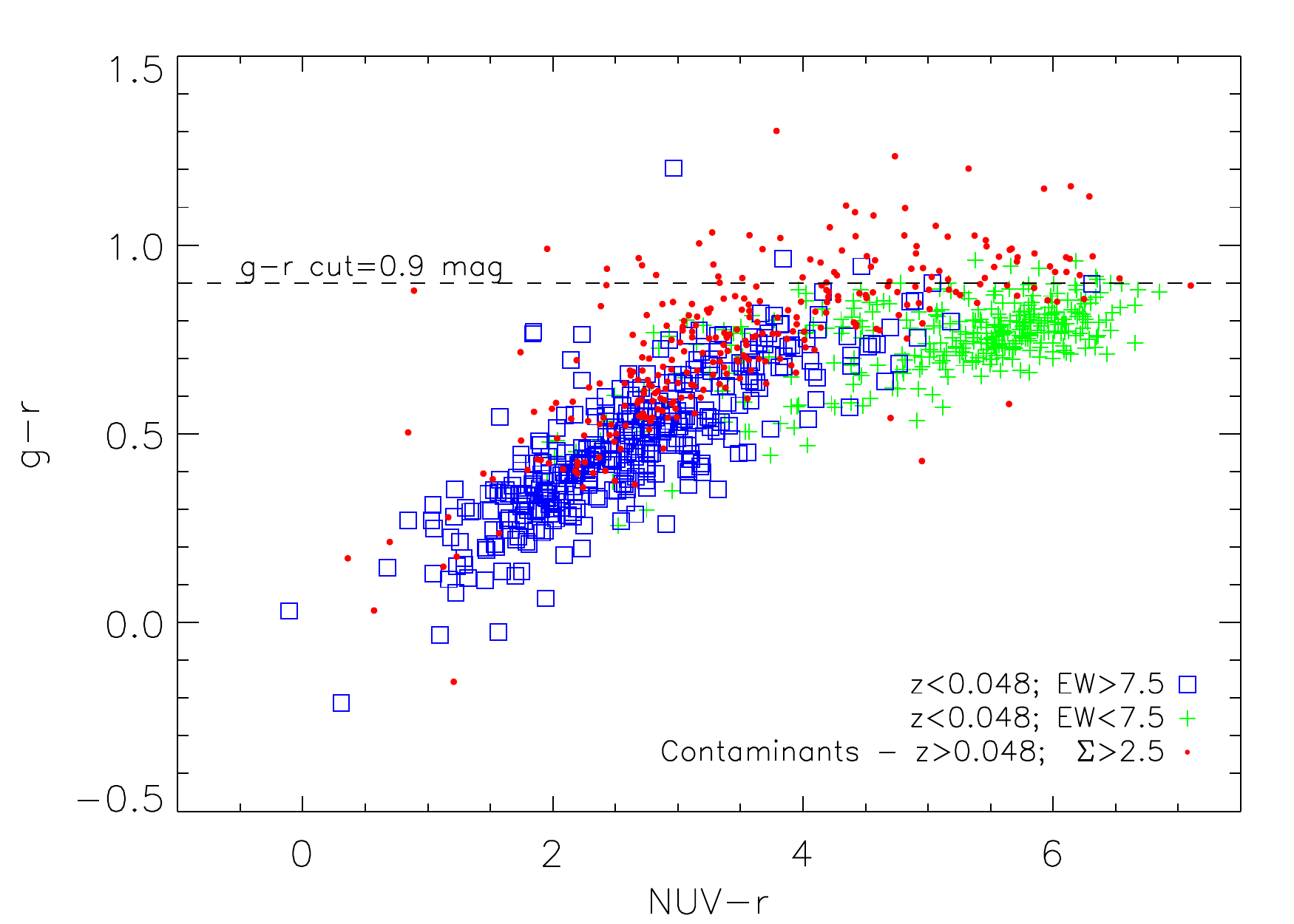}
  \caption{The $g-r$ versus NUV-r, color-color plot of all known galaxies in the preliminary fields, where the blue squares represent galaxies in our target volume with an \ha~EW greater than the $\Sigma=2.5$ limit (EW$\geq$7.5~\AA), the green pluses are all galaxies in our volume with small \ha~EWs (EW$<$7.5), and the red points are galaxy candidates outside of our target volume (z$>$0.047). A $g-r$ optical color cut removes contaminant galaxies while retaining the majority of galaxies in our volume with an EW greater than our selection limits (EW$\geq$7.5\AA). }
   \label{fig:colorcut}
   \end{center}
\end{figure}

Absorption lines in elliptical or red-sequence galaxies at intermediate redshifts are another source of contamination. Strong NaD absorption at redshifts of 0.11--0.17 can cause a lower flux in an \ha~filter relative to the flux in its pair filter resulting in a significant ``On-Off" color. Since red sequence galaxies tend to be redder than star-forming galaxies \citep[][]{salim07}, we can remove these contaminants via simple optical color cuts. Figure~\ref{fig:colorcut} shows the $g-r$ versus $NUV-r$ color-color diagram for galaxies in our preliminary fields with low and high \ha~EWs as measured via spectra from SDSS and CLU-\ha~follow-up.  We find that a $g-r$ color cut of $g-r$ greater than to 0.9~mag removes 60\% of red sequence galaxies at $z>0.047$ while removing only 4 galaxies with moderate EW and a redshift less than 0.047. 

\subsection{Spectroscopic Follow-up}
We have obtained spectra for 334 candidates  in the preliminary fields with no redshift information (except obvious stars, cosmic rays, and chip defects) with a $\Sigma$ value above 2.5. The spectra were taken with the 200 inch Hale telescope atop Palomar Mountain on multiple nights over 2016 and 2017 using the Double-Beam Spectrograph instrument \citep[DBSP;][]{dbsp} and on the 2.3 meter Wyoming Infra-Red Observatory (WIRO) with the Longslit spectrograph. The DBSP data were reduced using the \texttt{PyRAF}-based pipeline\footnote{https://github.com/ebellm/pyraf-dbsp} of \cite{bellem16} and the WIRO data were reduced with standard \texttt{IRAF} procedures. 

We took spectra of 334 galaxies where we confirm that 124 galaxies were indeed galaxies with redshifts less than 0.047, while the remainder were higher-redshift contaminants. In addition to finding new galaxies in the target volume, we also found 2 new galaxies (a Seyfert 1 galaxy and a green pea) at intermediate redshift via strong \oiii~emission lines redshifted into our filters (see \S\ref{sec:intcanddisc}).



\begin{figure*}
  \begin{center}
  \includegraphics[scale=0.65]{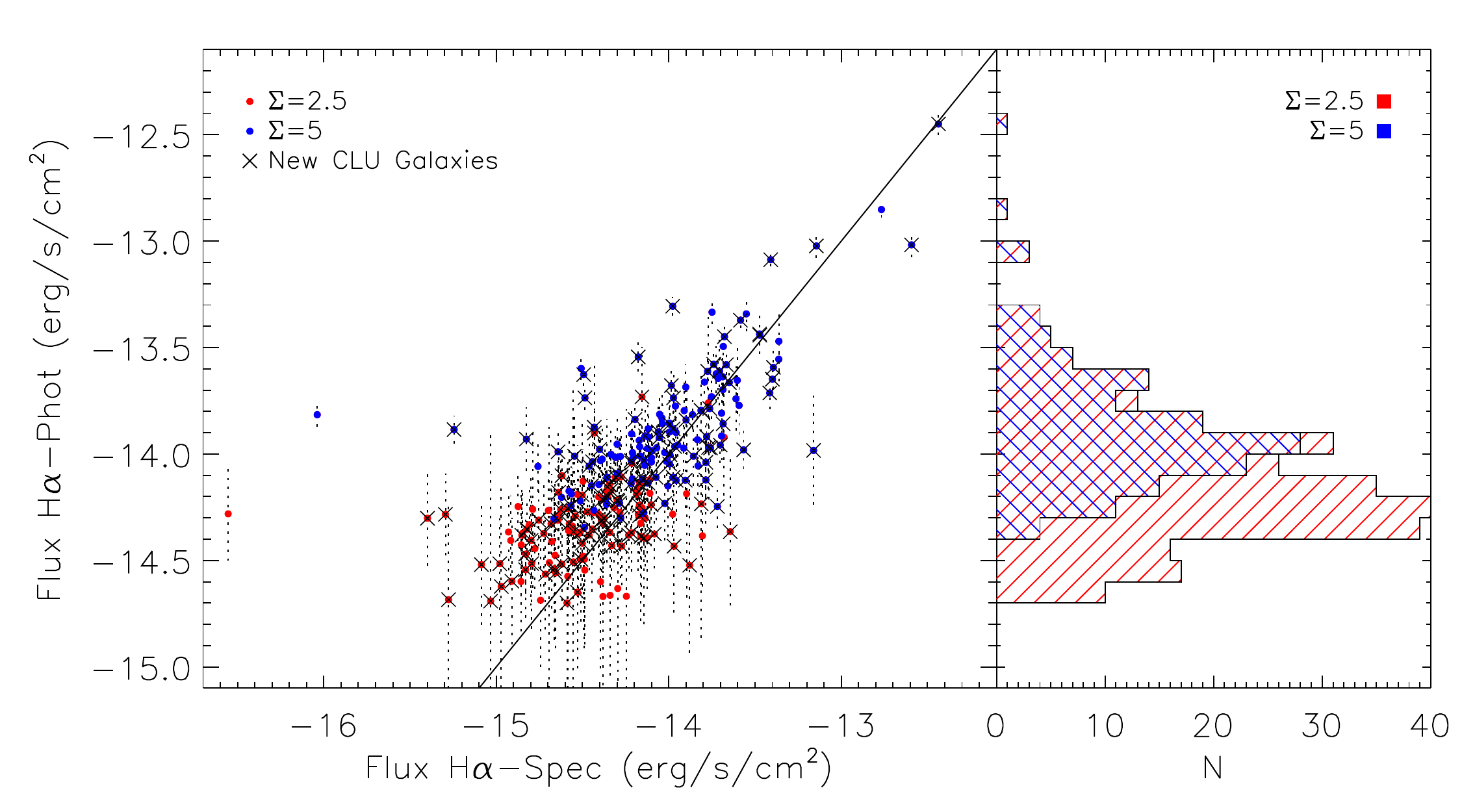}
  \caption{Left Panel) The photometrically-derived \ha~line flux versus the spectroscopically-derived \ha~line flux in the left panel. We find that the majority of the galaxies show good agreement for \ha~line fluxes between those derived from imaging and spectroscopy above a few$\times 10^{-15} \rm{erg~s^{-1}~cm^{-2}}$. Right Panel) The histogram of the $\Sigma=2.5$ and $\Sigma=5$ CLU-\ha~galaxy samples. }
   \label{fig:photspec}
   \end{center}
\end{figure*}

In the rest of this section we compare the photometric fluxes measured in our survey to with those derived from the spectroscopic measurements. In Figure~\ref{fig:photspec} we plot the photometric \ha~line flux versus the spectroscopic line flux in the left panel. The photometric \ha~line flux is measured via: 

\begin{equation}
    \rm{F_{line}(erg~s^{-1}~cm^{-2}) = \Delta\lambda_{On} (f_{On} - f_{Off})},
\label{eq:fline}
\end{equation}

\noindent where $\Delta\lambda_{\rm{On}}$ is the FWHM of the ``On" filter, f$_{\rm{On}}$ is the flux density of the ``On" filter, and f$_{\rm{Off}}$ is the flux density of the ``Off" filter. The flux densities of the ``On" and ``Off" filters are calculated via: 

\begin{equation}
    \rm{f_{on,off}(erg~s^{-1}~cm^{-2} \text{\AA}^{-1}) = \frac{c}{\lambda^2_{on,off}} 10^{-0.4(m_{on,off} + Zpt_{AB})} },
\label{eq:fden}
\end{equation}

\noindent where $c$ is the speed of light, Zpt$_{\rm{AB}}=48.59$, and $\lambda_{\rm{on,off}}$ are the central wavelengths of the ``On" and ``Off" filters. 

There is reasonable agreement for the majority of galaxies above a few$\times 10^{-15} \rm{erg~s^{-1}~cm^{-2}}$ for line fluxes derived from our \ha~photometry and those derived from spectroscopy. However, there is a small offset (a few tenths of a dex) at lower line fluxes where the photometrically derived fluxes are systematically higher than the spectroscopic fluxes. This offset is likely due to a mismatch in the apertures used to derive the line fluxes, and from contaminating flux from [NII] in the \ha~imaging. We use a 5$\arcsec$ aperture in the \ha~imaging, while the spectroscopic line fluxes are derived from either a 3$\arcsec$ fiber from SDSS or a 2$\arcsec$ slit from Palomar observations. In addition, the [NII]/\ha~ratio for star-forming galaxies in the local volume can range from 0.05 to 0.5 with an average of 0.2 \citep{kennicutt08}. These ratios translate into an average and max offset in Figure~\ref{fig:photspec} of 0.1 and 0.3~dex, respectively, and is likely the major source of the offsets seen in Figure~\ref{fig:photspec}. We also note that the strength of the [NII] contamination depends on the redshift of the galaxy where the [NII] lines can be shifted out of the ``On" filter if the \ha~line is near the filter edge.




\subsection{Comparison to Known Galaxies} \label{sec:galcomp}
In this section, we evaluate our selection criteria by crossmatching our galaxy candidates to currently cataloged galaxies with spectroscopic redshifts available from NED, SDSS, or our CLU-\ha~followup in the preliminary fields. Since our selection methods rely on the presence of a moderately strong \ha~emission line, we expect that we should recover a large fraction of the \ha~flux for galaxies in these fields, but recover a relatively low fraction by number of these galaxies due to the numerous elliptical galaxies in these fields (i.e., in the Coma cluster). 


First, we examine the composition of galaxy catalogs produced when using different $\Sigma$ cuts: 2.5, 3, 4, 5, 6, and 7. Table~\ref{tab:sigexp} shows the total number of galaxy candidates, contaminant galaxies (i.e., those with no emission line found in our narrow-band filters), galaxies found in the volume of our survey (z$<$0.047), galaxies with an emission line found in the correct filter (i.e., an emission line has been confirmed in the filter identified by our narrow-band colors), and galaxies with new distance measurements (i.e., with no previous distance measurements) now confirmed in the local Universe by our survey in the correct filter. 

We find that the percentage of contaminant galaxies falls significantly above a $\Sigma$ cut of 3, and that the percentage of galaxies with an emission line found in the correct filter is $\approx90\%$ above a $\Sigma$ cut of 5. Thus, we conclude that a $\Sigma$ cut above 5 produces a high fidelity galaxy catalog. However, a $\Sigma$ cut above 2.5 does produce a catalog with a higher number of galaxies with an emission line found in the correct filter, but with a high contamination percentage. In this and future publications we will release galaxy candidates with $\Sigma$ cuts above 2.5, and as a consequence we focus the text in the rest of this section and the paper on the galaxy catalogs from two extreme cuts ($\Sigma>2.5$ and $\Sigma>5$) but provide the statistics for all $\Sigma$ cuts in the tables.

\begin{table*}
\begin{center}
{Color Excess ($\Sigma$) Statistics}\\
\begin{tabular}{cccccccc}
\hline
\hline

$\Sigma$	& N Total	    & N Contaminant		& N In		   &\vline& N in Correct          & N New Distances	  \\
		    & Candidates	& Galaxies		    & Volume	   &\vline& \ha~Filter		      & \& in Correct \ha~Filter		  \\
(\#)		& (\#)		        & (\#,\% of candidates)	& (\#) &\vline& (\#,\% of candidates) & (\#)		  \\
\hline
$\geq$2.5	&663		&405~(61.1\%)	       &339	       &\vline&258~(38.9\%)		&90	 \\
$\geq$3.0	&399		&174~(43.6\%)	       &265	       &\vline&225~(56.4\%)		&78	 \\
$\geq$4.0	&218		&~49~(22.5\%)	       &180	       &\vline&169~(77.5\%)		&56	 \\
$\geq$5.0	&147		&~18~(12.2\%)	       &135	       &\vline&129~(87.8\%)		&41	 \\
$\geq$6.0	&106		&~~7~( 6.6\%)	       &100	       &\vline&~99~(93.4\%)		&32	 \\
$\geq$7.0	&~86		&~~3~( 3.5\%)	       &~83	       &\vline&~83~(96.5\%)		&23	 \\

\hline
\end{tabular}\\
\end{center}
\caption{The composition of our galaxy candidates when using different $\Sigma$ cuts, where the columns are from left-to-right: the minimum $\Sigma$ for the sample, the total number of all candidates, the number of galaxies with no emission-lines in our narrow-band filters, the number of galaxies found in the survey volume (z$<$0.047), the number of galaxies whose \ha~emission-line is located in the narrowband filter where the galaxy was identified, and the number of galaxies with no previous distance information found in CLU-\ha.}
\label{tab:sigexp}
\end{table*}

Next, we quantify the success rates of our selection methods both by number and by \ha~flux in Table~\ref{tab:knowngal} via a comparison to galaxies with known redshifts in our survey volume and spatially located in our preliminary fields. The simplest comparison we can make is by number where we successfully recover 25.1\% and 12.9\% for $\Sigma$ cuts of 2.5 and 5, respectively (column 2). However, There are 1226 known galaxies in our preliminary fields with known redshifts, where 722 have an \ha~EW lower than 7.5~\AA~($\approx$60\%). The majority of these galaxies are located in a single pointing that is coincident with the Coma cluster (i.e., `p3967'). Thus a low success rate by number is not surprising given the large fraction of elliptical galaxies in our preliminary fields. If we limit the sample to galaxies with an \ha~EW larger than the limits for the $\Sigma=2.5$, we find that we recover 54.5\% and 29.8\% by number for $\Sigma$ cuts of 2.5 and 5, respectively (column 4).

Since our selection methods are based on the strength of the \ha~emission line, it is more appropriate to quantify our success rates via \ha~flux. Column 3 of Table~\ref{tab:knowngal} shows the fraction of \ha~flux captured by each of the $\Sigma$ cuts, where we recover 82.5\% and 67.8\% for $\Sigma$ cuts of 2.5 and 5, respectively. If we limit the sample of known galaxies to those with a moderate \ha~EW ($>7.5$~\AA), we recover 86.7\% and 72.0\% by \ha~flux for $\Sigma$ cuts of 2.5 and 5, respectively. The success rates of other $\Sigma$ samples can be found in column 5 of Table~\ref{tab:knowngal}). 

\begin{table*}
\begin{center}
{Known Galaxies}\\
\begin{tabular}{cccccccc}
\hline
\hline

Sample		&\vline& N		& Sum \ha~Flux		&\vline& N 			& Sum \ha~Flux		\\
 		&\vline& (\#)		& (erg/s/cm$^2$)	&\vline& (\#)			& (erg/s/cm$^2$)	\\
		&\vline&        	&           		&\vline& (EW$>$7.5~\AA)		& (EW$>$7.5~\AA)	\\
		&\vline&(N,\% of N)	& (Flux,\% of Flux)	&\vline& (N,\% of N)		& (Flux,\% of Flux)	\\
\hline
All 		&\vline& 1173		& 4.05e-12		&\vline& 477			& 3.80e-12		\\
\hline
$\Sigma\geq2.5$	&\vline&295~(25.1\%)	&3.34e-12~(82.5\%)	&\vline&260~(54.5\%)		&3.30e-12~(86.7\%)	\\  
$\geq$3.0	&\vline&259~(22.1\%)	&3.23e-12~(80.0\%)	&\vline&232~(48.6\%)		&3.20e-12~(84.1\%)	\\
$\geq$4.0	&\vline&192~(16.4\%)	&2.93e-12~(72.5\%)	&\vline&179~(37.5\%)		&2.92e-12~(76.9\%)	\\
$\geq$5.0	&\vline&151~(12.9\%)	&2.74e-12~(67.8\%)	&\vline&142~(29.8\%)		&2.74e-12~(72.0\%)	\\
$\geq$6.0	&\vline&120~(10.2\%)	&2.55e-12~(63.1\%)	&\vline&112~(23.5\%)		&2.55e-12~(67.1\%)	\\
$\geq$7.0	&\vline&102~(~8.7\%)	&2.34e-12~(57.7\%)	&\vline&~96~(20.1\%)		&2.34e-12~(61.4\%)	\\

\hline
\end{tabular}\\
\end{center}
\caption{The success rates for galaxy catalogs generated with different $\Sigma$ values. The first column gives the $\Sigma$ values used to generate different galaxy samples, and the following columns give the success rates by number and \ha~flux. The vertical lines are used to separate the success rates given no EW cut and an EW$>$7.5~\AA. We find that our $\Sigma=2.5$ sample recovers $\sim$25\% by number and $\sim$82\% by \ha~flux compared to all galaxies, and that we recover a larger fraction by number and \ha~flux when comparing to galaxies with an EW greater our selection limits.}
\label{tab:knowngal}
\end{table*}

\begin{figure}
  \begin{center}
  \includegraphics[scale=0.5]{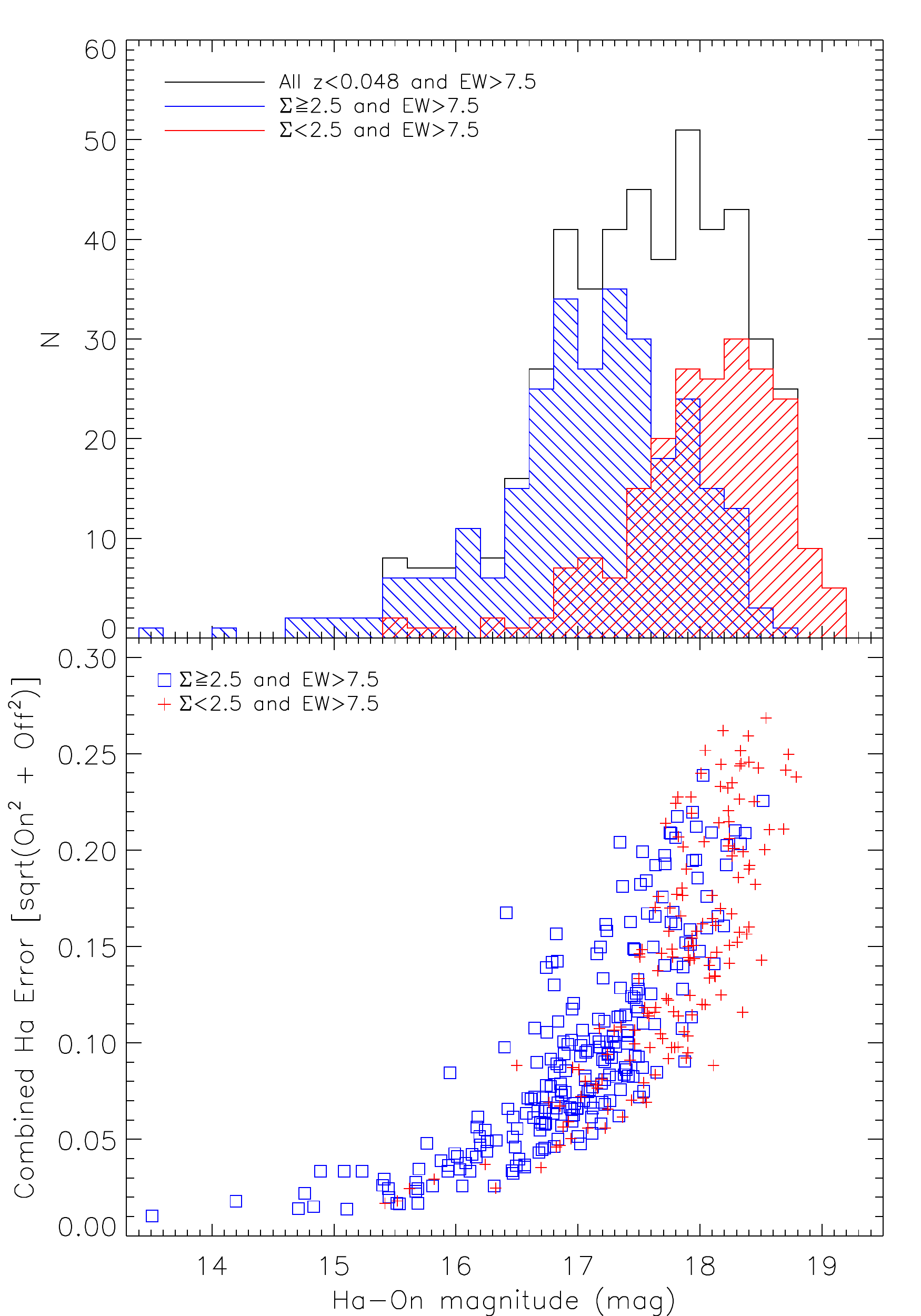}
  \caption{The flux properties all galaxies with a known redshift from NED, SDSS, or our spectroscopic follow up in our volume (z$\leq$0.047) with an \ha~EW$\geq$7.5~\AA. The blue symbols and histograms represent a subset with $\Sigma\geq2.5$ (candidates) and the red symbols and histograms represent a subset with $\Sigma<2.5$ (non-candidates). The top panel shows that the galaxies with a large enough EW to be selected in our survey but not selected as candidates tend to be the fainter objects. The bottom panel shows that these galaxies not selected as candidates also tend to have increased photometric scatter which require a stronger emission line to be selected by our signal-to-noise selection criteria.}
   \label{fig:haonallgal}
   \end{center}
\end{figure}

Next, we examine why galaxies with an EW greater than 7.5~\AA~are not selected as candidates (i.e., our false-negative rate). The top panel of Figure~\ref{fig:haonallgal} shows the \ha~``On" magnitude histogram for these galaxies with $\Sigma>=2.5$ and a $\Sigma<2.5$, and find that the galaxies missed in our survey that should be selected tend to be fainter. In these fainter galaxies, the significance of their narrow-band colors is reduced by the increased photometric noise and thus require a stronger EW line to be selected by our methods (See the bottom panel of Figure~\ref{fig:haonallgal}). We note that the photometric errors will be reduced in the final stacked CLU-\ha~images resulting in higher success rates in future analyses.

\subsection{Survey Limits}
In this section we explore the limits of our selection criteria and quantify them by \ha~flux. We find that our survey is 90\% complete at an \ha~flux of $1\times10^{-14}~\rm{erg~s^{-1}~cm^{-2}}$ and $4\times10^{-14}~\rm{erg~s^{-1}~cm^{-2}}$ for the $\Sigma$=2.5 and $\Sigma$=5, respectively.

\begin{figure}
  \begin{center}
  \includegraphics[scale=0.5]{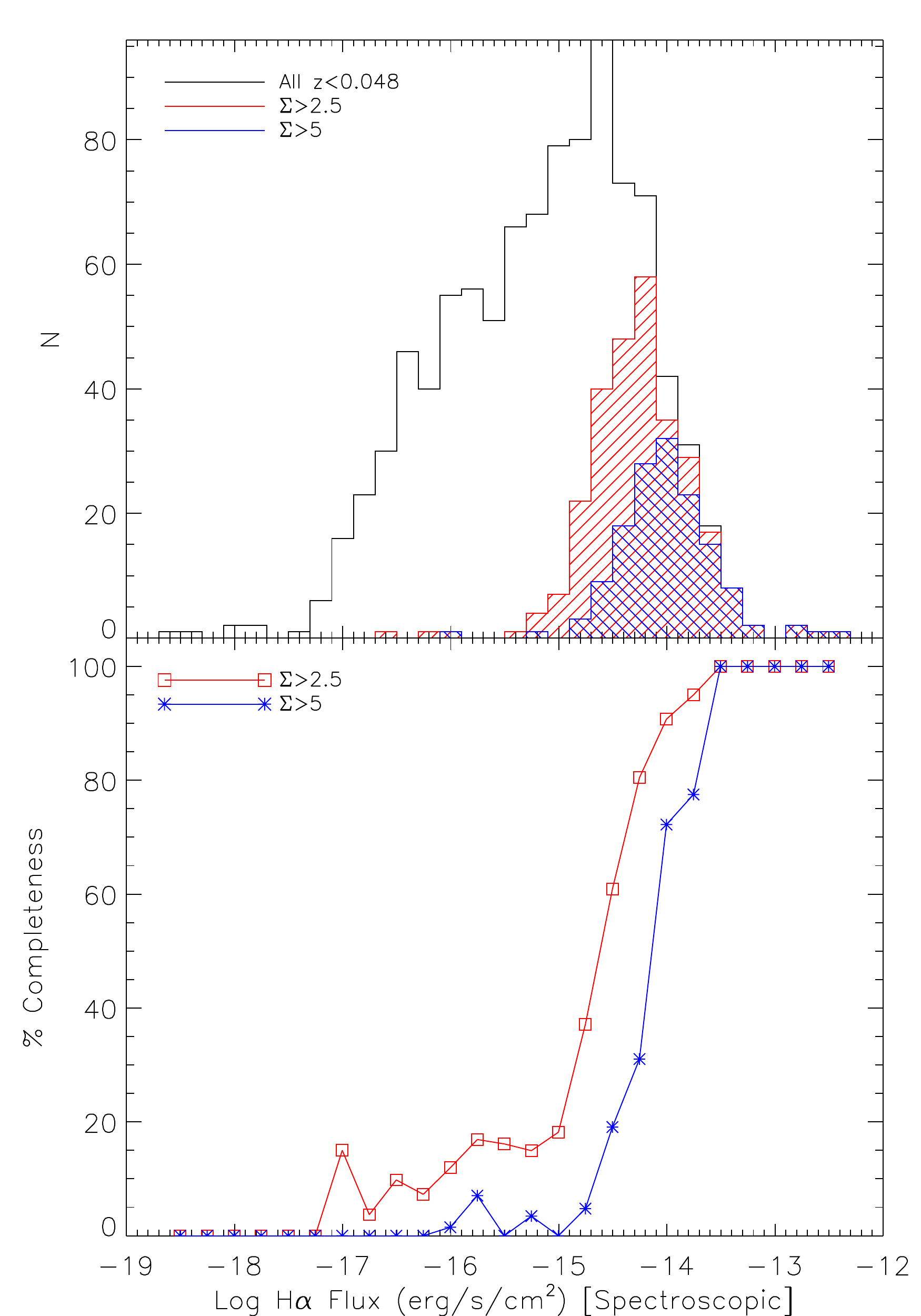}
  \caption{The \ha~flux histogram and completeness for the CLU-\ha~galaxies compared to all galaxies with a known redshift from NED, SDSS, or our spectroscopic follow up in the preliminary fields. In the top panel the unfilled, red-filled, and blue-filled histograms represents all galaxies in the volume, all galaxies in the volume with $\Sigma<2.5$ (non-candidates), CLU-\ha~galaxies ($\Sigma>=2.5$). The bottom panel shows the completeness percentage compared to all known galaxies in each bin of \ha~flux versus \ha~flux. The CLU-\ha~galaxies are 90\% complete at $\sim 1\times10^{-14}~\rm{erg~s^{-1}~cm^{-2}}$ and $\sim \times10^{-14}~\rm{erg~s^{-1}~cm^{-2}}$ for $\Sigma>2.5$ and  $\Sigma>5$, respectively.}
   \label{fig:fluxhist}
   \end{center}
\end{figure}

To quantify the limits of this survey, we present the \ha~flux histograms for our confirmed galaxy candidates and all galaxies in the preliminary fields with spectroscopic information from either SDSS or our own follow-up. In the top panel of Figure~\ref{fig:fluxhist}, the unfilled, red-filled, and blue-filled histograms represents all galaxies with spectroscopy and our CLU-\ha~galaxies with $\Sigma>2.5$ and with $\Sigma>5$. The completeness (defined as the fraction of all galaxies recovered in our survey in each line flux bin) is illustrated in the bottom panel of Figure~\ref{fig:fluxhist}. We find that we recover 90\% and 50\% of known galaxies at $F_{\rm{line}}=1\times10^{-14}$ and $3\times10^{-15}~\rm{erg~s^{-1}~cm^{-2}}$ for $\Sigma>2.5$ and $F_{\rm{line}}=4\times10^{-14}$ and $1\times10^{-14}~\rm{erg~s^{-1}~cm^{-2}}$ for $\Sigma>5$.

\begin{figure*}
  \begin{center}
  \includegraphics[scale=0.55]{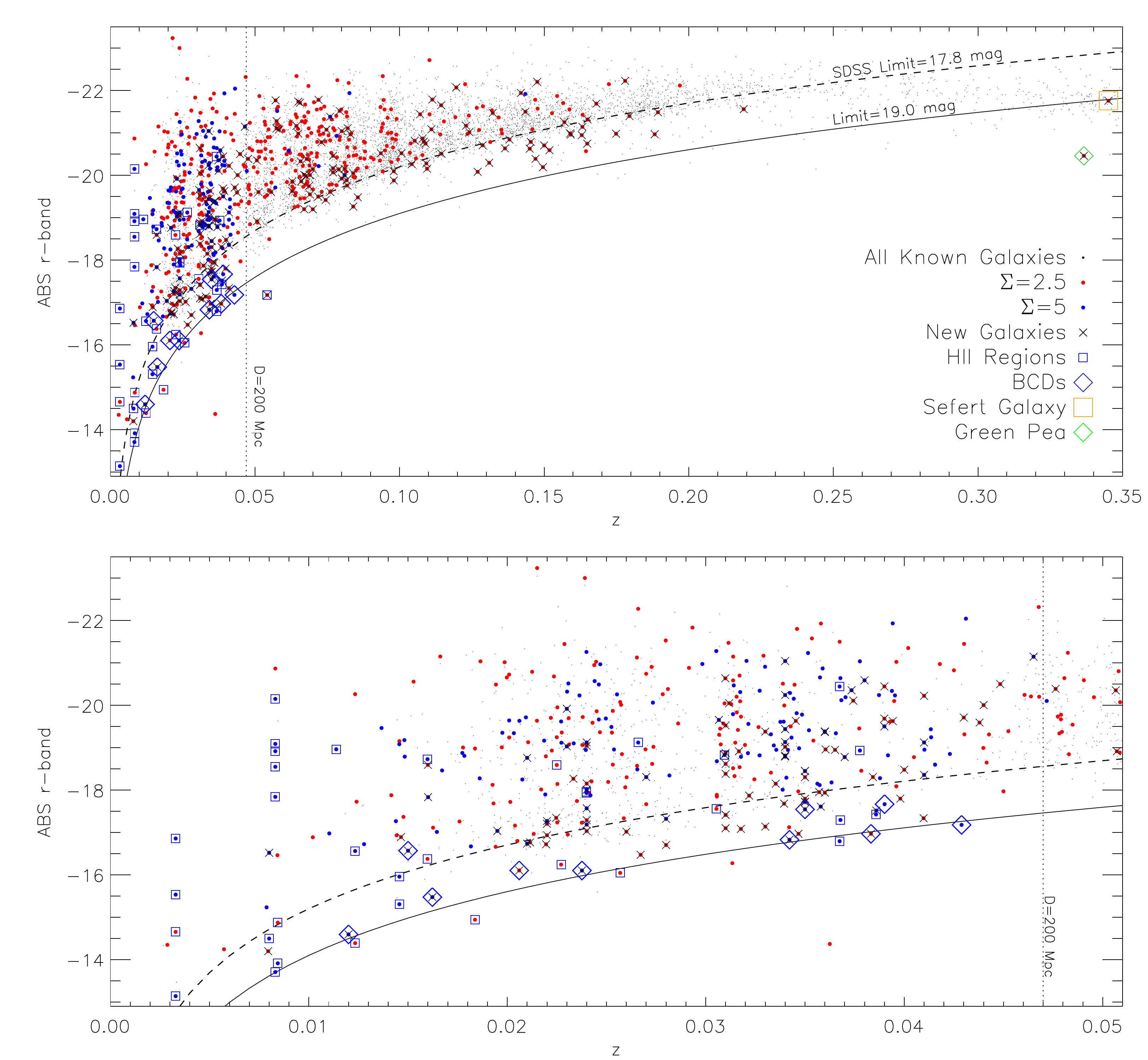}
  \caption{The distribution of absolute $M_{r}$-band magnitude and redshift for all galaxies with a known redshift from NED, SDSS, or our spectroscopic follow up and CLU-\ha~galaxies. The top panel presents the full redshift range of the CLU-\ha~candidates while the bottom panel is a zoom-in of the local Universe (z$<$0.05). The vertical dotted line represents the maximum redshift limit of our survey (i.e., z=0.047), the dashed- and solid-curved lines represent the relationship between an apparent magnitude limit of 17.8 (i.e., the limit of the SDSS spectroscopic galaxy survey) and 19~mag, respectively, with redshift. The small black, larger red, larger blue dots, and X's represent all known galaxies, $\Sigma=2.5$ CLU-\ha~galaxies, $\Sigma=5$ CLU-\ha~galaxies, and CLU-\ha~galaxies with no previous distance information, respectively.}
   \label{fig:absr_z}
   \end{center}
\end{figure*}

Next, we explore the redshift-magnitude distribution of the two CLU-\ha~galaxy catalogs in Figure~\ref{fig:absr_z} to illustrate the resulting CLU-\ha~galaxy catalogs and their magnitudes. The top panel presents the full redshift range of our galaxy candidates while the bottom panel is a zoom-in of the local Universe (z$<$0.05) In addition, the dashed- and solid-curved lines represent the relationship between an apparent magnitude limit of 17.8 (i.e., the limit of the SDSS spectroscopic galaxy survey) and 19~mag, respectively, with redshift.

The density of all galaxies with spectroscopic redshifts drops off significantly below the $r$-band magnitude limit of the SDSS spectroscopic galaxy survey (r=17.8~mag). In addition, the density of CLU-\ha~galaxies also significantly decreases near that of SDSS suggesting that our selection methods result in an effective magnitude limit similar to the SDSS limit. Our methods do not select bright galaxies with weak emission lines, but do find 69 CLU-\ha~galaxies with $r$-band magnitudes fainter than 18~mag with large \ha~EWs (a median value of 45~\AA) illustrating that a our selection methods can detect fainter galaxies with strong emission lines. We note that the "X" symbols in Figure~\ref{fig:absr_z} represent galaxies with no previous distance information in NED or SDSS.

\section{Results}
In this section we provide a description of the CLU-\ha~galaxies found in the preliminary fields and highlight two sub-samples composed of: 1) galaxy candidates with lower color significance ($\Sigma\geq$2.5) that will contain the majority of target galaxies (i.e., those in the survey volume with \ha~emission lines) but with high contamination; 2) galaxy candidates with higher color significance ($\Sigma\geq$5) that contain a reduced fraction of target galaxies but with low contamination.  First, we present examples of galaxies whose redshifts have been measured for the first time in this survey. Then, we examine the observable properties of the galaxies in our CLU-\ha~catalog and derive their physical properties. Finally, we compare the CLU-\ha~galaxies found in the preliminary fields to catalogs of emission-line galaxies from previous blind emission-line surveys.


We note that the \ha~galaxy catalogs and analyses presented here utilize single exposure images; not the final stacked images. The stacked CLU-\ha~images will produce a galaxy catalog with fainter detection limits and a higher completeness than the preliminary fields presented here. The stacking methods and analysis of the CLU-\ha~images are the subject of a future study and beyond the scope of this paper.


\subsection{Example New Galaxies}\label{sec:excand}

\begin{figure*}
  \begin{center}
  \includegraphics[scale=0.8]{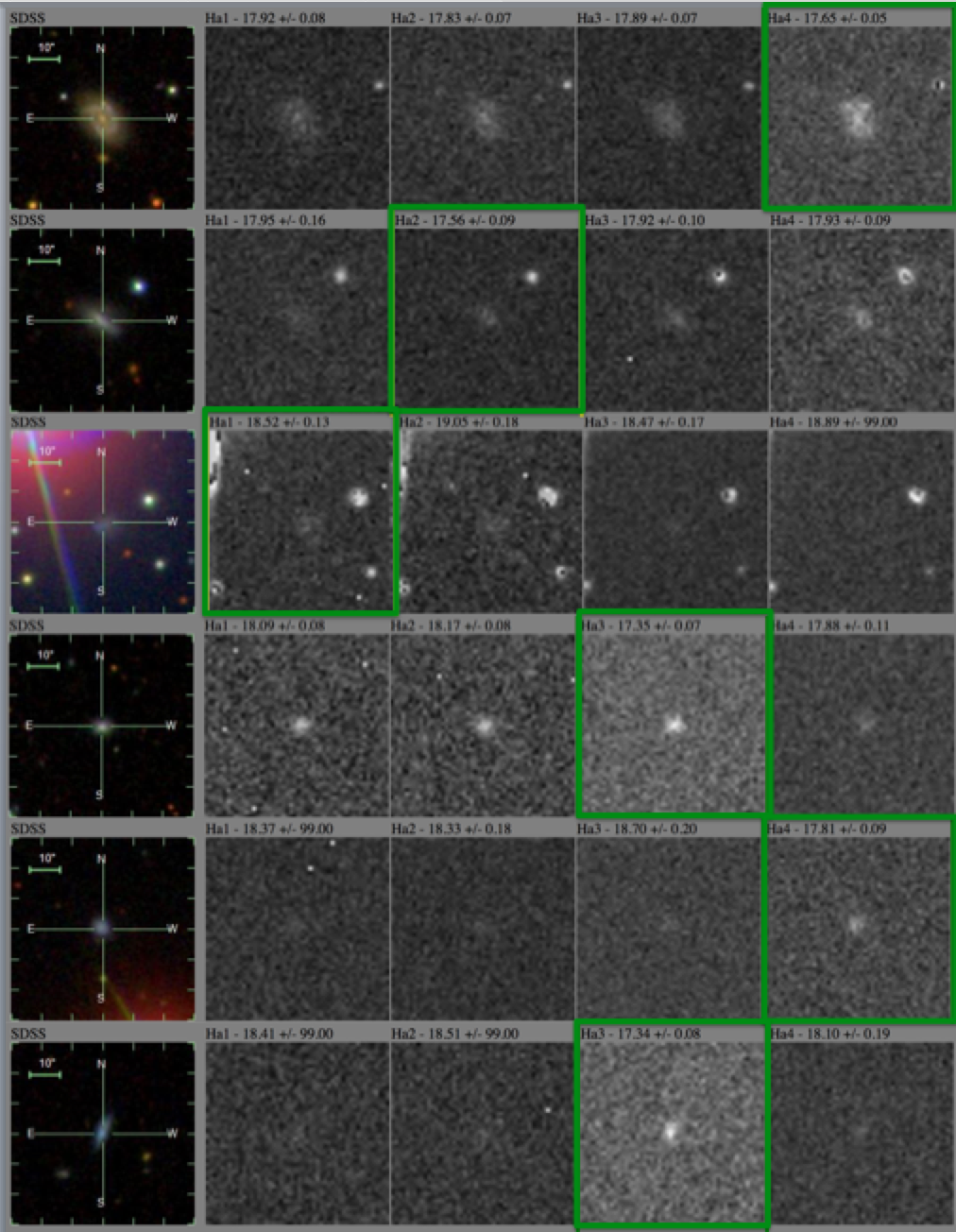}
  \caption{A mosaic of a representative sample of newly discovered galaxies in the local Universe, where (from left-to-right) the image cutouts are the SDSS gri composite and followed by the four \ha~filters (\ha1-\ha4). The \ha~magnitudes and errors are listed above each \ha~cutout and the green boxes indicate which filter the galaxy was identified (i.e., the brightest). In addition, the galaxies are sorted by \ha~EW, where the lowest EW galaxies are at the top (EW$\sim$10~\AA) and the highest are at the bottom (EW$\sim$100~\AA).}
   \label{fig:mosaic}
   \end{center}
\end{figure*}

Here we present examples of CLU-\ha~galaxies with no previous distance information that are now well constrained to be in the local Universe. Figure~\ref{fig:mosaic} presents a representative sample mosaic of these galaxies where the \ha~emission line has been spectroscopically confirmed in the filter identified by our narrow-band colors. The left image cutout in Figure~\ref{fig:mosaic} is the SDSS gri color composite, the four panels to the right are the cutouts for all four \ha~filters, and the green boxes highlight which filter the emission line has been confirmed. The galaxies in this mosaic span a range of \ha~EW, where EW increase towards the bottom of the figure from EW=10~\AA~at the top to EW=100~\AA~at the bottom. The galaxies also show a range of morphologies from compact to irregular to those showing spiral structure.

\subsection{Observable Properties of CLU-\ha~Galaxies} \label{sec:candidates}

\begin{figure*}
  \begin{center}
  \includegraphics[scale=0.6]{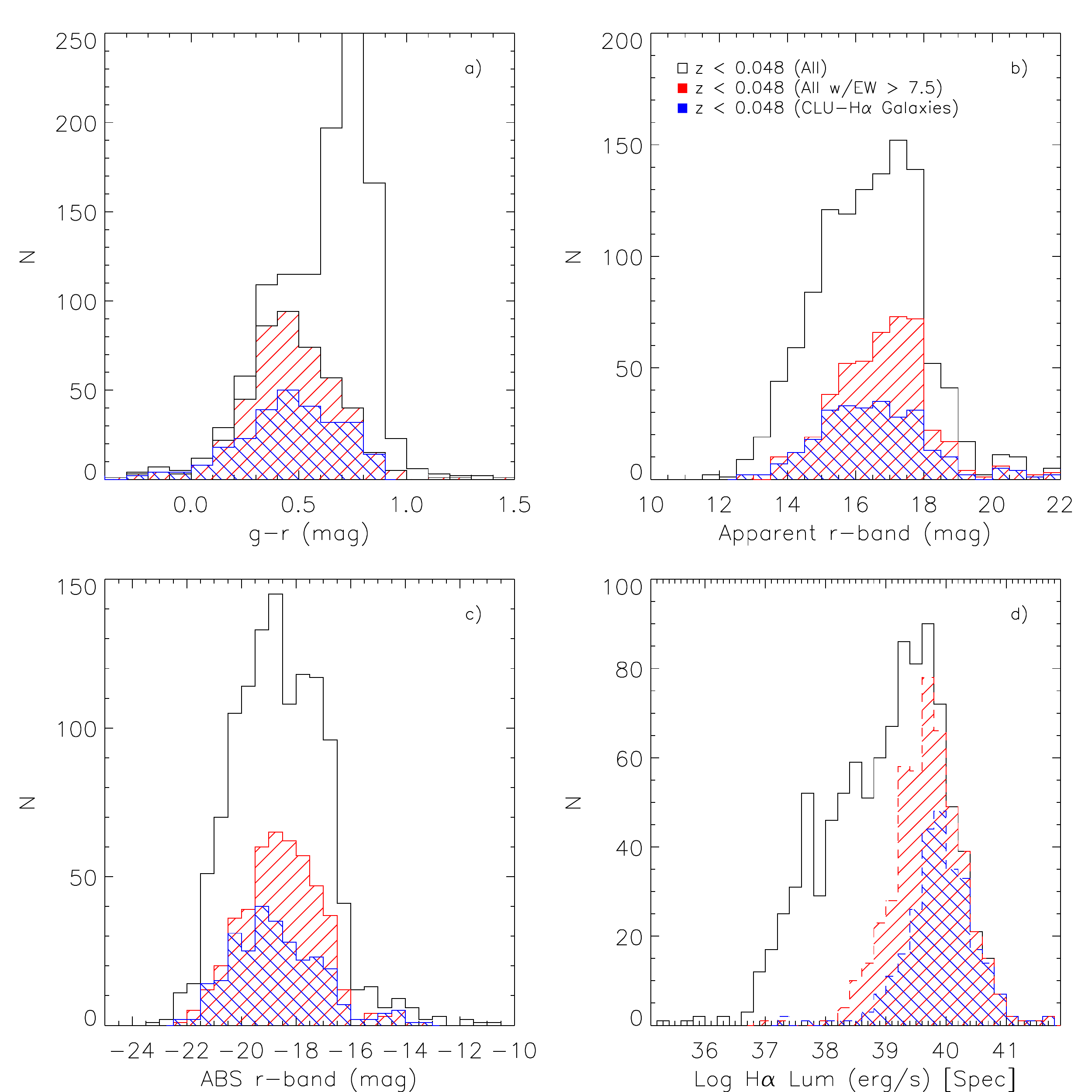}
  \caption{The observable property histograms of our $\Sigma=2.5$ galaxies in comparison to galaxies with \ha~EW information from SDSS or our CLU-\ha~followup. The unfilled, red-filled, and blue-filled histograms represent all galaxies with existing spectroscopy in our preliminary fields, a subset of these galaxies with an EW$>$7.5~\AA, and our $\Sigma=2.5$ galaxies, respectively. Panel a) shows the $g-r$ optical colors, panel b) shows the apparent $r$-band magnitudes, panel c) shows the absolute $r$-band magnitudes, and panel d) shows the \ha~derived luminosities. We find that the galaxies not selected by our methods with an EW greater than our selection limit have fainter apparent magnitudes near 18~mag. These fainter galaxies are not selected since our signal-to-noise criteria is an increasing function of ``On-Off" color towards fainter magnitudes. }
   \label{fig:genprop25sig}
   \end{center}
\end{figure*}

Here we explore the observable properties of our confirmed galaxy candidates in comparison to cataloged galaxies in our preliminary fields with spectroscopic information from SDSS or our CLU-\ha~followup. Figure~\ref{fig:genprop25sig} presents the SDSS $g-r$ color, apparent $r$-band magnitude, the absolute $r$-band magnitude, and the \ha~luminosity of all galaxies with spectroscopic information, a subset of these galaxies with \ha~EW greater than 7.5~\AA, and our CLU-\ha~$\Sigma=2.5$ galaxies. We do not show the observable properties for $\Sigma=5$ catalog since their distributions are similar to the $\Sigma=2.5$ catalog. The sub-sample of galaxies with an \ha~EW greater than 7.5~\AA~represent those above our minimum \ha~EW selection threshold for the $\Sigma=2.5$ galaxy list. Thus, these comparisons will reveal the properties of the galaxies with EW values above our limit that were not selected.

Panel `a)' of Figure~\ref{fig:genprop25sig} shows that the majority of galaxies in our preliminary fields have $g-r$ colors that peak around 0.75~mag largely due to the presence of early-type galaxies in the Coma field. However, both the sub-sample with \ha~EW$>7.5$\AA~and the CLU-\ha~galaxies show a bluer peak around 0.45~mag. The different color distributions show that our selection methods tend to select blue, star-forming galaxies; as expected. 

Panel `b)' of Figure~\ref{fig:genprop25sig} shows the apparent $r$-band magnitude where we find median values of 16.5, 16.9, and 16.6~mag for galaxies in our preliminary fields with spectroscopy, the sub-sample with an \ha~EW$>$7.5\AA, and CLU-\ha~galaxies, respectively. In addition, the distribution of the \ha~EW$>$7.5~\AA~sub-sample is peaked at fainter $r$-band magnitudes between 17--18~mag when compared to the CLU-\ha~galaxies. This suggests that a larger fraction of galaxies with moderate \ha~EWs that are missed by our selection methods have fainter apparent magnitudes with a median value at 17.2~mag. These fainter galaxies are not selected since the signal-to-noise criteria in our method is an increasing function of ``On-Off" color towards fainter magnitudes. In other words, the random ``On-Off" color scatter of sources due to the noise in the images is greater at fainter magnitudes and thus require a larger \ha~EW to be selected (See the signal-to-noise curves in Figures~\ref{fig:ha1cand}--\ref{fig:ha4cand}).

Panel `c)' of Figure~\ref{fig:genprop25sig} shows the absolute $M_{r}$ magnitude where we find median values of --18.7, --18.5, and --18.9~mag for galaxies in our preliminary fields with spectroscopy, the sub-sample with an \ha~EW$>$7.5\AA, and CLU-\ha~galaxies, respectively. We find that the subset of galaxies with an \ha~EW$>$7.5~\AA~and not selected in our survey span a range of magnitudes including both intrinsically faint and bright galaxies. These galaxies tend to have fainter apparent magnitudes due to their distances, where 75\% are at distances greater than 100~Mpc with a median $r$-band magnitude of 17.2~mag. Thus, these missed galaxies with moderate \ha~EWs would require stronger \ha~emission to be selected in our survey.

Panel `d)' of Figure~\ref{fig:genprop25sig} shows the \ha~luminosity where we find median Log $L_{\ha}$ (erg~s$^{-1}~cm^{2}$) values of 39.1, 39.7, and 39.9 for galaxies in our preliminary fields with spectroscopy, the sub-sample with an \ha~EW$>$7.5\AA, and CLU-\ha~galaxies, respectively. This panel clearly illustrates that a large fraction ($>$50\%) of galaxies with spectroscopy in our preliminary fields have weak \ha~emission (i.e., small \ha~EWs). We note that the early-type galaxies with \ha~line non-detections are not plotted in this panel. A comparison between the subset of galaxies with EW$>$7.5 and our CLU-\ha~galaxies shows that the CLU-\ha~galaxies have a larger median value by 0.2~dex. Thus, our methods tend to select galaxies with the intrinsically higher \ha~luminosities. We note that the moderate EW galaxies missed by our selection methods tend to be fainter with a median $r$-band magnitude of 17.1~mag, and would require larger \ha~emission to be selected.

\subsection{Physical Galaxy Properties} \label{sec:physprop}
Here we present the physical properties of the galaxy sample by cross-matching against the ALLWISE 3.4$\mu m$ and 22$\mu m$ fluxes. These data will provide stellar masses ($M_{\star}$) and extinction corrections due to dust. In addition, we use the spectroscopic \ha~fluxes of our survey to derive SFRs. The \ha~fluxes have been corrected for Milky Way extinction via the prescription of \cite{schlafly11}. 

The stellar masses are derived from mass-to-light ratios ($\Upsilon_{\star}$) using the WISE 3.4$\mu m$ fluxes. We utilize the fluxes derived from ALLWISE catalog profile fitting photometry, thus these fluxes should encompass each galaxies' full radial extent. The WISE 3.4$\mu m$ bandpasses provides a robust tracer of a galaxy's stellar mass as this light is dominated by an older stellar population (which make up the majority of a galaxy's stellar mass) and is less affected by attenuation from dust than shorter wavelengths. 

Many studies over the past few years have made comparisons between a variety of observationally derived stellar masses (e.g., baryonic Tully-Fisher relationship and resolved star color-magnitude diagrams) and luminosities in Spitzer 3.6$\mu m$ and WISE 3.4$\mu m$ bandpasses \citep[][]{oh08,eskew12,barnes14,mcgaugh14,meidt14,norris14,mcgaugh15,querejeta15}. The results of these comparisons show that a constant Spitzer $\Upsilon_{\star}^{3.6\mu m}$ of $0.4-0.55$ M$_{\odot}/$L$_{\odot}$ provides a robust estimation of a galaxies stellar mass with a relatively low error of $\sim$0.1 dex \citep[][]{meidt14,mcgaugh15}. A constant mass-to-light ratio of the same value for WISE 3.4$\mu m$ has been shown to yield similar stellar masses as Spitzer 3.6$\mu m$ \citep{jarret13,norris14}. We adopt the constant mass-to-light ratio of $\Upsilon_{\star}^{3.4\mu m}=0.5~$M$_{\odot}/$L$_{\odot,3.4\mu m}$, where M$_{\odot}/$L$_{ \odot,3.4\mu m}$ is the mass-to-light ratio in units of solar masses per the solar luminosity in the WISE 3.4$\mu m$ filter bandpass \citep[m$_{\odot,3.4\mu m}$=3.24~mag;  L$_{\odot,3.4\mu m}=1.58\times10^{32}$~erg~s$^{-1}$;][]{jarret13}.

Star formation rates (SFRs) for our galaxies can be estimated from many different luminosity tracers (e.g. \ha, FUV, etc.), where measured luminosities are transformed into SFRs via scaling prescriptions \citep[e.g., ][]{kennicutt98,murphy11}. We utilize the updated scaling relationships of \cite{murphy11} which assumes a Kroupa IMF \citep{kroupa01}. The \ha~SFRs are derived via a combination of \ha~and 22$\mu m$ luminosities which account for internal dust extinction \citep{calzetti10a,murphy11}:

\begin{equation}
  \rm{SFR}_{\ha,corr} (\rm{M}_{\odot}\rm{yr}^{-1}) = C \times \nu L_{\ha} + 0.031 \times \nu L_{22\mu m},
\end{equation}

\noindent where $\nu$L are the observed monochromatic luminosities of both \ha~and the WISE4 band at $22\mu m$ in ergs per second and C=5.37$\times10^{-42}$ \citep{murphy11}. 

\begin{figure*}
  \begin{center}
  \includegraphics[scale=0.65]{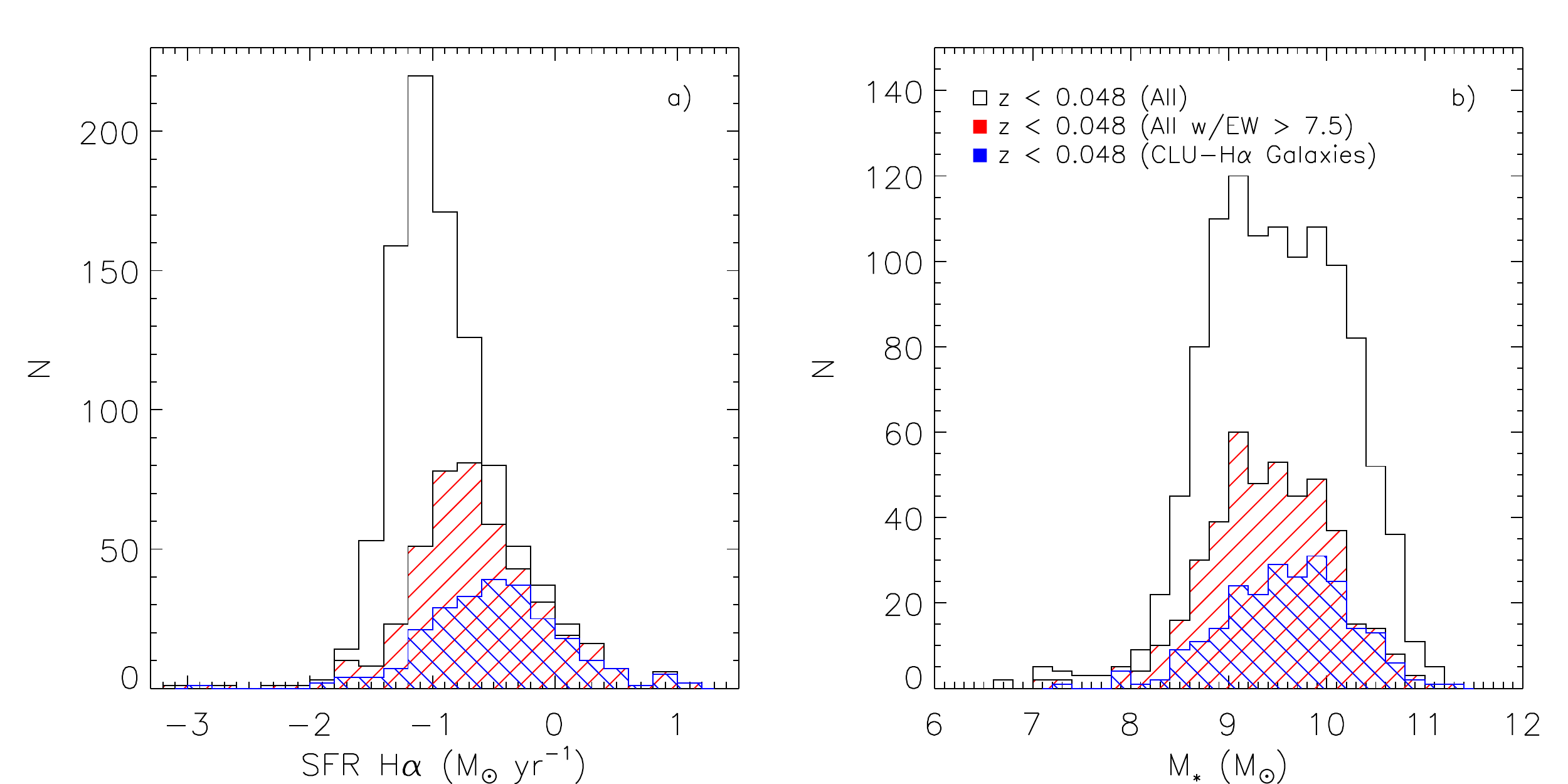}
  \caption{The physical property histograms of of our $\Sigma=2.5$ galaxies in comparison to galaxies with \ha~EW information from SDSS or our CLU-\ha~followup. The unfilled, red-filled, and blue-filled histograms represent all galaxies with existing spectroscopy in our preliminary fields, a subset of these galaxies with an EW$>$7.5~\AA, and our $\Sigma=2.5$ galaxies, respectively. We find that our selection methods tend select galaxies with higher SFRs, and can miss low-SFR or low-mass galaxies depending on their distance, and consequently their apparent magnitude. }
   \label{fig:physprop}
   \end{center}
\end{figure*}

The two panels of Figure~\ref{fig:physprop} show the \ha-derived SFR and stellar mass distributions for the same samples as in Figure~\ref{fig:genprop25sig}. Panel `a)' shows the SFR histograms where we find median Log SFR (M$_{\odot}~yr^{-1}$) values of --1.0, --0.7, and --0.5 for the galaxies in our preliminary fields with spectroscopy, the sub-sample with an \ha~EW$>$7.5\AA, and CLU-\ha~galaxies, respectively. Our selection methods tend to select galaxies with higher SFRs. However, we are able to recover some galaxies with moderately low SFRs near Log SFR (M$_{\odot}~yr^{-1}$)$\sim$$-2$.

Panel `b)' of Figure~\ref{fig:physprop} shows the stellar mass histograms where we find median Log $M_{\star}$ (M$_{\odot}$) values of 9.5, 9.4, and 9.6 for the galaxies in our preliminary fields with spectroscopy, the sub-sample with an \ha~EW$>$7.5\AA, and CLU-\ha~galaxies, respectively. The median values suggest that our methods select galaxies of roughly the same masses as other galaxies in the preliminary fields. We also find that the CLU-\ha~galaxies span a range of stellar mass ($8 \lesssim$ Log $M_{\star}$(M$_{\odot}$) to $\lesssim11$) suggesting that CLU-\ha~will recover some lower-mass dwarfs as well as larger spirals. However, CLU-\ha~will miss a significant fraction of dwarfs at greater distances.

\subsection{Comparison to Previous Blind Surveys for Active Galaxies}\label{sec:clucomp}

There have been several blind emission-line and UV-excess surveys undertaken prior to CLU-\ha~including the Markarian Survey (\citealt{markarian81} and references therein), the Case Northern Sky Survey \citep{case1}, the KPNO International Spectroscopic Survey \citep[KISS;][]{kiss00}, the University of Michingan survey \citep[UM;][]{UM1,UM2,UM3,UM4,UM5}, and the Hamburg/SAO Survey for Emission-Line Galaxies \citep[HSS;][]{hss99}.  Here we compare our galaxy catalogs ($\Sigma>2.5$ and $\Sigma>5$) to the objects in the first three surveys since there is no overlap between our preliminary fields and both UM and HSS. 

The Markarian \citep[i.e., the first Byurakan;][]{markarian83,markarian89,petrosian07} survey used an objective prism and photographic plates to find a total of 1544 UV-excess galaxies via a visual search for steep UV continuum slopes in their low-dispersion spectra. This was the first systematic search for active galaxies and covered 17,000 square degrees down to a continuum brightness of 17.5~mag in the V-band. The Case survey used a combination of steep UV continuum slopes and emission line selection ([OIII]) to find more galaxies per area resulting in 2339 galaxies in $\sim1200$ square degrees down to 18~mag. Finally, the KISS survey used an objective prism and line selection (\ha) but implemented modern CCD detectors to survey fainter sources (20-21~mag) and found 2425 galaxies in $\sim200$ square degrees. The use of CCDs and line selection methods facilitated deeper surveys and the detection of galaxies with a wider range of properties since the \ha~line (compared to UV slope and [OIII] line selection) will be stronger in more massive galaxies.

The CLU-\ha~survey utilizes CCD detectors and narrow-band imaging for low resolution line selection (\ha) to increase the survey area (3$\pi$~for the full survey) and to ensure the detection of galaxies with a wide range of properties. Using single image exposures in preliminary fields, we found 258 confirmed galaxies with $\Sigma>2.5$~in 100 deg$^2$ and can detect the continuum of sources down to 18.5~mag. Thus, the CLU survey depth and galaxy density are in between previous surveys that used photographic plates (Markarian and Case) and modern CCDs (KISS).  

\begin{figure*}
  \begin{center}
  \includegraphics[scale=0.55]{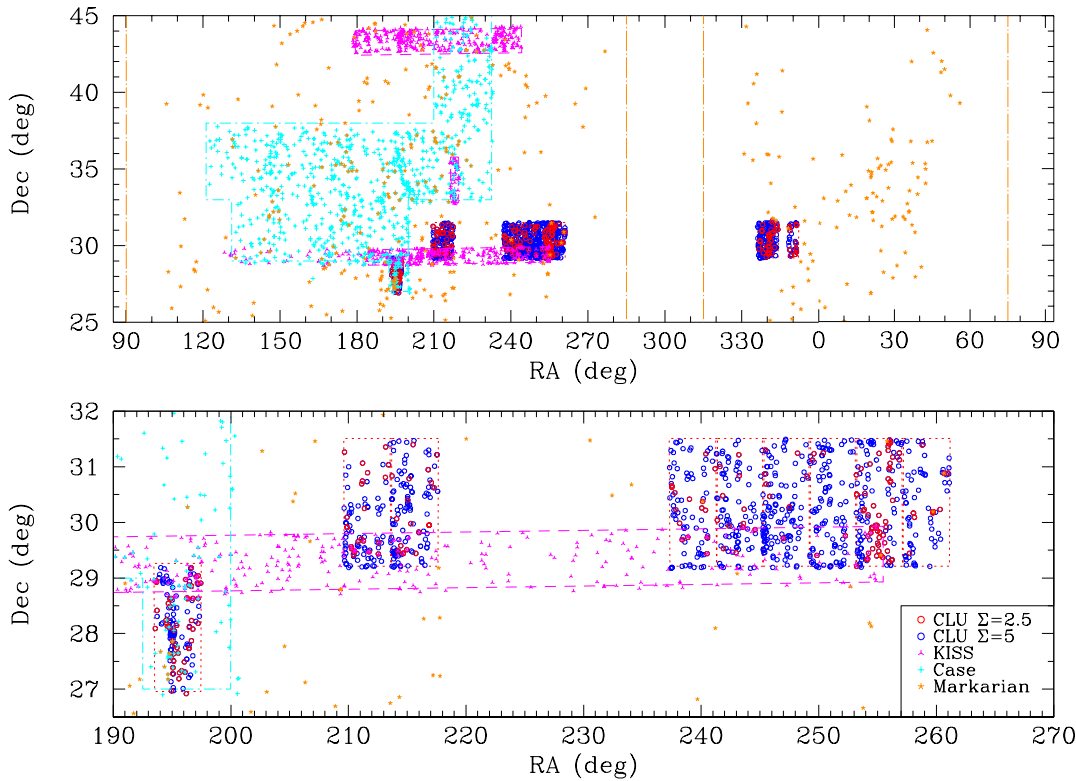}
  \caption{\textbf{Top:} The locations of catalogued galaxies below z=0.047 in the CLU-\ha preliminary fields as well as the KISS, Case, and Markarian surveys.  \textbf{Bottom:} Here we highlight a smaller region of the spring sky to show more clearly where our preliminary fields overlap with the KISS, Case, and Markarian surveys.}
  \label{fig:skyarea}
  \end{center}
\end{figure*}

Figure~\ref{fig:skyarea} shows the sky regions covered by the CLU-\ha~preliminary fields presented in this paper and each of the three comparison surveys with spatial overlap: Markarian \citep{petrosian07}, Case \citep{case1,case2,case3,case4,case5,case8,case9,case11,case12}, and KISS \citep{kiss00,kiss01,kiss04,kiss05}. In addition, the bottom panel of Figure~\ref{fig:skyarea} is a zoomed-in version covering our preliminary fields in more detail between an RA of 180--270 degrees. We first spatially crossmatched CLU-\ha~to the galaxies in each of the catalogs that are spatially coincident with our preliminary fields (within 4$\arcsec$), and have limited these catalogs to galaxies with a redshift below 0.047 (i.e., redshift probed by our survey). We also note that one of our CCD chips (\#3) has been nonfunctional for all of our pointings since the beginning of the survey. In addition, the two lower-left chips for the field 'p3967' (center near RA=195, Dec=28) in the \ha4 filter had poor data quality and could not be reduced. We remove galaxies that overlap with chip 3 in all pointings and the poor data chips for 'p3967' in this comparison. The number of galaxies from the three comparison catalogs with a redshift less than 0.047 and inside our preliminary fields is 8, 18, and 80 for the Markarian, Case, and KISS surveys, respectively.

A comparison with the Markarian galaxies shows that we successfully recover 7 of the 8 galaxies (88\%). However, CLU-\ha~found an additional 247 objects in the overlapping volume. The one Markarian galaxy not selected in CLU-\ha~has a star within 1$\arcsec$ of the galaxy center and was mislabeled as a star in our catalog. We note that this galaxy still exhibits a $\Sigma$ value of 15. 

A comparison with the Case galaxies shows that we successfully recover 14 of the 18 galaxies (78\%). However, CLU-\ha~found an additional 30 objects in the overlapping survey regions. Two of the Case galaxies missed by CLU-\ha~have small EW values below 2~\AA. The third galaxy missed by CLU-\ha~is an extended galaxy whose central region shows a small $\Sigma$ value; however, we did find several of this galaxy's \hii~regions with high $\Sigma$ values. The last galaxy missed by CLU-\ha~shows two nearby \hii~regions where our photometric aperture is located between two clumps and thus is missing some of the \ha~flux. 

A comparison with the KISS galaxies shows that we successfully recover 42 of the 80 galaxies (53\%). However, CLU-\ha~found an additional 24 objects in the overlapping survey regions. Nearly a third of the KISS galaxies missed by CLU-\ha~(N=12) were below our detection limits of 18.5~mag. The remaining KISS galaxies were missed by CLU for the following reasons: 8 had low \ha~EWs less than 7.5~\AA, 6 were labeled as stars, and 12 had moderate \ha~EWs combined with fainter magnitudes between 17.5--18.5~mag. The 12 missed galaxies with moderate EWs were cut in our survey since their fainter magnitudes required higher EWs to exhibit a significant narrow-band color. 

The CLU-\ha~survey is able to recover the majority of the galaxies in both the Markarian and Case surveys, and was able to find additional galaxies. Our comparison with the KISS survey shows that CLU-\ha~recovers roughly half as many emission-line galaxies as KISS due to our brighter limits. However, CLU-\ha~will cover a much larger area of the sky (3$\pi$~sr) compared to all previous emission-line galaxy surveys to a moderate depth. Thus, the CLU-\ha~survey will discover many emission-line galaxies across the sky.

\section{Discussion}
In this section we examine interesting candidates found in the preliminary fields of the \ha~survey where we find new BCDs, a newly discovered green pea, a new Seyfert 1 galaxy, and a known planetary nebula. We also put the CLU-\ha~galaxy properties into the context of previously established relationships between different galaxy physical properties (e.g., the star-forming \ms). The majority of the CLU-\ha~galaxies show physical properties similar to normal star-forming galaxies; however, several extreme galaxies (i.e., BCDs) show deviations from previously established galaxy trends. We end this section with a discussion of how the CLU-\ha~survey can help focus the search for the electromagnetic counterparts to gravitational wave events.


\subsection{Interesting Candidates}\label{sec:intcanddisc}

\begin{figure*}
  \begin{center}
  \includegraphics[scale=0.65]{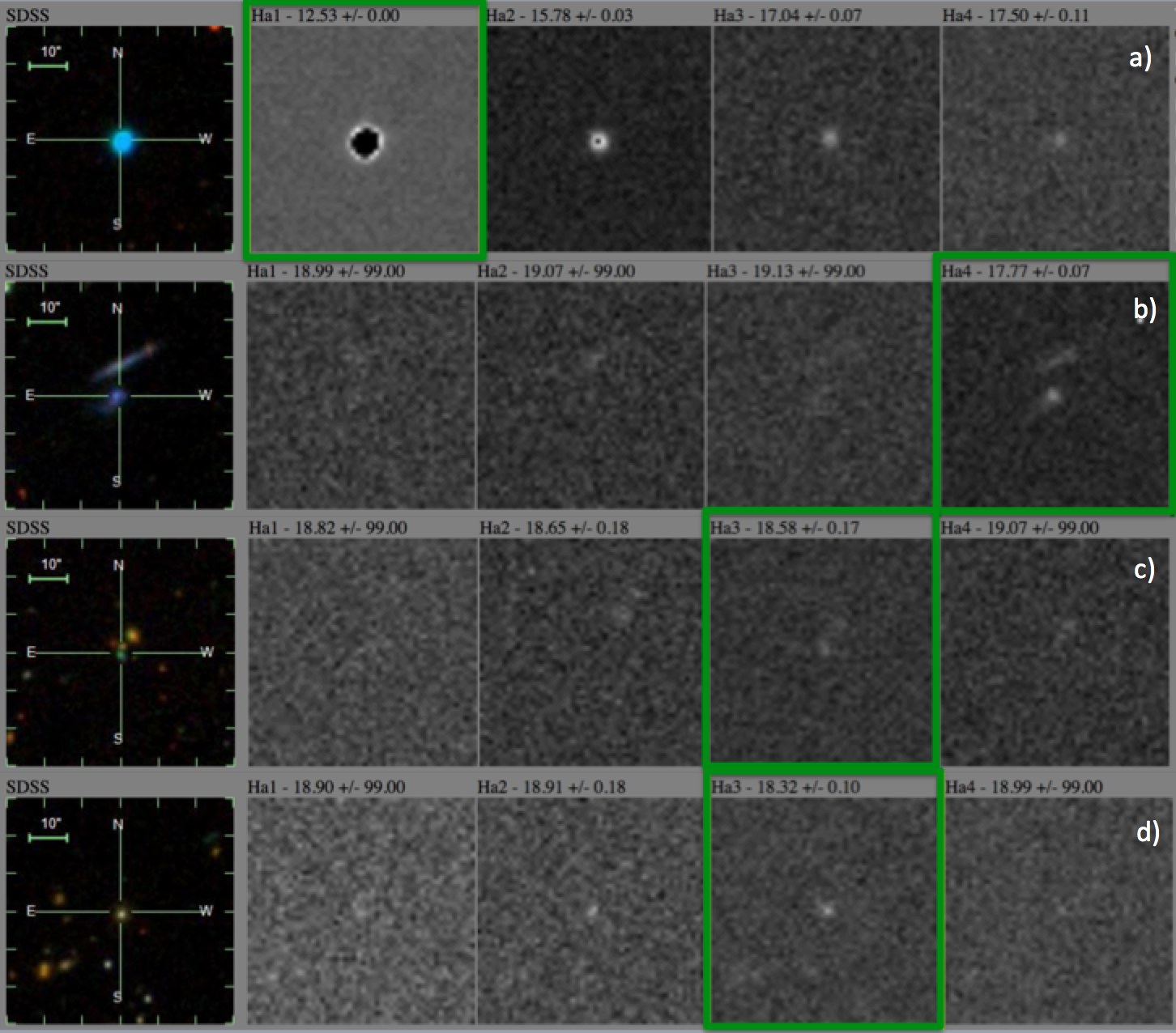}
  \caption{A mosaic of some interesting candidates found in our preliminary fields similar to Figure~\ref{fig:mosaic}. Panels a-d shows the known planetary nebula, an example of a new BCD, the new green pea, and the new Seyfert 1 galaxy, respectively. In addition, the thin galaxy just above center in panel b) is also newly discovered CLU-\ha~galaxy with an \ha~EW=80~\AA. }
   \label{fig:intcands}
   \end{center}
\end{figure*}

Amongst our emission-line candidates we find interesting extreme objects some of which have no previous distance information. In figure~\ref{fig:intcands} we present a mosaic of four interesting candidates, where panel a) shows the known planetary nebula, panel b) shows one of the BCDs, panel c) shows the green pea, and panel d) shows Seyfert 1 galaxy. Due to the large ``On-Off'' magnitudes of the planetary nebula, the central pixels in the top panel of Figure~\ref{fig:intcands} in the first \ha~filter were masked to provide a more consistent background and visual comparison across the four \ha~images. 

\begin{figure*}
  \begin{center}
  \includegraphics[scale=0.6]{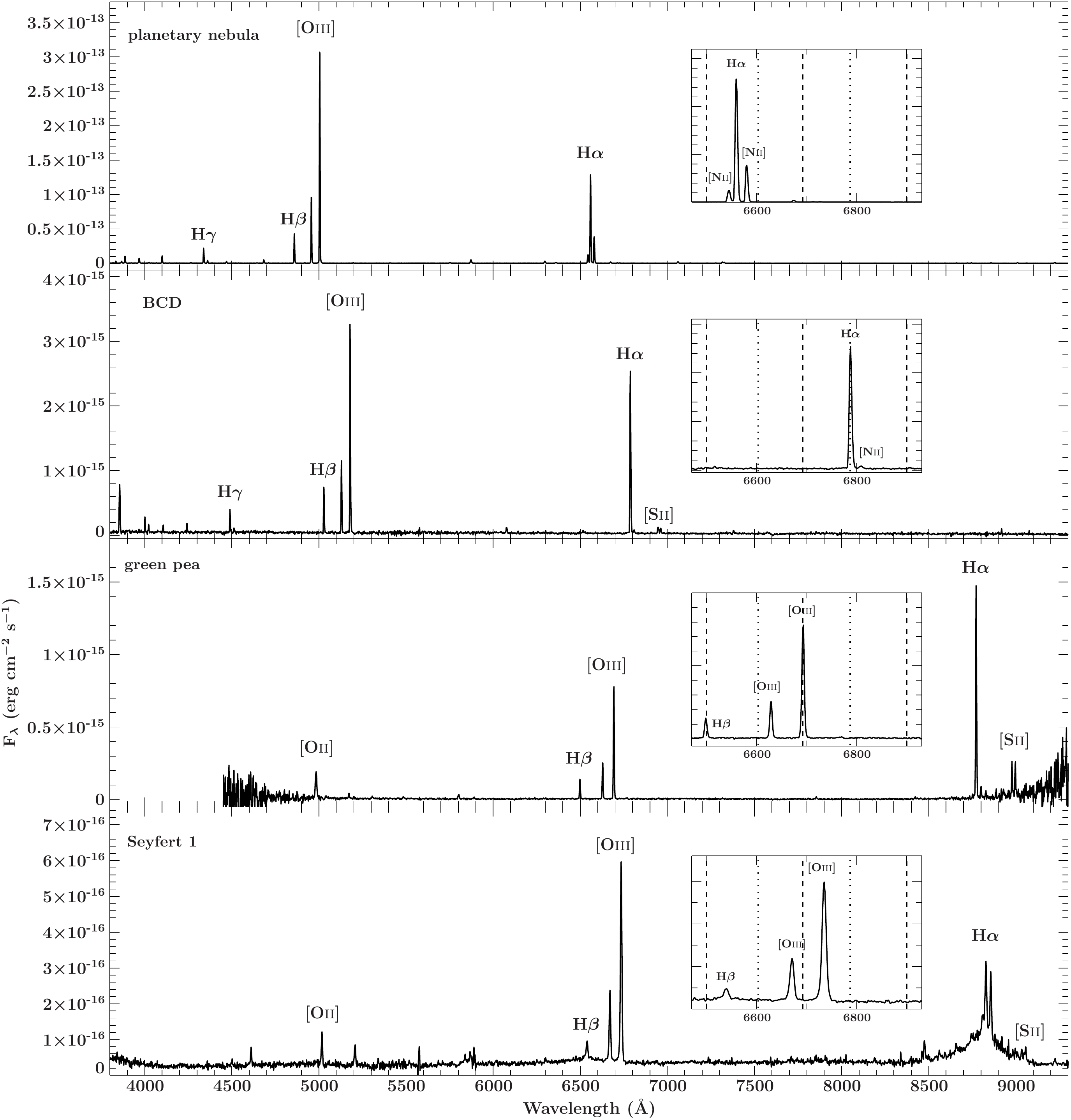}
  \caption{The Palomar DBSP spectra for the interesting candidates found in our preliminary fields where (from top-to-bottom) the panels display the known planetary nebula, an example BCD, the green pea, and the Seyfert 1 galaxy. Strong nebular lines are labeled in each panel. The inserts in each panel show a zoomed-in window where the vertical dashed lines represent the wavelength range covered by our narrow-band filter pairs (6500$-$6692~\AA~and 6692$-$6900~\AA). The dotted lines represent the separation between the filters within the filter pairs.}
   \label{fig:IntCandSpec}
   \end{center}
\end{figure*}

The planetary nebula is PN H 4-1 and is located at a high Galactic lattitude: $b=88.14757^{\circ}$, $l=49.3065^{\circ}$ \citep[][]{haro51}. Panel a) of Figure~\ref{fig:intcands} shows the SDSS gri color and our 4 \ha~filters (from left-to-right, respectively). PN H 4-1 shows a large flux excess in \ha1 compared to \ha2 (i.e., ``On-Off") equal to three magnitudes ($\Delta \rm{mag} = -3.05$). The measured spectroscopic \ha~line flux and EW are $1.21 \times 10^{-12} \rm{erg~s^{-1} cm^{-2}}$ and 1200 \AA, respectively. The spectrum of this PN is shown in top panel of Figure~\ref{fig:IntCandSpec}, and exhibits emission lines similar to other known planetary nebulae.

The detection of this planetary nebula is interesting since our choice of preliminary fields avoided the Galactic plane and PN H 4-1 is located at a high Galactic latitude ($b\sim$88$^{\circ}$). We anticipate that the CLU-\ha~survey can be used as a discovery engine for new planetary nebulae at intermediate galactic latitudes. This expectation is based on the galactic latitude limits of previous emission-line galaxy and Galactic Plane surveys in the northern sky. Previous galaxy surveys (Markarian, Case, UM, KISS, HSS, etc.) avoided the galactic plane above $|b|>20\degr$, while the largest-area galactic plane survey in the northern hemisphere was limited to $|b|<5\degr$~\citep[The INT Photometric H$\alpha$~Survey, IPHAS;][]{iphas}. Thus, there is a gap at intermediate galactic latitudes that has not been uniformly searched for emission-line sources. In addition, the large-area galactic plane search for planetary nebulae in the southern hemisphere \citep[The Macquarie/AAO/Strasbourg \ha, MASH;][]{parker06} found a few hundred newly discovered planetary nebulae at galactic latitudes of $5<|b|<15$~degrees ($\approx4000~\rm{deg}^{2}$) with a detection limit of $5\times10^{-16}\rm{erg~s^{-1} cm^{-2}}$. Despite the higher selection limit of our survey ($1\times10^{-14}\rm{erg~s^{-1} cm^{-2}}$), we anticipate finding 10s-to-100 new PNe in the area not covered by IPHAS and MASH ($\approx3500~\rm{deg}^{2}$) at intermediate galactic latitudes assuming the same distribution of PNe as in the southern hemisphere.

In the CLU-\ha~sample, there are 9 BCDs where 7 have distances measured for the first time in this survey. Panel b) of Figure~\ref{fig:intcands} shows the image cutouts of an example BCD with new distance measurements, where the source is brightest in the fourth \ha~filter. The second panel of Figure~\ref{fig:IntCandSpec} shows the spectrum of the BCD where the redshifted wavelength of the strong \ha~line confirms the identification in the fourth filter. We note that the \ha~emission line has a wavelength that is only a few angstroms greater than the separation between the third and fourth filter, and is clearly brighter in the fourth filter image. The spectrum of the example BCD is representative of the other BCDs, where both the \oiii~and \ha~lines exhibit strong emission lines: EWs of 418~\AA~and 446~\AA~for \oiii~and \ha, respectively. We measure the 12+Log(O/H) metallicity of 8.02 from O3N2 methods \citep{pp04} and a dust corrected SFR of 0.41 M$_{\odot}$ yr$^{-1}$. 

The SDSS and CLU-\ha~images of the green pea are shown in panel c) of Figure~\ref{fig:intcands}. The spectrum of this object is shown in the third panel of Figure~\ref{fig:IntCandSpec} where the \oiii~lines confirm the correct identification in the \ha3 filter. Both the \oiii~and \ha~lines exhibit strong emission lines: EWs of 510~\AA~and 560~\AA~ for \oiii~and \ha, respectively. We have measured metallicity of 12+log(O/H) = 8.09 via the O3N2 method. The \ha~flux of $9. \times 10^{-15} \rm{erg~s^{-1}~cm^{-2}}$ which corresponds to a dust corrected SFR of 24 M$_{\odot}$ yr$^{-1}$ given the measured redshift of 0.337 and H$_0$ of 72 km/s/Mpc. 

The image cutouts of the new Seyfert 1 galaxy found in our survey of preliminary fields is shown in panel d) of Figure~\ref{fig:intcands} where the object is brightest in the third \ha~filter. In addition, the spectrum of this object is shown in the bottom panel of Figure~\ref{fig:IntCandSpec} which shows strong \oiii~lines confirming the correct identification in our third filter. Furthermore, the broadened \ha~and \hb~emission lines clearly indicate the presence of a strong AGN which allows us to classify this object as a Seyfert 1 galaxy.

  
\subsection{Galaxy Trends}
In this section we show where the CLU-\ha~galaxies are located on established trends of star-forming galaxies. Previous studies of star-forming galaxies have found a relatively tight correlation between the current SFR and the total stellar mass for both local Universe and higher redshift galaxy samples: the galaxy \msp This trend can provide insights into how galaxies evolve over time since the normalization increases with redshift \citep[e.g.,][]{heinis14}. We use this relationship to illustrate the distribution of CLU-\ha~galaxy properties.


\begin{figure}
  \begin{center}
  \includegraphics[scale=0.5]{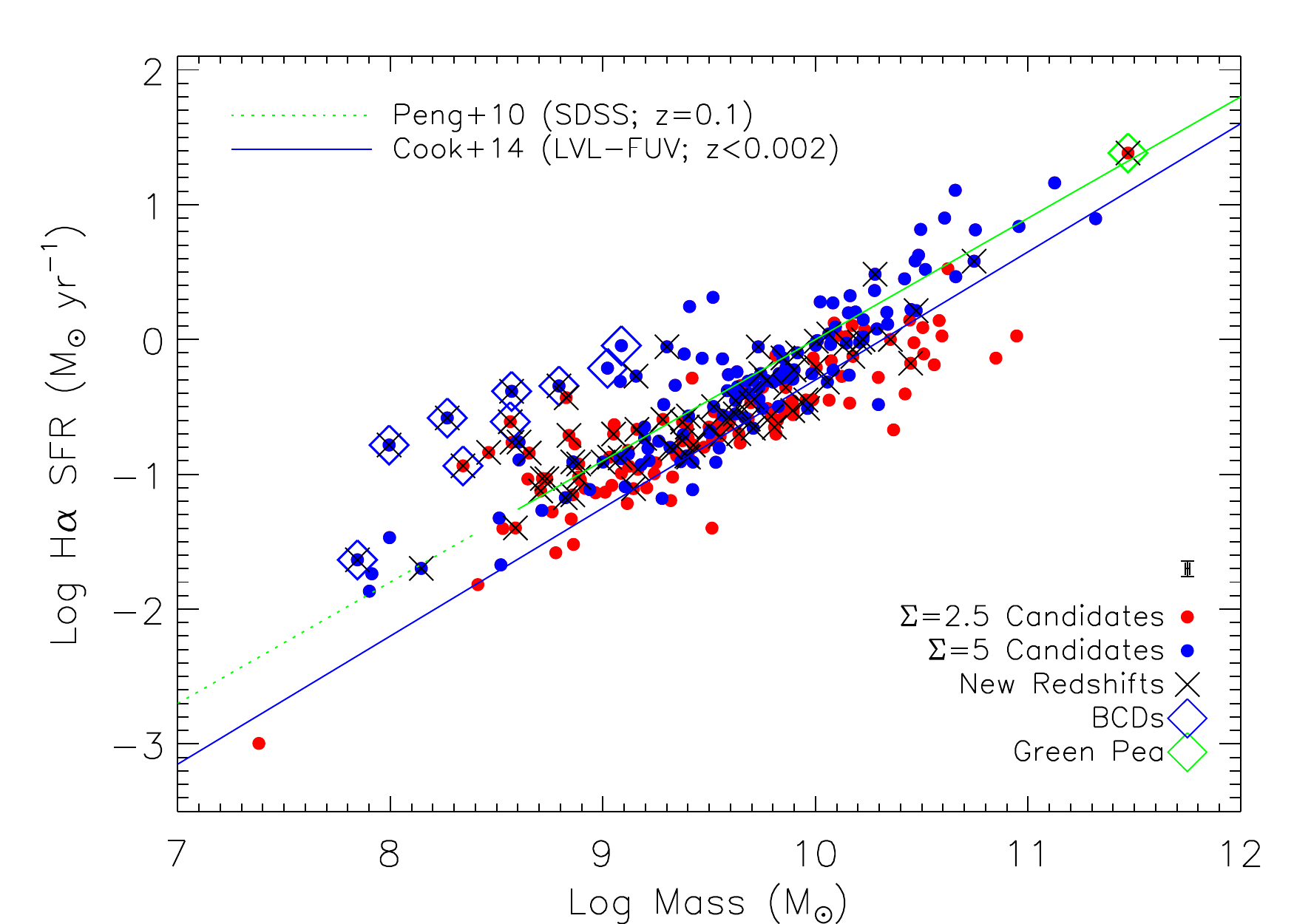}
  \caption{The \ms~of star-forming galaxies for CLU-\ha~galaxies in the preliminary fields. The red dots, blue dots, large X's represent $\Sigma=2.5$ CLU-\ha~galaxies, $\Sigma=5$ CLU-\ha~galaxies, and CLU-\ha~galaxies with no previous distance information. The blue and green lines represents the relationships found in the LVL sample \citep[D$<$11~Mpc;][]{cook14c} and SDSS sample \citep[z$\sim0.1$;][]{peng10}. The green solid–dotted line transition represents the lower end of the stellar mass range probed by the SDSS survey. The error bars near the legend represent the median errors of all data points. We find that the majority of the CLU-\ha~galaxies show agreement with the local Universe \ms. In addition, the blue and green diamonds represent the BCD and green pea galaxies found in our preliminary fields. These extreme galaxies show higher SFRs for their given stellar mass.}
   \label{fig:ms}
   \end{center}
\end{figure}

Figure~\ref{fig:ms} shows the dust-corrected \ha-SFR versus stellar mass (M$_{\star}$) for the CLU-\ha~galaxies, where the symbols are the same as those in Figure~\ref{fig:absr_z}. The blue and green lines represents the relationships found in the LVL sample \citep[D$<$11~Mpc;][]{cook14c} and SDSS sample \citep[z$\sim0.1$;][]{peng10}. The majority of the CLU-\ha~galaxies follow the previous ``Main Sequence'' relationship and span a wide range in both SFR and stellar mass from low-mass, low-SFR dwarfs to high-mass, high-SFR spirals. 


The BCDs found in the CLU-\ha~sample show high SFRs given their stellar masses illutstrating that these galaxies are undergoing an episode of enhanced star formation. The new green pea, on the other hand, shows good agreement with the SDSS \ms relationship suggesting that this galaxy is not undergoing an enhanced period of star formation; this is contradictory to previous studies which find that green peas show signs of significant star formation activity \citep{cardamone09,izotov11}. However we do not have enough statistics in this study to draw meaningful conclusions as to the connection between BCDs and green peas. Such an analysis is left for a future CLU-\ha~study that will collect a larger sample of both types of objects.

\begin{figure}
  \begin{center}
  \includegraphics[scale=0.5]{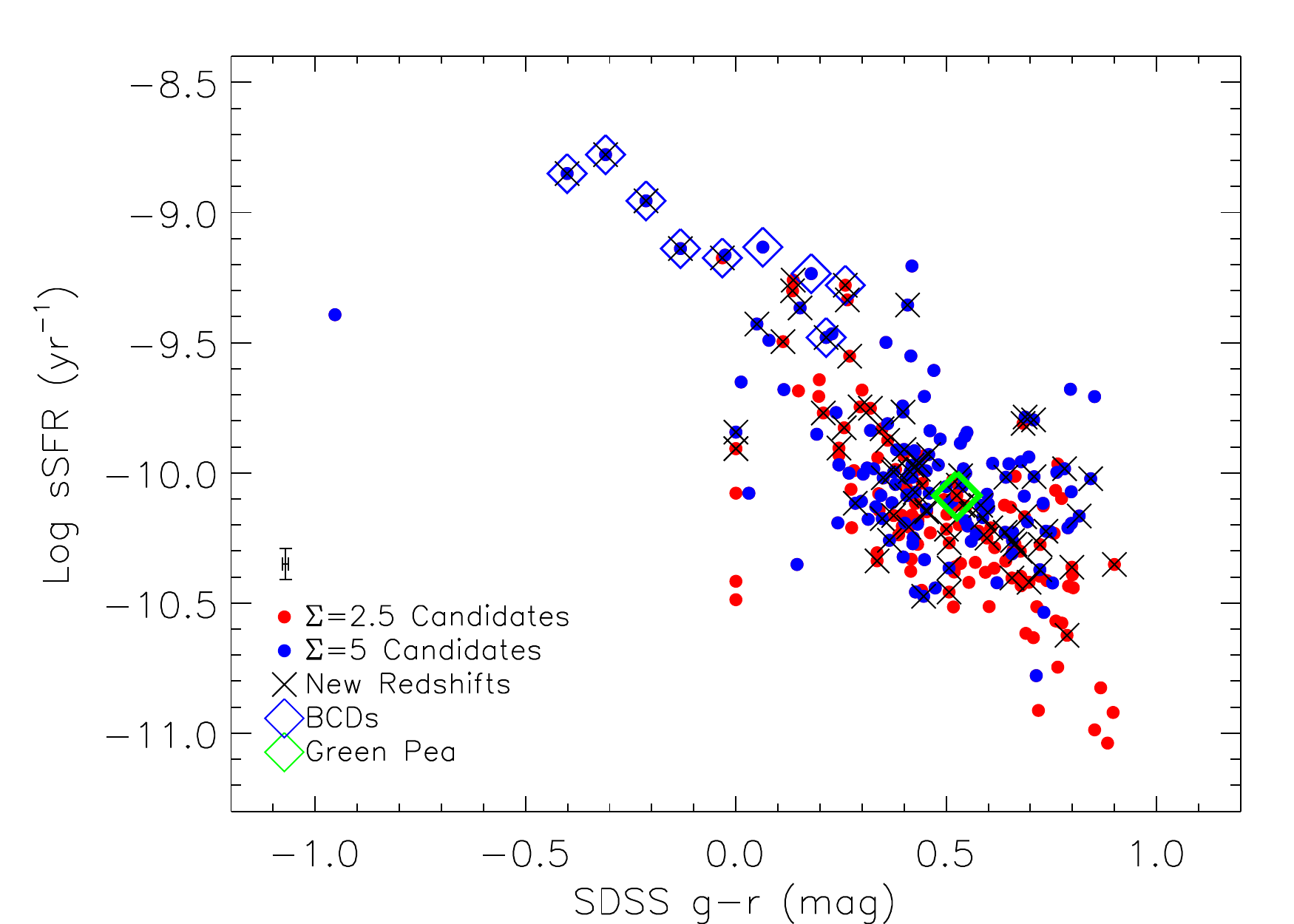}
  \caption{The specific star formation rate (sSFR$\equiv$SFR/M$_{\star}$) versus $g-r$ optical color, where sSFR is an indicator for the intensity of star formation. The error bars near the legend represent the median errors of all data points. We find that the BCDs show the highest star formation rate intensities (sSFR) and tend to have the bluest colors. The new green pea galaxy shows a sSFR similar to other normal star-forming galaxies.}
   \label{fig:sSFRcolor}
   \end{center}
\end{figure}

We also explore a relationship between physical and observed properties of galaxies, where local Universe studies have found that the specific SFR (sSFR$\equiv$SFR/M$_{\star}$) forms a relatively tight trend with optical colors \citep[e.g.,][]{cook14c,schawinski14}. Figure~\ref{fig:sSFRcolor} shows the sSFR versus the SDSS $g-r$ color of the CLU-\ha~galaxies, where the symbols are the same as that in Figure~\ref{fig:ms}. We find that the CLU-\ha~galaxies follow a similar trend to the LVL galaxies \citep{cook14c}, but that several CLU-\ha~BCDs tend to have sSFRs (Log(sSFR) as high as $-8.7$) and bluest optical colors due to the large volume sampled. The CLU-\ha~green pea galaxy shows a sSFR and color similar to that of the other CLU-\ha~galaxies. A future CLU-\ha~study will include larger numbers of BCDs and green pea galaxies to study this relationship.


\subsection{Electromagnetic Counterparts to Gravitational Wave Events}

The detection of gravitational waves via the Laser Interferometer Gravitational Wave Observatory \citep[LIGO;][]{abbott16} allows for a new way of observing the Universe and thus provides new tools with which to understand fundamental physics. Associating gravitational waves and electromagnetic counterparts will have dramatic impacts on our understanding of the Universe. However, the size of the 90\% sky localizations of LIGO GW events have ranged from 30-1000 square degrees on the sky making the search for EM counterparts a challenge. 

Significant efforts of large area surveys \citep[e.g.,][]{kasliwal16,coulter17,evans17,hallinan17,kasliwal17,troja17} have scanned the most probable sections of gravitational wave localizations in search of an electromagnetic counterpart. However, the efficiency of this effort can be greatly enhanced by utilizing the locations of known galaxies in the LIGO sensitivity volume for neutron star mergers \citep[D$\leq$200~Mpc;][]{aligo}. Targeted follow-up observations of likely host galaxies can narrow down the search area by a factor of 100 \citep{nissanke13,gehrels16}. 

Previous efforts have been made to provide lists of galaxies specifically designed for EMGW follow-up \citep{kopparapu08,white11}, where the latest catalog from \cite{white11} shows a $B$-band luminosity completeness of 65\% at 100~Mpc. However, the galaxies in both previous catalogs were limited to 100~Mpc which is an eighth of the LIGO sensitivity volume out to 200~Mpc. In addition, several new or updated spectroscopic galaxy surveys have published secure distances via redshifts for new galaxies in the local Universe \citep[e.g., SDSS, ALFALFA, 6dFRGS;][]{sdss12,haynes11,jones09}; thus, leaving previous EMGW galaxy catalogs less complete than previously estimated. There are efforts which utilize increased numbers of galaxies with photometric redshifts (e.g., GLADE\footnote{http://aquarius.elte.hu/glade/}); however, we focus on galaxy catalogs with secure distance measurements. The lack of an EMGW catalog that extends to the full LIGO sensitivity volume and the existence of new galaxies found in the local Universe motivate the construction of a new compiled galaxy catalog to be used in the search for EM counterparts to GW events. 


In an effort to provide the most complete list of galaxies with measured distances in the LIGO sensitivity volume, our team has compiled a catalog of all known galaxies out to 200~Mpc, hereafter referred to as CLU-compiled. The galaxies were taken from existing galaxy databases: NASA/IPAC Extragalactic Database (NED)\footnote{https://ned.ipac.caltech.edu}, Hyperleda\footnote{http://leda.univ-lyon1.fr} \citep{hyperleda}, Extragalactic Distance Database\footnote{http://edd.ifa.hawaii.edu} \citep[EDD;][]{EDD}, the Sloan digital sky survey DR12 \citep[SDSS;][]{sdss12}, The 2dF Galaxy Redshift Survey \citep[6dFRGS;][]{jones09}, and The Arecibo Legacy Fast ALFA \citep[ALFALFA][]{haynes11}. Distances based on Tully-Fischer methods were favored over kinematic (i.e., redshift) distances; however, the majority of the distances are based upon redshift information. The catalog contains $\sim$234,500 galaxies with existing distances less than 200~Mpc.

In addition to distances, the catalog also contains compiled photometric information. We have cross-matched (within a 4$\arcsec$ separation) the CLU-compiled catalog with GALEX all sky \citep{galex}, WISE all sky \citep{wise}, and SDSS DR12 \citep{sdss12} surveys to obtain fluxes from the ultraviolet (UV) to the infrared (IR). We find $\sim$154,200 matches for GALEX FUV, $\sim$216,000 for WISE 3.4 and 22$\mu m$, and $\sim$100,400 for SDSS $r$-band. With the UV and IR fluxes, we have also derived physical properties (SFRs, stellar masses, and dust extinction) for the CLU-compiled galaxies via similar methods as the CLU-\ha~survey in \S\ref{sec:physprop}.

The CLU galaxy catalog is being utilized by the GROWTH (Global Relay of Observatories Watching Transients Happen\footnote{http://growth.caltech.edu}) collaboration to search for EM counterparts to gravitational wave (GW) events across the electromagnetic spectrum. On August 17th 2017 GWs from two merging neutron stars were observed in both the LIGO and Virgo detectors \citep[hereafter GW170817;][]{ligodet170817}  and is the first GW event to have an electromagnetic counterpart found \citep{GW170817}. Shortly after the GW event, several collaborations around the world began a large multi-wavelength observational campaign to search for the electromagnetic counterpart \citep[e.g.,][]{coulter17,evans17,hallinan17,kasliwal17,troja17}. Our team crossmatched the CLU-compiled catalog against the LIGO-Virgo volume \citep[initially 31 square degrees on the sky at a distance of 40$\pm$8~Mpc;][]{GW170817} and found 49 galaxies \citep[see;][]{kasliwal17}.  In addition, since the physical properties of these galaxies were determined in advance by our team, we were able to promptly prioritize these galaxies by stellar mass and publish this list on the Gamma-ray Coordinate Network \citep[GCN;][]{GCN21519}. The EM counterpart was later found in NGC4993 \citep{coulter17}, which was the third highest priority galaxy on our published list. We note that the CLU galaxy catalog produced the only published GCN to correctly identify NGC4993 in the localization volume of GW170817. Future efforts to search for the EM counterparts to GW events using CLU will include new galaxies found in the CLU-\ha~survey. 

\begin{figure*}
  \begin{center}
  \includegraphics[scale=0.5]{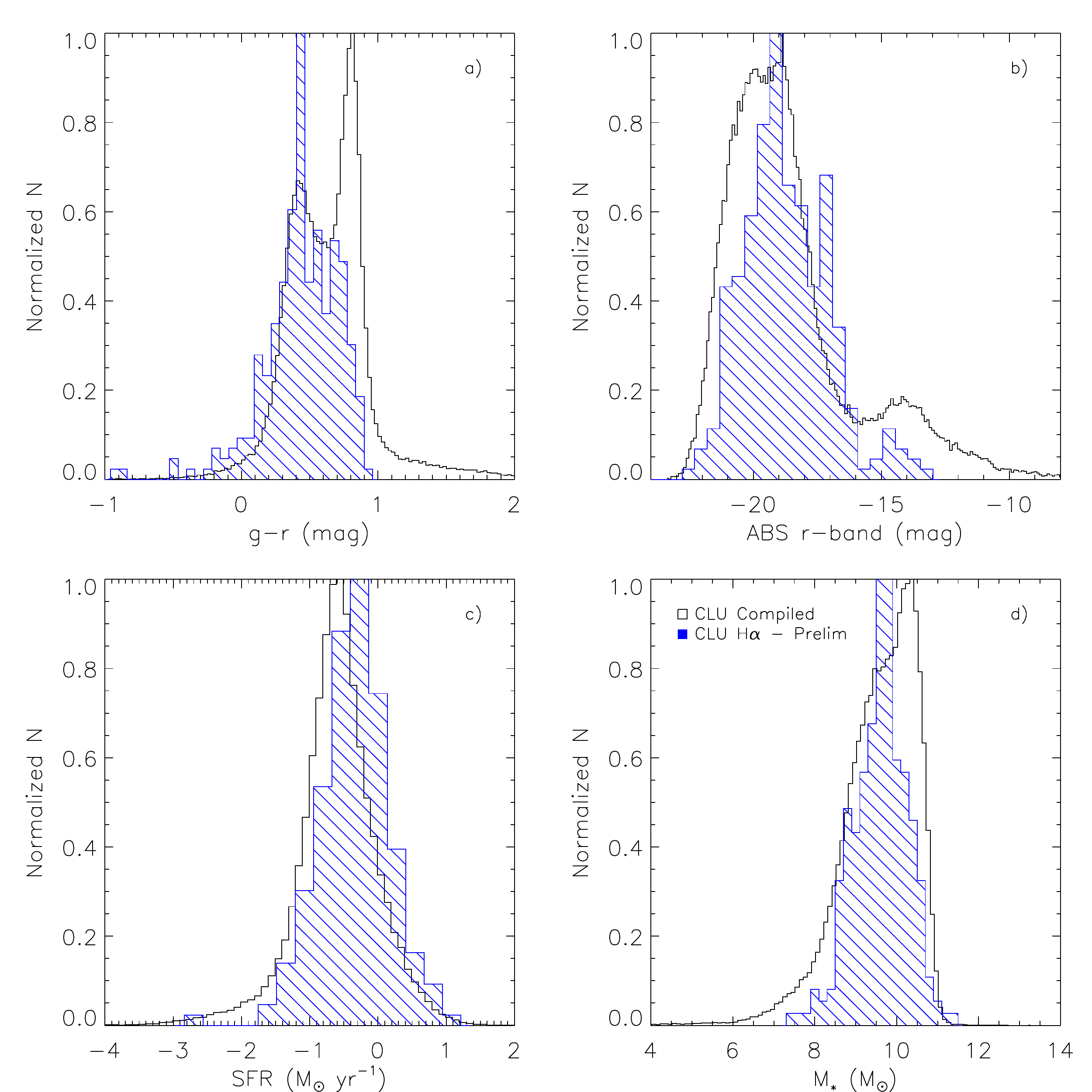}
  \caption{A Comparison observable and physical properties between the CLU-compiled and CLU-\ha~galaxy samples. Panels a, b, c, and d show the histograms of $g-r$ color, absolute $M_r$-band magnitude, FUV-derived SFR, and stellar mass, respectively. We find that CLU-\ha~will likely add galaxies that tend to be blue in optical color and high in SFR to existing catalogs of galaxies; however, will miss more massive galaxies that are red in optical color (i.e., red-sequcence galaxies).}
   \label{fig:clugenprop}
   \end{center}
\end{figure*}

Next, we examine the basic properties of our CLU-compiled catalog and investigate how the new galaxies from the CLU-\ha~survey will contribute to the full CLU catalog. Figure~\ref{fig:clugenprop} shows the observable and physical property histograms of both the CLU-compiled galaxies (open histogram) and CLU-\ha~galaxies in the preliminary fields (blue-filled histogram), where all histograms have been normalized to the peak. 

Panel a) shows the SDSS $g-r$ color histograms, where the CLU-compiled sample shows a double peak while the CLU-\ha~sample shows a single peak that overlaps with the CLU-compiled blue peak. The double peak is due to the dichotomy between blue star-forming galaxies and red early-type galaxies with little-to-no star formation. The absence of a red peak in the CLU-\ha~sample shows that the CLU-\ha~survey is not sensitive to early-type galaxies. This can also be seen in panel c) which shows the FUV-derived SFRs, where the CLU-\ha~galaxies are peaked at 0.3~dex higher than that of CLU-compiled. Thus, the galaxies added by the CLU-\ha~survey will be biased towards bluer galaxies with recent star formation. 

Panel b) shows the absolute magnitude $M_{\rm r}$ histogram, where both CLU-compiled and CLU-\ha~samples show two peaks: one for intrinsically bright galaxies ($M_{\rm r}\sim$-19~mag) and one for fainter galaxies ($M_{\rm r}\sim-14$~mag). The reduced number of low luminosity CLU-\ha~galaxies suggests that the CLU-\ha~survey will not probe the same number densities of dwarfs found in the volume out to 200~Mpc. There is also a deficit of intrinsically bright CLU-\ha~galaxies ($M_{\rm r} \lesssim -20$~mag) compared to CLU-compiled. This can also be seen in panel d) which shows the stellar mass histograms, where the CLU-compiled sample peaks at 0.5 dex higher in M$_{\star}$. An examination of the morphologies of the galaxies brighter than --20~mag shows that 60\% have RC3 T-types less than 1 (or S0 and earlier morphologies). Thus, the CLU-\ha~survey will not add massive early-type galaxies to existing catalogs. However, the majority of these luminous galaxies are already in CLU-compiled given the inclusion of SDSS galaxy spectra.




We will add CLU-\ha~galaxies found in our narrow-band survey with no previous distance information to our CLU-compiled catalog to produce a more complete census of the local Universe. This combined catalog will then be utilized to focus the search for future electromagnetic counterparts to gravitational wave events.

\subsection{Future Improvements}
Future improvements to maximize the completeness and minimize the contamination of the CLU-\ha~catalog include:
(i) Stacking all \ha~images to produce a deeper \ha~source catalog and a resulting deeper galaxy catalog. These data will not only provide deeper images from which to find more galaxies, but will also facilitate removal of cosmic rays and chip defects. 
(ii) Extended source photometry;
(iii) Machine learning for star-galaxy separation;
(iv) Image subtraction of ``On'' and ``Off'' filters. 

In addition, efforts to use machine learning for candidate selection are currently underway. The preliminary CLU-\ha~survey fields are critical to our machine learning efforts as they establish a training set of objects with known redshifts.  We have run preliminary tests on how well our machine learning algorithms perform the tasks of classifying objects as simply within 200~Mpc and classifying objects as within one of our narrow-band filters.  Preliminary results are promising, showing that we can increase our completeness by using our current metrics along with additional information (e.g., Pan-STARRS, WISE, GALEX magnitudes) as features in our machine learning algorithms (Zhang et al. 2018; in prep).




\section{Summary}
In this paper we have presented the Census of the Local Universe (CLU) emission-line (\ha) galaxy survey. The CLU-\ha~survey has imaged 26,470 square degrees of the northern sky above $-20^{\circ}$ declination using four narrow-band filters with a FWHM of 75$-$90~\AA~and a wavelength range of 6525$-$6878~\AA~(out to z=0.047). The observations utilize 3 spatially staggered grids where each grid has 3626 fields. The first two filters cover the galactic plane and the last two filters avoid the galactic plane ($|b|\gtrsim 3^{\circ}$). The analysis of CLU-\ha~fields in this study utilize only 14 preliminary fields ($\approx$100~deg$^2$) in one of the spatially staggered grids; however, future studies will examine the full $\approx$3$\pi$ area with 3 stacked images on nearly every point in the CLU-\ha~survey area.

In the 14 preliminary fields we implemented a widely used signal-to-noise selection method to quantify the presence of an emission-line via narrow-band color excess significance ($\Sigma$). In addition, we undertook a spectroscopic follow-up campaign at Palomar Observatory and WIRO to obtain redshifts of all galaxy candidates with $\Sigma\geq$2.5 and no previous redshift information (N=334). These redshifts allow us to explore the composition of our galaxy candidates at different $\Sigma$ cuts. There are 290 galaxies whose \ha~emission line has been confirmed in our filters with 61\% contamination for $\Sigma\geq$2.5, and 151 confirmed galaxies and 12\% contamination for $\Sigma\geq$5. We conclude that $\Sigma=2.5$ cut produces a more complete catalog of galaxies in our local volume but has higher contamination while a $\Sigma=5$ cut produces a high fidelity galaxy catalog with little contamination. We release a table of all candidates above a $\Sigma=2.5$ cut in the electronic version of this publication.

In addition, we use the 14 preliminary fields to examine the limits of CLU-\ha~survey. The median 5$\sigma$ detection limit for a point source in the narrow-band imaging is 18.5~mag. A comparison of \ha~fluxes derived from narrow-band imaging and spectroscopy show good agreement for the majority of galaxies, but show some contamination from [NII] lines near the \ha~lines. A comparison to galaxies with spectroscopic redshifts and fluxes in our preliminary fields shows that we recover 90\% of the \ha~flux at $1\times10^{-14}~\rm{erg~s^{-1}~cm^{-2}}$  and $4\times10^{-14}~\rm{erg~s^{-1}~cm^{-2}}$ for $\Sigma$ cuts above 2.5 and 5, respectively.  

The CLU-\ha~survey is able to recover the majority of the galaxies in both the Markarian and Case surveys, and was able to find an additional $\sim$250 galaxies when compared to the Markarian survey. Our comparison with the KISS survey shows that CLU-\ha~will recover roughly half as many emission-line galaxies as KISS due to our brighter limits. However, CLU-\ha~will cover a much larger area of the sky (3$\pi$~sr) compared to all previous emission-line galaxy surveys to a moderate depth. Thus, the CLU-\ha~survey can be used as for the discovery of many emission-line galaxies across the sky.

We have cross-matched the resulting CLU-\ha~galaxies to GALEX and WISE to derive the physical properties (extinction corrected SFR and stellar mass). We find that the CLU-\ha~galaxies span a range in both SFR and stellar mass including dwarf galaxies (i.e., M$_{\star}\sim10^8$~M$_{\odot}$ and SFR$\sim10^{-2}$~M$_{\odot}~\rm{yr}^{-1}$) as well as larger spirals (i.e., M$_{\star}\sim10^{11}$~M$_{\odot}$ and SFR$\sim10^{1}$~M$_{\odot}~\rm{yr}^{-1}$). We find that the majority of the CLU-\ha~galaxies show agreement with galaxy \ms~trends found in the local Universe. However, we do find some extreme galaxies that lie above the \ms~trend and some of the highest sSFRs in the CLU-\ha~sample.

Within the 14 preliminary fields, we find several interesting objects with no previous distance information: 7 blue compact dwarfs, 1 green pea, and a Seyfert 1 galaxy; we also identified a known planetary nebula. The existence of these objects in our preliminary fields (in just 0.3\% of the full survey) exemplifies that the full CLU-\ha~survey will serve as a discovery machine for a wide variety of objects in our own Galaxy and for extreme galaxies out to intermediate redshifts. 

Finally, the CLU-\ha~galaxies with newly constrained distances will be added to existing galaxy catalogs to focus the search for the electromagnetic counterparts to gravitational wave events. We have constructed a list of galaxies with distances out to 200~Mpc (CLU-compiled) and will add the CLU-\ha~galaxies to this list. Our team has already successfully applied the CLU-compiled catalog towards this goal by publishing a prioritized list of galaxies within the neutron star merger event detected by the LIGO and Virgo detectors on August 17th 2017 (GW170817). Within hours of the LIGO trigger, our galaxy list was published on the Gamma-ray Coordinates Network (GCN) and the electromagnetic counterpart to GW170817 was later found in the third highest priority galaxy on our list.

\bibliographystyle{tex/mn2e.bst}   
\bibliography{tex/all}  

\begin{thebibliography}{115}
\expandafter\ifx\csname natexlab\endcsname\relax\def\natexlab#1{#1}\fi

\bibitem[{{Aasi} {et~al}\mbox{.}(2015){Aasi}, {Abbott}, {Abbott}, {Abbott},
  {Abernathy}, {Ackley}, {Adams}, {Adams}, {Addesso}, \& et~al.}]{aligo}
{Aasi} J. {et~al.}, 2015, Classical and Quantum Gravity, 32, 074001

\bibitem[{{Abazajian} {et~al}\mbox{.}(2009){Abazajian}, {Adelman-McCarthy},
  {Ag{\"u}eros}, {Allam}, {Allende Prieto}, {An}, {Anderson}, {Anderson},
  {Annis}, {Bahcall}, \& et~al.}]{sdss7}
{Abazajian} K.~N. {et~al.}, 2009, \apjs, 182, 543

\bibitem[{{Abbott} {et~al}\mbox{.}(2016){Abbott}, {Abbott}, {Abbott},
  {Abernathy}, {Acernese}, {Ackley}, {Adams}, {Adams}, {Addesso}, {Adhikari},
  \& et~al.}]{abbott16}
{Abbott} B.~P. {et~al.}, 2016, Physical Review Letters, 116, 061102

\bibitem[{{Abbott} {et~al}\mbox{.}(2017{\natexlab{a}}){Abbott}, {Abbott},
  {Abbott}, {Acernese}, {Ackley}, {Adams}, {Adams}, {Addesso}, {Adhikari},
  {Adya}, \& et~al.}]{ligodet170817}
{Abbott} B.~P. {et~al.}, 2017{\natexlab{a}}, Physical Review Letters, 119,
  161101

\bibitem[{{Abbott} {et~al}\mbox{.}(2017{\natexlab{b}}){Abbott}, {Abbott},
  {Abbott}, {Acernese}, {Ackley}, {Adams}, {Adams}, {Addesso}, {Adhikari},
  {Adya}, \& et~al.}]{GW170817}
{Abbott} B.~P. {et~al.}, 2017{\natexlab{b}}, Physical Review Letters, 119,
  161101

\bibitem[{{Alam} {et~al}\mbox{.}(2015){Alam}, {Albareti}, {Allende Prieto},
  {Anders}, {Anderson}, {Anderton}, {Andrews}, {Armengaud}, {Aubourg},
  {Bailey}, \& et~al.}]{sdss12}
{Alam} S. {et~al.}, 2015, \apjs, 219, 12

\bibitem[{{Amor{\'{\i}}n}, {P{\'e}rez-Montero} \&
  {V{\'{\i}}lchez}(2010){Amor{\'{\i}}n}, {P{\'e}rez-Montero}, \&
  {V{\'{\i}}lchez}}]{amorin10}
{Amor{\'{\i}}n} R.~O., {P{\'e}rez-Montero} E., {V{\'{\i}}lchez} J.~M., 2010,
  \apjl, 715, L128

\bibitem[{{Barnes} {et~al}\mbox{.}(2014){Barnes}, {van Zee}, {Dale},
  {Staudaher}, {Bullock}, {Calzetti}, {Chandar}, \& {Dalcanton}}]{barnes14}
{Barnes} K.~L., {van Zee} L., {Dale} D.~A., {Staudaher} S., {Bullock} J.~S.,
  {Calzetti} D., {Chandar} R., {Dalcanton} J.~J., 2014, \apj, 789, 126

\bibitem[{{Bellm} \& {Sesar}(2016)}]{bellem16}
{Bellm} E.~C., {Sesar} B., 2016, {pyraf-dbsp: Reduction pipeline for the
  Palomar Double Beam Spectrograph}. Astrophysics Source Code Library

\bibitem[{{Bertin}(2006)}]{scamp}
{Bertin} E., 2006, in Astronomical Society of the Pacific Conference Series,
  Vol. 351, Astronomical Data Analysis Software and Systems XV, {Gabriel} C.,
  {Arviset} C., {Ponz} D., {Enrique} S., eds., p. 112

\bibitem[{{Bertin} \& {Arnouts}(1996)}]{sex96}
{Bertin} E., {Arnouts} S., 1996, \aaps, 117, 393

\bibitem[{{Bianchi}, {Conti} \& {Shiao}(2014){Bianchi}, {Conti}, \&
  {Shiao}}]{galexgr67}
{Bianchi} L., {Conti} A., {Shiao} B., 2014, VizieR Online Data Catalog, 2335

\bibitem[{{Brinchmann} {et~al}\mbox{.}(2004){Brinchmann}, {Charlot}, {White},
  {Tremonti}, {Kauffmann}, {Heckman}, \& {Brinkmann}}]{brinchmann04}
{Brinchmann} J., {Charlot} S., {White} S.~D.~M., {Tremonti} C., {Kauffmann} G.,
  {Heckman} T., {Brinkmann} J., 2004, \mnras, 351, 1151

\bibitem[{{Bunker} {et~al}\mbox{.}(1995){Bunker}, {Warren}, {Hewett}, \&
  {Clements}}]{bunker95}
{Bunker} A.~J., {Warren} S.~J., {Hewett} P.~C., {Clements} D.~L., 1995, \mnras,
  273, 513

\bibitem[{{Calzetti} {et~al}\mbox{.}(2010){Calzetti}, {Wu}, {Hong},
  {Kennicutt}, {Lee}, {Dale}, {Engelbracht}, {van Zee}, {Draine}, {Hao},
  {Gordon}, {Moustakas}, {Murphy}, {Regan}, {Begum}, {Block}, {Dalcanton},
  {Funes}, {Gil de Paz}, {Johnson}, {Sakai}, {Skillman}, {Walter}, {Weisz},
  {Williams}, \& {Wu}}]{calzetti10a}
{Calzetti} D. {et~al.}, 2010, \apj, 714, 1256

\bibitem[{{Cardamone} {et~al}\mbox{.}(2009){Cardamone}, {Schawinski}, {Sarzi},
  {Bamford}, {Bennert}, {Urry}, {Lintott}, {Keel}, {Parejko}, {Nichol},
  {Thomas}, {Andreescu}, {Murray}, {Raddick}, {Slosar}, {Szalay}, \&
  {Vandenberg}}]{cardamone09}
{Cardamone} C. {et~al.}, 2009, \mnras, 399, 1191

\bibitem[{{Chambers} {et~al}\mbox{.}(2016){Chambers}, {Magnier}, {Metcalfe},
  {Flewelling}, {Huber}, {Waters}, {Denneau}, {Draper}, {Farrow}, {Finkbeiner},
  {Holmberg}, {Koppenhoefer}, {Price}, {Saglia}, {Schlafly}, {Smartt},
  {Sweeney}, {Wainscoat}, {Burgett}, {Grav}, {Heasley}, {Hodapp}, {Jedicke},
  {Kaiser}, {Kudritzki}, {Luppino}, {Lupton}, {Monet}, {Morgan}, {Onaka},
  {Stubbs}, {Tonry}, {Banados}, {Bell}, {Bender}, {Bernard}, {Botticella},
  {Casertano}, {Chastel}, {Chen}, {Chen}, {Cole}, {Deacon}, {Frenk},
  {Fitzsimmons}, {Gezari}, {Goessl}, {Goggia}, {Goldman}, {Grebel}, {Hambly},
  {Hasinger}, {Heavens}, {Heckman}, {Henderson}, {Henning}, {Holman}, {Hopp},
  {Ip}, {Isani}, {Keyes}, {Koekemoer}, {Kotak}, {Long}, {Lucey}, {Liu},
  {Martin}, {McLean}, {Morganson}, {Murphy}, {Nieto-Santisteban}, {Norberg},
  {Peacock}, {Pier}, {Postman}, {Primak}, {Rae}, {Rest}, {Riess}, {Riffeser},
  {Rix}, {Roser}, {Schilbach}, {Schultz}, {Scolnic}, {Szalay}, {Seitz},
  {Shiao}, {Small}, {Smith}, {Soderblom}, {Taylor}, {Thakar}, {Thiel},
  {Thilker}, {Urata}, {Valenti}, {Walter}, {Watters}, {Werner}, {White},
  {Wood-Vasey}, \& {Wyse}}]{ps1}
{Chambers} K.~C. {et~al.}, 2016, ArXiv e-prints

\bibitem[{{Cook} {et~al}\mbox{.}(2017){Cook}, {Van Sistine}, {Singer}, \&
  {Kasliwal}}]{GCN21519}
{Cook} D., {Van Sistine} A., {Singer} L., {Kasliwal} M., 2017, Gamma Ray
  Coordinates Network Circular, 21519

\bibitem[{{Cook} {et~al}\mbox{.}(2014{\natexlab{a}}){Cook}, {Dale}, {Johnson},
  {Van Zee}, {Lee}, {Kennicutt}, {Calzetti}, {Staudaher}, \&
  {Engelbracht}}]{cook14a}
{Cook} D.~O. {et~al.}, 2014{\natexlab{a}}, \mnras, 445, 881

\bibitem[{{Cook} {et~al}\mbox{.}(2014{\natexlab{b}}){Cook}, {Dale}, {Johnson},
  {Van Zee}, {Lee}, {Kennicutt}, {Calzetti}, {Staudaher}, \&
  {Engelbracht}}]{cook14c}
{Cook} D.~O. {et~al.}, 2014{\natexlab{b}}, \mnras, 445, 899

\bibitem[{{Coulter} {et~al}\mbox{.}(2017){Coulter}, {Foley}, {Kilpatrick},
  {Drout}, {Piro}, {Shappee}, {Siebert}, {Simon}, {Ulloa}, {Kasen}, {Madore},
  {Murguia-Berthier}, {Pan}, {Prochaska}, {Ramirez-Ruiz}, {Rest}, \&
  {Rojas-Bravo}}]{coulter17}
{Coulter} D.~A. {et~al.}, 2017, ArXiv e-prints

\bibitem[{{Daddi} {et~al}\mbox{.}(2007){Daddi}, {Dickinson}, {Morrison},
  {Chary}, {Cimatti}, {Elbaz}, {Frayer}, {Renzini}, {Pope}, {Alexander},
  {Bauer}, {Giavalisco}, {Huynh}, {Kurk}, \& {Mignoli}}]{daddi07}
{Daddi} E. {et~al.}, 2007, \apj, 670, 156

\bibitem[{{Dale} {et~al}\mbox{.}(2009){Dale}, {Cohen}, {Johnson}, {Schuster},
  {Calzetti}, {Engelbracht}, {Gil de Paz}, {Kennicutt}, {Lee}, {Begum},
  {Block}, {Dalcanton}, {Funes}, {Gordon}, {Johnson}, {Marble}, {Sakai},
  {Skillman}, {van Zee}, {Walter}, {Weisz}, {Williams}, {Wu}, \& {Wu}}]{dale09}
{Dale} D.~A. {et~al.}, 2009, \apj, 703, 517

\bibitem[{{Dale} {et~al}\mbox{.}(1999){Dale}, {Giovanelli}, {Haynes},
  {Campusano}, \& {Hardy}}]{dale99}
{Dale} D.~A., {Giovanelli} R., {Haynes} M.~P., {Campusano} L.~E., {Hardy} E.,
  1999, \aj, 118, 1489

\bibitem[{{Drew} {et~al}\mbox{.}(2005){Drew}, {Greimel}, {Irwin},
  {Aungwerojwit}, {Barlow}, {Corradi}, {Drake}, {G{\"a}nsicke}, {Groot},
  {Hales}, {Hopewell}, {Irwin}, {Knigge}, {Leisy}, {Lennon}, {Mampaso},
  {Masheder}, {Matsuura}, {Morales-Rueda}, {Morris}, {Parker}, {Phillipps},
  {Rodriguez-Gil}, {Roelofs}, {Skillen}, {Sokoloski}, {Steeghs}, {Unruh},
  {Viironen}, {Vink}, {Walton}, {Witham}, {Wright}, {Zijlstra}, \&
  {Zurita}}]{iphas}
{Drew} J.~E. {et~al.}, 2005, \mnras, 362, 753

\bibitem[{{Elbaz} {et~al}\mbox{.}(2007){Elbaz}, {Daddi}, {Le Borgne},
  {Dickinson}, {Alexander}, {Chary}, {Starck}, {Brandt}, {Kitzbichler},
  {MacDonald}, {Nonino}, {Popesso}, {Stern}, \& {Vanzella}}]{elbaz07}
{Elbaz} D. {et~al.}, 2007, \aap, 468, 33

\bibitem[{{Eskew}, {Zaritsky} \& {Meidt}(2012){Eskew}, {Zaritsky}, \&
  {Meidt}}]{eskew12}
{Eskew} M., {Zaritsky} D., {Meidt} S., 2012, \aj, 143, 139

\bibitem[{{Evans} {et~al}\mbox{.}(2017){Evans}, {Cenko}, {Kennea}, {Emery},
  {Kuin}, {Korobkin}, {Wollaeger}, {Fryer}, {Madsen}, {Harrison}, {Xu},
  {Nakar}, {Hotokezaka}, {Lien}, {Campana}, {Oates}, {Troja}, {Breeveld},
  {Marshall}, {Barthelmy}, {Beardmore}, {Burrows}, {Cusumano}, {D'Ai},
  {D'Avanzo}, {D'Elia}, {de Pasquale}, {Even}, {Fontes}, {Forster}, {Garcia},
  {Giommi}, {Grefenstette}, {Gronwall}, {Hartmann}, {Heida}, {Hungerford},
  {Kasliwal}, {Krimm}, {Levan}, {Malesani}, {Melandri}, {Miyasaka}, {Nousek},
  {O'Brien}, {Osborne}, {Pagani}, {Page}, {Palmer}, {Perri}, {Pike}, {Racusin},
  {Rosswog}, {Siegel}, {Sakamoto}, {Sbarufatti}, {Tagliaferri}, {Tanvir}, \&
  {Tohuvavohu}}]{evans17}
{Evans} P.~A. {et~al.}, 2017, ArXiv e-prints

\bibitem[{{Finkelstein} {et~al}\mbox{.}(2012){Finkelstein}, {Papovich}, {Ryan},
  {Pawlik}, {Dickinson}, {Ferguson}, {Finlator}, {Koekemoer}, {Giavalisco},
  {Cooray}, {Dunlop}, {Faber}, {Grogin}, {Kocevski}, \&
  {Newman}}]{finkelstein12}
{Finkelstein} S.~L. {et~al.}, 2012, \apj, 758, 93

\bibitem[{Fujita {et~al}\mbox{.}(2003)Fujita, Ajiki, Shioya, Nagao, Murayama,
  Taniguchi, Umeda, Yamada, Yagi, Okamura, \& Komiyama}]{fujita03}
Fujita S.~S. {et~al.}, 2003, The Astrophysical Journal Letters, 586, L115

\bibitem[{{Fukugita} {et~al}\mbox{.}(2007){Fukugita}, {Nakamura}, {Okamura},
  {Yasuda}, {Barentine}, {Brinkmann}, {Gunn}, {Harvanek}, {Ichikawa}, {Lupton},
  {Schneider}, {Strauss}, \& {York}}]{fukugita07}
{Fukugita} M. {et~al.}, 2007, \aj, 134, 579

\bibitem[{{Geach} {et~al}\mbox{.}(2008){Geach}, {Smail}, {Best}, {Kurk},
  {Casali}, {Ivison}, \& {Coppin}}]{geach08}
{Geach} J.~E., {Smail} I., {Best} P.~N., {Kurk} J., {Casali} M., {Ivison}
  R.~J., {Coppin} K., 2008, \mnras, 388, 1473

\bibitem[{{Gehrels} {et~al}\mbox{.}(2016){Gehrels}, {Cannizzo}, {Kanner},
  {Kasliwal}, {Nissanke}, \& {Singer}}]{gehrels16}
{Gehrels} N., {Cannizzo} J.~K., {Kanner} J., {Kasliwal} M.~M., {Nissanke} S.,
  {Singer} L.~P., 2016, \apj, 820, 136

\bibitem[{{Gronwall} {et~al}\mbox{.}(2004){Gronwall}, {Salzer}, {Sarajedini},
  {Jangren}, {Chomiuk}, {Moody}, {Frattare}, \& {Boroson}}]{kiss04}
{Gronwall} C., {Salzer} J.~J., {Sarajedini} V.~L., {Jangren} A., {Chomiuk} L.,
  {Moody} J.~W., {Frattare} L.~M., {Boroson} T.~A., 2004, \aj, 127, 1943

\bibitem[{{Hallinan} {et~al}\mbox{.}(2017){Hallinan}, {Corsi}, {Mooley},
  {Hotokezaka}, {Nakar}, {Kasliwal}, {Kaplan}, {Frail}, {Myers}, {Murphy},
  {De}, {Dobie}, {Allison}, {Bannister}, {Bhalerao}, {Chandra}, {Clarke},
  {Giacintucci}, {Ho}, {Horesh}, {Kassim}, {Kulkarni}, {Lenc}, {Lockman},
  {Lynch}, {Nichols}, {Nissanke}, {Palliyaguru}, {Peters}, {Piran}, {Rana},
  {Sadler}, \& {Singer}}]{hallinan17}
{Hallinan} G. {et~al.}, 2017, ArXiv e-prints

\bibitem[{{Haro}(1951)}]{haro51}
{Haro} G., 1951, Publications of the Astronomical Society of the Pacific, 63,
  144

\bibitem[{{Haynes} {et~al}\mbox{.}(2011){Haynes}, {Giovanelli}, {Martin},
  {Hess}, {Saintonge}, {Adams}, {Hallenbeck}, {Hoffman}, {Huang}, {Kent},
  {Koopmann}, {Papastergis}, {Stierwalt}, {Balonek}, {Craig}, {Higdon},
  {Kornreich}, {Miller}, {O'Donoghue}, {Olowin}, {Rosenberg}, {Spekkens},
  {Troischt}, \& {Wilcots}}]{haynes11}
{Haynes} M.~P. {et~al.}, 2011, \aj, 142, 170

\bibitem[{{Heinis} {et~al}\mbox{.}(2014){Heinis}, {Buat}, {B{\'e}thermin},
  {Bock}, {Burgarella}, {Conley}, {Cooray}, {Farrah}, {Ilbert}, {Magdis},
  {Marsden}, {Oliver}, {Rigopoulou}, {Roehlly}, {Schulz}, {Symeonidis},
  {Viero}, {Xu}, \& {Zemcov}}]{heinis14}
{Heinis} S. {et~al.}, 2014, \mnras, 437, 1268

\bibitem[{{Hirschauer} {et~al}\mbox{.}(2016){Hirschauer}, {Salzer}, {Skillman},
  {Berg}, {McQuinn}, {Cannon}, {Gordon}, {Haynes}, {Giovanelli}, {Adams},
  {Janowiecki}, {Rhode}, {Pogge}, {Croxall}, \& {Aver}}]{hirschauer16}
{Hirschauer} A.~S. {et~al.}, 2016, \apj, 822, 108

\bibitem[{{Izotov}, {Guseva} \& {Thuan}(2011){Izotov}, {Guseva}, \&
  {Thuan}}]{izotov11}
{Izotov} Y.~I., {Guseva} N.~G., {Thuan} T.~X., 2011, \apj, 728, 161

\bibitem[{{Izotov} {et~al}\mbox{.}(2006){Izotov}, {Papaderos}, {Guseva},
  {Fricke}, \& {Thuan}}]{izotov06}
{Izotov} Y.~I., {Papaderos} P., {Guseva} N.~G., {Fricke} K.~J., {Thuan} T.~X.,
  2006, \aap, 454, 137

\bibitem[{{Izotov} \& {Thuan}(2007)}]{izotov07}
{Izotov} Y.~I., {Thuan} T.~X., 2007, \apj, 665, 1115

\bibitem[{{Izotov}, {Thuan} \& {Guseva}(2005){Izotov}, {Thuan}, \&
  {Guseva}}]{izotov05}
{Izotov} Y.~I., {Thuan} T.~X., {Guseva} N.~G., 2005, \apj, 632, 210

\bibitem[{{Izotov} {et~al}\mbox{.}(2018){Izotov}, {Thuan}, {Guseva}, \&
  {Liss}}]{izotov18}
{Izotov} Y.~I., {Thuan} T.~X., {Guseva} N.~G., {Liss} S.~E., 2018, \mnras, 473,
  1956

\bibitem[{{Jangren} {et~al}\mbox{.}(2005){Jangren}, {Salzer}, {Sarajedini},
  {Gronwall}, {Werk}, {Chomiuk}, {Moody}, \& {Boroson}}]{kiss05}
{Jangren} A., {Salzer} J.~J., {Sarajedini} V.~L., {Gronwall} C., {Werk} J.~K.,
  {Chomiuk} L., {Moody} J.~W., {Boroson} T.~A., 2005, \aj, 130, 2571

\bibitem[{{Jarrett} {et~al}\mbox{.}(2013){Jarrett}, {Masci}, {Tsai}, {Petty},
  {Cluver}, {Assef}, {Benford}, {Blain}, {Bridge}, {Donoso}, {Eisenhardt},
  {Koribalski}, {Lake}, {Neill}, {Seibert}, {Sheth}, {Stanford}, \&
  {Wright}}]{jarret13}
{Jarrett} T.~H. {et~al.}, 2013, \aj, 145, 6

\bibitem[{{Jones} {et~al}\mbox{.}(2009){Jones}, {Read}, {Saunders}, {Colless},
  {Jarrett}, {Parker}, {Fairall}, {Mauch}, {Sadler}, {Watson}, {Burton},
  {Campbell}, {Cass}, {Croom}, {Dawe}, {Fiegert}, {Frankcombe}, {Hartley},
  {Huchra}, {James}, {Kirby}, {Lahav}, {Lucey}, {Mamon}, {Moore}, {Peterson},
  {Prior}, {Proust}, {Russell}, {Safouris}, {Wakamatsu}, {Westra}, \&
  {Williams}}]{jones09}
{Jones} D.~H. {et~al.}, 2009, \mnras, 399, 683

\bibitem[{{Kasliwal} {et~al}\mbox{.}(2016){Kasliwal}, {Cenko}, {Singer},
  {Corsi}, {Cao}, {Barlow}, {Bhalerao}, {Bellm}, {Cook}, {Duggan}, {Ferretti},
  {Frail}, {Horesh}, {Kendrick}, {Kulkarni}, {Lunnan}, {Palliyaguru}, {Laher},
  {Masci}, {Manulis}, {Miller}, {Nugent}, {Perley}, {Prince}, {Rana},
  {Rebbapragada}, {Sesar}, {Singhal}, {Surace}, \& {Van Sistine}}]{kasliwal16}
{Kasliwal} M.~M. {et~al.}, 2016, ArXiv e-prints

\bibitem[{{Kasliwal} {et~al}\mbox{.}(2017){Kasliwal}, {Nakar}, {Singer},
  {Kaplan}, {Cook}, {Van Sistine}, {Lau}, {Fremling}, {Gottlieb}, {Jencson},
  {Adams}, {Feindt}, {Hotokezaka}, {Ghosh}, {Perley}, {Yu}, {Piran}, {Allison},
  {Anupama}, {Balasubramanian}, {Bannister}, {Bally}, {Barnes}, {Barway},
  {Bellm}, {Bhalerao}, {Bhattacharya}, {Blagorodnova}, {Bloom}, {Brady},
  {Cannella}, {Chatterjee}, {Cenko}, {Cobb}, {Copperwheat}, {Corsi}, {De},
  {Dobie}, {Emery}, {Evans}, {Fox}, {Frail}, {Frohmaier}, {Goobar}, {Hallinan},
  {Harrison}, {Helou}, {Hinderer}, {Ho}, {Horesh}, {Ip}, {Itoh}, {Kasen},
  {Kim}, {Kuin}, {Kupfer}, {Lynch}, {Madsen}, {Mazzali}, {Miller}, {Mooley},
  {Murphy}, {Ngeow}, {Nichols}, {Nissanke}, {Nugent}, {Ofek}, {Qi}, {Quimby},
  {Rosswog}, {Rusu}, {Sadler}, {Schmidt}, {Sollerman}, {Steele}, {Williamson},
  {Xu}, {Yan}, {Yatsu}, {Zhang}, \& {Zhao}}]{kasliwal17}
{Kasliwal} M.~M. {et~al.}, 2017, ArXiv e-prints

\bibitem[{{Kennicutt} \& {Evans}(2012)}]{kennicutt12}
{Kennicutt} R.~C., {Evans} N.~J., 2012, \araa, 50, 531

\bibitem[{{Kennicutt}(1998)}]{kennicutt98}
{Kennicutt}, Jr. R.~C., 1998, \araa, 36, 189

\bibitem[{{Kennicutt} {et~al}\mbox{.}(2008){Kennicutt}, {Lee}, {Funes},
  {Sakai}, \& {Akiyama}}]{kennicutt08}
{Kennicutt}, Jr. R.~C., {Lee} J.~C., {Funes}, Jos{\'e}~G. S.~J., {Sakai} S.,
  {Akiyama} S., 2008, \apjs, 178, 247

\bibitem[{{Kopparapu} {et~al}\mbox{.}(2008){Kopparapu}, {Hanna}, {Kalogera},
  {O'Shaughnessy}, {Gonz{\'a}lez}, {Brady}, \& {Fairhurst}}]{kopparapu08}
{Kopparapu} R.~K., {Hanna} C., {Kalogera} V., {O'Shaughnessy} R.,
  {Gonz{\'a}lez} G., {Brady} P.~R., {Fairhurst} S., 2008, \apj, 675, 1459

\bibitem[{{Kroupa}(2001)}]{kroupa01}
{Kroupa} P., 2001, \mnras, 322, 231

\bibitem[{{Laher} {et~al}\mbox{.}(2014){Laher}, {Surace}, {Grillmair}, {Ofek},
  {Levitan}, {Sesar}, {van Eyken}, {Law}, {Helou}, {Hamam}, {Masci},
  {Mattingly}, {Jackson}, {Hacopeans}, {Mi}, {Groom}, {Teplitz}, {Desai},
  {Hale}, {Smith}, {Walters}, {Quimby}, {Kasliwal}, {Horesh}, {Bellm},
  {Barlow}, {Waszczak}, {Prince}, \& {Kulkarni}}]{laher14}
{Laher} R.~R. {et~al.}, 2014, \pasp, 126, 674

\bibitem[{Law {et~al}\mbox{.}(2009)Law, Kulkarni, Dekany, Ofek, Quimby, Nugent,
  Surace, Grillmair, Bloom, Kasliwal, Bildsten, Brown, Cenko, Ciardi, Croner,
  Djorgovski, van Eyken, Filippenko, Fox, Gal-Yam, Hale, Hamam, Helou, Henning,
  Howell, Jacobsen, Laher, Mattingly, McKenna, Pickles, Poznanski, Rahmer, Rau,
  Rosing, Shara, Smith, Starr, Sullivan, Velur, Walters, \& Zolkower}]{law09}
Law N.~M. {et~al.}, 2009, Publications of the Astronomical Society of the
  Pacific, 121, 1395

\bibitem[{{Lee} {et~al}\mbox{.}(2011){Lee}, {Gil de Paz}, {Kennicutt},
  {Bothwell}, {Dalcanton}, {Jos{\'e} G.~Funes S.}, {Johnson}, {Sakai},
  {Skillman}, {Tremonti}, \& {van Zee}}]{lee11}
{Lee} J.~C. {et~al.}, 2011, \apjs, 192, 6

\bibitem[{{Lee} {et~al}\mbox{.}(2012){Lee}, {Ly}, {Spitler}, {Labb{\'e}},
  {Salim}, {Persson}, {Ouchi}, {Dale}, {Monson}, \& {Murphy}}]{jlee12}
{Lee} J.~C. {et~al.}, 2012, \pasp, 124, 782

\bibitem[{{Lintott} {et~al}\mbox{.}(2008){Lintott}, {Schawinski}, {Slosar},
  {Land}, {Bamford}, {Thomas}, {Raddick}, {Nichol}, {Szalay}, {Andreescu},
  {Murray}, \& {Vandenberg}}]{lintott08}
{Lintott} C.~J. {et~al.}, 2008, \mnras, 389, 1179

\bibitem[{Ly {et~al}\mbox{.}(2010)Ly, Lee, Dale, Momcheva, Salim, Staudaher,
  Moore, \& Finn}]{ly10}
Ly C., Lee J.~C., Dale D.~A., Momcheva I., Salim S., Staudaher S., Moore C.~A.,
  Finn R., 2010, The Astrophysical Journal, 726, 109

\bibitem[{{MacAlpine} \& {Lewis}(1978)}]{UM4}
{MacAlpine} G.~M., {Lewis} D.~W., 1978, \apjs, 36, 587

\bibitem[{{MacAlpine}, {Lewis} \& {Smith}(1977){MacAlpine}, {Lewis}, \&
  {Smith}}]{UM3}
{MacAlpine} G.~M., {Lewis} D.~W., {Smith} S.~B., 1977, \apjs, 35, 203

\bibitem[{{MacAlpine}, {Smith} \& {Lewis}(1977{\natexlab{a}}){MacAlpine},
  {Smith}, \& {Lewis}}]{UM1}
{MacAlpine} G.~M., {Smith} S.~B., {Lewis} D.~W., 1977{\natexlab{a}}, \apjs, 34,
  95

\bibitem[{{MacAlpine}, {Smith} \& {Lewis}(1977{\natexlab{b}}){MacAlpine},
  {Smith}, \& {Lewis}}]{UM2}
{MacAlpine} G.~M., {Smith} S.~B., {Lewis} D.~W., 1977{\natexlab{b}}, \apjs, 35,
  197

\bibitem[{{MacAlpine} \& {Williams}(1981)}]{UM5}
{MacAlpine} G.~M., {Williams} G.~A., 1981, \apjs, 45, 113

\bibitem[{{Mahajan}, {Haines} \& {Raychaudhury}(2010){Mahajan}, {Haines}, \&
  {Raychaudhury}}]{mahajan10}
{Mahajan} S., {Haines} C.~P., {Raychaudhury} S., 2010, \mnras, 404, 1745

\bibitem[{{Makarov} {et~al}\mbox{.}(2014){Makarov}, {Prugniel}, {Terekhova},
  {Courtois}, \& {Vauglin}}]{hyperleda}
{Makarov} D., {Prugniel} P., {Terekhova} N., {Courtois} H., {Vauglin} I., 2014,
  \aap, 570, A13

\bibitem[{{Markarian}, {Lipovetskii} \& {Stepanian}(1981){Markarian},
  {Lipovetskii}, \& {Stepanian}}]{markarian81}
{Markarian} B.~E., {Lipovetskii} V.~A., {Stepanian} D.~A., 1981, Astrofizika,
  17, 619

\bibitem[{{Markarian} {et~al}\mbox{.}(1989){Markarian}, {Lipovetsky},
  {Stepanian}, {Erastova}, \& {Shapovalova}}]{markarian89}
{Markarian} B.~E., {Lipovetsky} V.~A., {Stepanian} J.~A., {Erastova} L.~K.,
  {Shapovalova} A.~I., 1989, Soobshcheniya Spetsial'noj Astrofizicheskoj
  Observatorii, 62, 5

\bibitem[{{Markarian} \& {Stepanian}(1983)}]{markarian83}
{Markarian} B.~E., {Stepanian} D.~A., 1983, Astrofizika, 19, 639

\bibitem[{{Martin} {et~al}\mbox{.}(2005){Martin}, {Fanson}, {Schiminovich},
  {Morrissey}, {Friedman}, {Barlow}, {Conrow}, {Grange}, {Jelinsky},
  {Milliard}, {Siegmund}, {Bianchi}, {Byun}, {Donas}, {Forster}, {Heckman},
  {Lee}, {Madore}, {Malina}, {Neff}, {Rich}, {Small}, {Surber}, {Szalay},
  {Welsh}, \& {Wyder}}]{galex}
{Martin} D.~C. {et~al.}, 2005, \apjl, 619, L1

\bibitem[{{McGaugh} \& {Schombert}(2014)}]{mcgaugh14}
{McGaugh} S.~S., {Schombert} J.~M., 2014, \aj, 148, 77

\bibitem[{{McGaugh} \& {Schombert}(2015)}]{mcgaugh15}
{McGaugh} S.~S., {Schombert} J.~M., 2015, \apj, 802, 18

\bibitem[{{Meidt} {et~al}\mbox{.}(2014){Meidt}, {Schinnerer}, {van de Ven},
  {Zaritsky}, {Peletier}, {Knapen}, {Sheth}, {Regan}, {Querejeta},
  {Mu{\~n}oz-Mateos}, {Kim}, {Hinz}, {Gil de Paz}, {Athanassoula}, {Bosma},
  {Buta}, {Cisternas}, {Ho}, {Holwerda}, {Skibba}, {Laurikainen}, {Salo},
  {Gadotti}, {Laine}, {Erroz-Ferrer}, {Comer{\'o}n}, {Men{\'e}ndez-Delmestre},
  {Seibert}, \& {Mizusawa}}]{meidt14}
{Meidt} S.~E. {et~al.}, 2014, \apj, 788, 144

\bibitem[{Mickaelian(2014)}]{mickaelian14}
Mickaelian A.~M., 2014, Multiwavelength AGN Surveys and Studies, 304, 1

\bibitem[{{Monet} {et~al}\mbox{.}(2003){Monet}, {Levine}, {Canzian}, {Ables},
  {Bird}, {Dahn}, {Guetter}, {Harris}, {Henden}, {Leggett}, {Levison},
  {Luginbuhl}, {Martini}, {Monet}, {Munn}, {Pier}, {Rhodes}, {Riepe}, {Sell},
  {Stone}, {Vrba}, {Walker}, {Westerhout}, {Brucato}, {Reid}, {Schoening},
  {Hartley}, {Read}, \& {Tritton}}]{usnob1}
{Monet} D.~G. {et~al.}, 2003, \aj, 125, 984

\bibitem[{{Murphy} {et~al}\mbox{.}(2011){Murphy}, {Condon}, {Schinnerer},
  {Kennicutt}, {Calzetti}, {Armus}, {Helou}, {Turner}, {Aniano}, {Beir{\~a}o},
  {Bolatto}, {Brandl}, {Croxall}, {Dale}, {Donovan Meyer}, {Draine},
  {Engelbracht}, {Hunt}, {Hao}, {Koda}, {Roussel}, {Skibba}, \&
  {Smith}}]{murphy11}
{Murphy} E.~J. {et~al.}, 2011, \apj, 737, 67

\bibitem[{{Nissanke}, {Kasliwal} \& {Georgieva}(2013){Nissanke}, {Kasliwal}, \&
  {Georgieva}}]{nissanke13}
{Nissanke} S., {Kasliwal} M., {Georgieva} A., 2013, \apj, 767, 124

\bibitem[{{Noeske} {et~al}\mbox{.}(2007){Noeske}, {Weiner}, {Faber},
  {Papovich}, {Koo}, {Somerville}, {Bundy}, {Conselice}, {Newman},
  {Schiminovich}, {Le Floc'h}, {Coil}, {Rieke}, {Lotz}, {Primack}, {Barmby},
  {Cooper}, {Davis}, {Ellis}, {Fazio}, {Guhathakurta}, {Huang}, {Kassin},
  {Martin}, {Phillips}, {Rich}, {Small}, {Willmer}, \& {Wilson}}]{noeske07}
{Noeske} K.~G. {et~al.}, 2007, \apjl, 660, L43

\bibitem[{{Norris} {et~al}\mbox{.}(2014){Norris}, {Meidt}, {Van de Ven},
  {Schinnerer}, {Groves}, \& {Querejeta}}]{norris14}
{Norris} M.~A., {Meidt} S., {Van de Ven} G., {Schinnerer} E., {Groves} B.,
  {Querejeta} M., 2014, \apj, 797, 55

\bibitem[{Ofek {et~al}\mbox{.}(2012)Ofek, Laher, Surace, Levitan, Sesar,
  Horesh, Law, van Eyken, Kulkarni, Prince, Nugent, Sullivan, Yaron, Pickles,
  Ag{\"u}eros, Arcavi, Bildsten, Bloom, Cenko, Gal-Yam, Grillmair, Helou,
  Kasliwal, Poznanski, \& Quimby}]{ofek12}
Ofek E.~O. {et~al.}, 2012, Publications of the Astronomical Society of the
  Pacific, 124, 854

\bibitem[{{Oh} {et~al}\mbox{.}(2008){Oh}, {de Blok}, {Walter}, {Brinks}, \&
  {Kennicutt}}]{oh08}
{Oh} S.-H., {de Blok} W.~J.~G., {Walter} F., {Brinks} E., {Kennicutt}, Jr.
  R.~C., 2008, \aj, 136, 2761

\bibitem[{{Oke} \& {Gunn}(1982)}]{dbsp}
{Oke} J.~B., {Gunn} J.~E., 1982, PASP, 94, 586

\bibitem[{Parker {et~al}\mbox{.}(2006)Parker, Acker, Frew, Hartley, Peyaud,
  Ochsenbein, Phillipps, Russeil, Beaulieu, Cohen, Koppen, Miszalski, Morgan,
  Morris, Pierce, \& Vaughan}]{parker06}
Parker Q.~A. {et~al.}, 2006, Monthly Notices of the Royal Astronomical Society,
  373, 79

\bibitem[{Pascual {et~al}\mbox{.}(2001)Pascual, Gallego, Arag~n Salamanca, \&
  Zamorano}]{pascual01}
Pascual S., Gallego J., Arag~n Salamanca A., Zamorano J., 2001, Astronomy and
  Astrophysics, 379, 798

\bibitem[{{Peng} {et~al}\mbox{.}(2010){Peng}, {Lilly}, {Kova{\v c}},
  {Bolzonella}, {Pozzetti}, {Renzini}, {Zamorani}, {Ilbert}, {Knobel},
  {Iovino}, {Maier}, {Cucciati}, {Tasca}, {Carollo}, {Silverman}, {Kampczyk},
  {de Ravel}, {Sanders}, {Scoville}, {Contini}, {Mainieri}, {Scodeggio},
  {Kneib}, {Le F{\`e}vre}, {Bardelli}, {Bongiorno}, {Caputi}, {Coppa}, {de la
  Torre}, {Franzetti}, {Garilli}, {Lamareille}, {Le Borgne}, {Le Brun},
  {Mignoli}, {Perez Montero}, {Pello}, {Ricciardelli}, {Tanaka}, {Tresse},
  {Vergani}, {Welikala}, {Zucca}, {Oesch}, {Abbas}, {Barnes}, {Bordoloi},
  {Bottini}, {Cappi}, {Cassata}, {Cimatti}, {Fumana}, {Hasinger}, {Koekemoer},
  {Leauthaud}, {Maccagni}, {Marinoni}, {McCracken}, {Memeo}, {Meneux}, {Nair},
  {Porciani}, {Presotto}, \& {Scaramella}}]{peng10}
{Peng} Y.-j. {et~al.}, 2010, \apj, 721, 193

\bibitem[{{Pesch} \& {Sanduleak}(1983)}]{case1}
{Pesch} P., {Sanduleak} N., 1983, \apjs, 51, 171

\bibitem[{{Pesch} \& {Sanduleak}(1986)}]{case3}
{Pesch} P., {Sanduleak} N., 1986, \apjs, 60, 543

\bibitem[{{Pesch} \& {Sanduleak}(1988)}]{case5}
{Pesch} P., {Sanduleak} N., 1988, \apjs, 66, 297

\bibitem[{{Pesch} \& {Sanduleak}(1989)}]{case8}
{Pesch} P., {Sanduleak} N., 1989, \apjs, 70, 163

\bibitem[{{Pesch}, {Sanduleak} \& {Stephenson}(1991){Pesch}, {Sanduleak}, \&
  {Stephenson}}]{case12}
{Pesch} P., {Sanduleak} N., {Stephenson} C.~B., 1991, \apjs, 76, 1043

\bibitem[{{Petrosian} {et~al}\mbox{.}(2007){Petrosian}, {McLean}, {Allen}, \&
  {MacKenty}}]{petrosian07}
{Petrosian} A., {McLean} B., {Allen} R.~J., {MacKenty} J.~W., 2007, \apjs, 170,
  33

\bibitem[{{Pettini} \& {Pagel}(2004)}]{pp04}
{Pettini} M., {Pagel} B.~E.~J., 2004, \mnras, 348, L59

\bibitem[{{Querejeta} {et~al}\mbox{.}(2015){Querejeta}, {Meidt}, {Schinnerer},
  {Cisternas}, {Mu{\~n}oz-Mateos}, {Sheth}, {Knapen}, {van de Ven}, {Norris},
  {Peletier}, {Laurikainen}, {Salo}, {Holwerda}, {Athanassoula}, {Bosma},
  {Groves}, {Ho}, {Gadotti}, {Zaritsky}, {Regan}, {Hinz}, {Gil de Paz},
  {Menendez-Delmestre}, {Seibert}, {Mizusawa}, {Kim}, {Erroz-Ferrer}, {Laine},
  \& {Comer{\'o}n}}]{querejeta15}
{Querejeta} M. {et~al.}, 2015, \apjs, 219, 5

\bibitem[{{Rau} {et~al}\mbox{.}(2009){Rau}, {Kulkarni}, {Law}, {Bloom},
  {Ciardi}, {Djorgovski}, {Fox}, {Gal-Yam}, {Grillmair}, {Kasliwal}, {Nugent},
  {Ofek}, {Quimby}, {Reach}, {Shara}, {Bildsten}, {Cenko}, {Drake},
  {Filippenko}, {Helfand}, {Helou}, {Howell}, {Poznanski}, \&
  {Sullivan}}]{rau09}
{Rau} A. {et~al.}, 2009, \pasp, 121, 1334

\bibitem[{{Salim} {et~al}\mbox{.}(2007){Salim}, {Rich}, {Charlot},
  {Brinchmann}, {Johnson}, {Schiminovich}, {Seibert}, {Mallery}, {Heckman},
  {Forster}, {Friedman}, {Martin}, {Morrissey}, {Neff}, {Small}, {Wyder},
  {Bianchi}, {Donas}, {Lee}, {Madore}, {Milliard}, {Szalay}, {Welsh}, \&
  {Yi}}]{salim07}
{Salim} S. {et~al.}, 2007, \apjs, 173, 267

\bibitem[{{Salzer} {et~al}\mbox{.}(2000){Salzer}, {Gronwall}, {Lipovetsky},
  {Kniazev}, {Moody}, {Boroson}, {Thuan}, {Izotov}, {Herrero}, \&
  {Frattare}}]{kiss00}
{Salzer} J.~J. {et~al.}, 2000, \aj, 120, 80

\bibitem[{{Salzer} {et~al}\mbox{.}(2001){Salzer}, {Gronwall}, {Lipovetsky},
  {Kniazev}, {Moody}, {Boroson}, {Thuan}, {Izotov}, {Herrero}, \&
  {Frattare}}]{kiss01}
{Salzer} J.~J. {et~al.}, 2001, \aj, 121, 66

\bibitem[{{Sanduleak} \& {Pesch}(1984)}]{case2}
{Sanduleak} N., {Pesch} P., 1984, \apjs, 55, 517

\bibitem[{{Sanduleak} \& {Pesch}(1987)}]{case4}
{Sanduleak} N., {Pesch} P., 1987, \apjs, 63, 809

\bibitem[{{Sanduleak} \& {Pesch}(1989)}]{case9}
{Sanduleak} N., {Pesch} P., 1989, \apjs, 70, 173

\bibitem[{{Sanduleak} \& {Pesch}(1990)}]{case11}
{Sanduleak} N., {Pesch} P., 1990, \apjs, 72, 291

\bibitem[{{Schawinski} {et~al}\mbox{.}(2014){Schawinski}, {Urry}, {Simmons},
  {Fortson}, {Kaviraj}, {Keel}, {Lintott}, {Masters}, {Nichol}, {Sarzi},
  {Skibba}, {Treister}, {Willett}, {Wong}, \& {Yi}}]{schawinski14}
{Schawinski} K. {et~al.}, 2014, \mnras

\bibitem[{{Schlafly} \& {Finkbeiner}(2011)}]{schlafly11}
{Schlafly} E.~F., {Finkbeiner} D.~P., 2011, \apj, 737, 103

\bibitem[{Sobral {et~al}\mbox{.}(2009)Sobral, Best, Geach, Smail, Kurk,
  Cirasuolo, Casali, Ivison, Coppin, \& Dalton}]{sobral09}
Sobral D. {et~al.}, 2009, Monthly Notices of the Royal Astronomical Society,
  398, 75

\bibitem[{{Sobral} {et~al}\mbox{.}(2012){Sobral}, {Best}, {Matsuda}, {Smail},
  {Geach}, \& {Cirasuolo}}]{sobral12}
{Sobral} D., {Best} P.~N., {Matsuda} Y., {Smail} I., {Geach} J.~E., {Cirasuolo}
  M., 2012, \mnras, 420, 1926

\bibitem[{{Stroe} \& {Sobral}(2015)}]{stroe15}
{Stroe} A., {Sobral} D., 2015, \mnras, 453, 242

\bibitem[{{Thuan} \& {Izotov}(2005)}]{thuan05}
{Thuan} T.~X., {Izotov} Y.~I., 2005, The Astrophysical Journal Supplement
  Series, 161, 240

\bibitem[{{Troja} {et~al}\mbox{.}(2017){Troja}, {Piro}, {van Eerten},
  {Wollaeger}, {Im}, {Fox}, {Butler}, {Cenko}, {Sakamoto}, {Fryer}, {Ricci},
  {Lien}, {Ryan}, {Korobkin}, {Lee}, {Burgess}, {Lee}, {Watson}, {Choi},
  {Covino}, {D' Avanzo}, {Fontes}, {Becerra Gonzalez}, {Khandrika}, {Kim},
  {Kim}, {Lee}, {Lee}, {Kutyrev}, {Lim}, {Sanchez Ramirez}, {Veilleux},
  {Wieringa}, \& {Yoon}}]{troja17}
{Troja} E. {et~al.}, 2017, ArXiv e-prints

\bibitem[{{Tully} {et~al}\mbox{.}(2009){Tully}, {Rizzi}, {Shaya}, {Courtois},
  {Makarov}, \& {Jacobs}}]{EDD}
{Tully} R.~B., {Rizzi} L., {Shaya} E.~J., {Courtois} H.~M., {Makarov} D.~I.,
  {Jacobs} B.~A., 2009, \aj, 138, 323

\bibitem[{{Ugryumov} {et~al}\mbox{.}(1999){Ugryumov}, {Engels}, {Lipovetsky},
  {Hagen}, {Hopp}, {Pustilnik}, {Kniazev}, {Richter}, {Izotov}, \&
  {Popescu}}]{hss99}
{Ugryumov} A.~V. {et~al.}, 1999, \aaps, 135, 511

\bibitem[{{White} {et~al}\mbox{.}(2011){White}, {Blanton}, {Bolton},
  {Schlegel}, {Tinker}, {Berlind}, {da Costa}, {Kazin}, {Lin}, {Maia},
  {McBride}, {Padmanabhan}, {Parejko}, {Percival}, {Prada}, {Ramos}, {Sheldon},
  {de Simoni}, {Skibba}, {Thomas}, {Wake}, {Zehavi}, {Zheng}, {Nichol},
  {Schneider}, {Strauss}, {Weaver}, \& {Weinberg}}]{white11}
{White} M. {et~al.}, 2011, \apj, 728, 126

\bibitem[{{Wright} {et~al}\mbox{.}(2010){Wright}, {Eisenhardt}, {Mainzer},
  {Ressler}, {Cutri}, {Jarrett}, {Kirkpatrick}, {Padgett}, {McMillan},
  {Skrutskie}, {Stanford}, {Cohen}, {Walker}, {Mather}, {Leisawitz}, {Gautier},
  {McLean}, {Benford}, {Lonsdale}, {Blain}, {Mendez}, {Irace}, {Duval}, {Liu},
  {Royer}, {Heinrichsen}, {Howard}, {Shannon}, {Kendall}, {Walsh}, {Larsen},
  {Cardon}, {Schick}, {Schwalm}, {Abid}, {Fabinsky}, {Naes}, \& {Tsai}}]{wise}
{Wright} E.~L. {et~al.}, 2010, \aj, 140, 1868

\bibitem[{{York} {et~al}\mbox{.}(2000){York}, {Adelman}, {Anderson},
  {Anderson}, {Annis}, {Bahcall}, {Bakken}, {Barkhouser}, {Bastian}, {Berman},
  {Boroski}, {Bracker}, {Briegel}, {Briggs}, {Brinkmann}, {Brunner}, {Burles},
  {Carey}, {Carr}, {Castander}, {Chen}, {Colestock}, {Connolly}, {Crocker},
  {Csabai}, {Czarapata}, {Davis}, {Doi}, {Dombeck}, {Eisenstein}, {Ellman},
  {Elms}, {Evans}, {Fan}, {Federwitz}, {Fiscelli}, {Friedman}, {Frieman},
  {Fukugita}, {Gillespie}, {Gunn}, {Gurbani}, {de Haas}, {Haldeman}, {Harris},
  {Hayes}, {Heckman}, {Hennessy}, {Hindsley}, {Holm}, {Holmgren}, {Huang},
  {Hull}, {Husby}, {Ichikawa}, {Ichikawa}, {Ivezi{\'c}}, {Kent}, {Kim},
  {Kinney}, {Klaene}, {Kleinman}, {Kleinman}, {Knapp}, {Korienek}, {Kron},
  {Kunszt}, {Lamb}, {Lee}, {Leger}, {Limmongkol}, {Lindenmeyer}, {Long},
  {Loomis}, {Loveday}, {Lucinio}, {Lupton}, {MacKinnon}, {Mannery}, {Mantsch},
  {Margon}, {McGehee}, {McKay}, {Meiksin}, {Merelli}, {Monet}, {Munn},
  {Narayanan}, {Nash}, {Neilsen}, {Neswold}, {Newberg}, {Nichol}, {Nicinski},
  {Nonino}, {Okada}, {Okamura}, {Ostriker}, {Owen}, {Pauls}, {Peoples},
  {Peterson}, {Petravick}, {Pier}, {Pope}, {Pordes}, {Prosapio},
  {Rechenmacher}, {Quinn}, {Richards}, {Richmond}, {Rivetta}, {Rockosi},
  {Ruthmansdorfer}, {Sandford}, {Schlegel}, {Schneider}, {Sekiguchi}, {Sergey},
  {Shimasaku}, {Siegmund}, {Smee}, {Smith}, {Snedden}, {Stone}, {Stoughton},
  {Strauss}, {Stubbs}, {SubbaRao}, {Szalay}, {Szapudi}, {Szokoly}, {Thakar},
  {Tremonti}, {Tucker}, {Uomoto}, {Vanden Berk}, {Vogeley}, {Waddell}, {Wang},
  {Watanabe}, {Weinberg}, {Yanny}, {Yasuda}, \& {SDSS Collaboration}}]{york00}
{York} D.~G. {et~al.}, 2000, \aj, 120, 1579

\bibitem[{{Zacharias} {et~al}\mbox{.}(2010){Zacharias}, {Finch}, {Girard},
  {Hambly}, {Wycoff}, {Zacharias}, {Castillo}, {Corbin}, {DiVittorio}, {Dutta},
  {Gaume}, {Gauss}, {Germain}, {Hall}, {Hartkopf}, {Hsu}, {Holdenried},
  {Makarov}, {Martinez}, {Mason}, {Monet}, {Rafferty}, {Rhodes}, {Siemers},
  {Smith}, {Tilleman}, {Urban}, {Wieder}, {Winter}, \& {Young}}]{ucac3}
{Zacharias} N. {et~al.}, 2010, \aj, 139, 2184

\end{thebibliography}

\section*{acknowledgements}
We thank the anonymous referee for their rigorous and helpful feedback. This work was supported by the GROWTH (Global Relay of Observatories Watching Transients Happen) project funded by the National Science Foundation under PIRE Grant No 1545949. D. H. Reitze gratefully acknowledges the support of the NSF (award PHY-0757058). We thank Adam Miller for helpful comments.

The Intermediate Palomar Transient Factory project is a scientific collaboration among the California Institute of Technology, Los Alamos National Laboratory, the University of Wisconsin, Milwaukee, the Oskar Klein Center, the Weizmann Institute of Science, the TANGO Program of the University System of Taiwan, and the Kavli Institute for the Physics and Mathematics of the Universe.

This research has made use of the NASA/IPAC Extragalactic Database (NED) which is operated by the Jet Propulsion Laboratory, California Institute of Technology, under contract with the National Aeronautics and Space Administration. This publication makes use of data products from the Wide-field Infrared Survey Explorer, which is a joint project of the University of California, Los Angeles, and the Jet Propulsion Laboratory/California Institute of Technology, funded by the National Aeronautics and Space Administration. 

The Pan-STARRS1 Surveys (PS1) and the PS1 public science archive have been made possible through contributions by the Institute for Astronomy, the University of Hawaii, the Pan-STARRS Project Office, the Max-Planck Society and its participating institutes, the Max Planck Institute for Astronomy, Heidelberg and the Max Planck Institute for Extraterrestrial Physics, Garching, The Johns Hopkins University, Durham University, the University of Edinburgh, the Queen's University Belfast, the Harvard-Smithsonian Center for Astrophysics, the Las Cumbres Observatory Global Telescope Network Incorporated, the National Central University of Taiwan, the Space Telescope Science Institute, the National Aeronautics and Space Administration under Grant No. NNX08AR22G issued through the Planetary Science Division of the NASA Science Mission Directorate, the National Science Foundation Grant No. AST-1238877, the University of Maryland, Eotvos Lorand University (ELTE), the Los Alamos National Laboratory, and the Gordon and Betty Moore Foundation.   

Funding for SDSS-III has been provided by the Alfred P. Sloan Foundation, the Participating Institutions, the National Science Foundation, and the U.S. Department of Energy Office of Science. The SDSS-III web site is http://www.sdss3.org/.

SDSS-III is managed by the Astrophysical Research Consortium for the Participating Institutions of the SDSS-III Collaboration including the University of Arizona, the Brazilian Participation Group, Brookhaven National Laboratory, Carnegie Mellon University, University of Florida, the French Participation Group, the German Participation Group, Harvard University, the Instituto de Astrofisica de Canarias, the Michigan State/Notre Dame/JINA Participation Group, Johns Hopkins University, Lawrence Berkeley National Laboratory, Max Planck Institute for Astrophysics, Max Planck Institute for Extraterrestrial Physics, New Mexico State University, New York University, Ohio State University, Pennsylvania State University, University of Portsmouth, Princeton University, the Spanish Participation Group, University of Tokyo, University of Utah, Vanderbilt University, University of Virginia, University of Washington, and Yale University.

Based on observations obtained at the Hale Telescope,
Palomar Observatory as part of a continuing collaboration between the California Institute of Technology,
NASA/JPL, Yale University, and the National Astronomical Observatories of China.

\onecolumn{\begin{longtable}{cccccccc}	 
{Preliminary Fields Properties}\\ 
\hline			     
\hline			     
Field	& RA		& DEC 		&Filter &Obs-Date	&5$\sigma$ Detection	&EW Limit     &EW Limit	   \\ 
	& center	& center	&       &		&Limit			&$\Sigma$=2.5 &$\Sigma$=5  \\ 
(name) & (J2000)	& (J2000)	&(name) &		&(AB mag)		& (\AA)	      & (\AA)	   \\ 
\hline 
\endfirsthead 
{Table Continued}\\ 
\hline 
\endhead 
\hline 
\endfoot 
\caption{The basic properties of the 14 preliminary fields analyzed here. The first three columns list the name and central coordinates of the 14 preliminary fields followed by the basic information of the four \ha~fitlers in each pointing. The information listed for the four \ha~filters are from left-to-right: the filter name, the observation date, the median and standard deviation 5$\sigma$ detection limits for the 11 chips in the pointing, the EW limit for the $\Sigma=2.5$ and $\Sigma=5$ limits set by the scatter in bright continuum sources as seen in the color-magnitude diagrams of Figures~\ref{fig:ha1cand}-\ref{fig:ha4cand}.} \\ 
\endlastfoot 
p3967    &   13:01:56.1   &   28:07:30.0   \\ 
 & & &  \ha1  &   2012-05-09   &	18.59   $\pm$    0.12   &   7.74 &   16.27 \\ 
 & & &  \ha2  &   2012-05-09   &	18.73   $\pm$    0.12   &   7.97 &   16.76 \\ 
 & & &  \ha3  &   2015-06-02   &	18.63   $\pm$    0.17   &   9.41 &   19.78 \\ 
 & & &  \ha4  &   2015-06-02   &	18.54   $\pm$    0.14   &   9.44 &   19.85 \\ 
\hline 
p4063    &   14:06:35.6   &   30:22:30.0   \\ 
 & & &  \ha1  &   2014-02-15   &	18.61   $\pm$    0.18   &   7.06 &   14.77 \\ 
 & & &  \ha2  &   2014-02-15   &	18.61   $\pm$    0.18   &   7.27 &   15.22 \\ 
 & & &  \ha3  &   2015-02-04   &	18.87   $\pm$    0.19   &   9.27 &   19.47 \\ 
 & & &  \ha4  &   2015-02-04   &	18.92   $\pm$    0.19   &   9.30 &   19.54 \\ 
\hline 
p4064    &   14:22:25.1   &   30:22:30.0   \\ 
 & & &  \ha1  &   2014-02-15   &	18.43   $\pm$    0.18   &   8.09 &   17.03 \\ 
 & & &  \ha2  &   2014-02-15   &	18.61   $\pm$    0.17   &   8.33 &   17.54 \\ 
 & & &  \ha3  &   2015-02-04   &	18.88   $\pm$    0.18   &   7.87 &   16.42 \\ 
 & & &  \ha4  &   2015-02-04   &	18.96   $\pm$    0.20   &   7.90 &   16.47 \\ 
\hline 
p4070    &   15:57:21.8   &   30:22:30.0   \\ 
 & & &  \ha1  &   2013-03-26   &	18.65   $\pm$    0.19   &   6.17 &   12.85 \\ 
 & & &  \ha2  &   2013-03-26   &	18.97   $\pm$    0.19   &   6.36 &   13.24 \\ 
 & & &  \ha3  &   2015-07-01   &	18.65   $\pm$    0.13   &   6.75 &   13.99 \\ 
 & & &  \ha4  &   2015-07-01   &	18.62   $\pm$    0.14   &   6.77 &   14.03 \\ 
\hline 
p4071    &   16:13:11.2   &   30:22:30.0   \\ 
 & & &  \ha1  &   2013-03-26   &	18.95   $\pm$    0.20   &   6.05 &   12.59 \\ 
 & & &  \ha2  &   2013-03-26   &	19.07   $\pm$    0.20   &   6.24 &   12.97 \\ 
 & & &  \ha3  &   2015-07-01   &	18.45   $\pm$    0.13   &   8.27 &   17.29 \\ 
 & & &  \ha4  &   2015-07-01   &	18.56   $\pm$    0.14   &   8.30 &   17.34 \\ 
\hline 
p4072    &   16:29: 0.7   &   30:22:30.0   \\ 
 & & &  \ha1  &   2013-03-26   &	19.00   $\pm$    0.20   &   5.91 &   12.27 \\ 
 & & &  \ha2  &   2013-03-26   &	19.07   $\pm$    0.20   &   6.08 &   12.64 \\ 
 & & &  \ha3  &   2015-09-28   &	19.28   $\pm$    0.28   &   9.51 &   20.00 \\ 
 & & &  \ha4  &   2015-09-28   &	19.06   $\pm$    0.27   &   9.54 &   20.07 \\ 
\hline 
p4073    &   16:44:50.1   &   30:22:30.0   \\ 
 & & &  \ha1  &   2013-03-26   &	19.01   $\pm$    0.21   &   5.79 &   12.02 \\ 
 & & &  \ha2  &   2013-03-26   &	18.91   $\pm$    0.19   &   5.97 &   12.38 \\ 
 & & &  \ha3  &   2015-09-27   &	18.86   $\pm$    0.13   &   7.80 &   16.27 \\ 
 & & &  \ha4  &   2015-09-27   &	18.84   $\pm$    0.15   &   7.83 &   16.32 \\ 
\hline 
p4074    &   17:00:39.6   &   30:22:30.0   \\ 
 & & &  \ha1  &   2013-04-29   &	18.88   $\pm$    0.19   &   6.65 &   13.87 \\ 
 & & &  \ha2  &   2013-04-29   &	18.96   $\pm$    0.18   &   6.85 &   14.29 \\ 
 & & &  \ha3  &   2015-08-01   &	19.01   $\pm$    0.16   &   8.05 &   16.82 \\ 
 & & &  \ha4  &   2015-08-01   &	19.12   $\pm$    0.16   &   8.08 &   16.87 \\ 
\hline 
p4075    &   17:16:29.0   &   30:22:30.0   \\ 
 & & &  \ha1  &   2013-04-29   &	18.75   $\pm$    0.20   &   7.27 &   15.23 \\ 
 & & &  \ha2  &   2013-04-29   &	18.88   $\pm$    0.17   &   7.49 &   15.69 \\ 
 & & &  \ha3  &   2015-08-01   &	19.01   $\pm$    0.14   &   10.39 &   21.95 \\ 
 & & &  \ha4  &   2015-08-01   &	18.89   $\pm$    0.24   &   10.42 &   22.02 \\ 
\hline 
p4095    &   22:32:58.0   &   30:22:30.0   \\ 
 & & &  \ha1  &   2013-05-24   &	18.33   $\pm$    0.17   &   6.58 &   13.72 \\ 
 & & &  \ha2  &   2013-05-24   &	18.47   $\pm$    0.18   &   6.77 &   14.13 \\ 
 & & &  \ha3  &   2015-09-29   &	18.77   $\pm$    0.15   &   7.98 &   16.64 \\ 
 & & &  \ha4  &   2015-09-29   &	18.78   $\pm$    0.17   &   8.00 &   16.70 \\ 
\hline 
p4096    &   22:48:47.5   &   30:22:30.0   \\ 
 & & &  \ha1  &   2013-06-21   &	19.00   $\pm$    0.31   &   6.74 &   14.07 \\ 
 & & &  \ha2  &   2013-06-21   &	19.14   $\pm$    0.29   &   6.94 &   14.49 \\ 
 & & &  \ha3  &   2015-10-26   &	18.61   $\pm$    0.15   &   6.82 &   14.15 \\ 
 & & &  \ha4  &   2015-10-26   &	18.57   $\pm$    0.16   &   6.84 &   14.20 \\ 
\hline 
p4098    &   23:20:26.4   &   30:22:30.0   \\ 
 & & &  \ha1  &   2013-06-21   &	19.05   $\pm$    0.29   &   6.52 &   13.61 \\ 
 & & &  \ha2  &   2013-06-21   &	19.03   $\pm$    0.30   &   6.72 &   14.02 \\ 
 & & &  \ha3  &   2015-10-29   &	18.67   $\pm$    0.13   &   8.15 &   17.01 \\ 
 & & &  \ha4  &   2015-10-29   &	18.70   $\pm$    0.14   &   8.17 &   17.07 \\ 
\hline 
p4192    &   00:41:22.8   &   34:52:30.0   \\ 
 & & &  \ha1  &   2014-02-14   &	18.41   $\pm$    0.19   &   6.92 &   14.47 \\ 
 & & &  \ha2  &   2014-02-14   &	18.58   $\pm$    0.17   &   7.13 &   14.91 \\ 
 & & &  \ha3  &   2015-09-26   &	18.84   $\pm$    0.19   &   7.41 &   15.43 \\ 
 & & &  \ha4  &   2015-09-26   &	18.85   $\pm$    0.20   &   7.44 &   15.48 \\ 
\hline 
p4193    &   00:57:55.9   &   34:52:30.0   \\ 
 & & &  \ha1  &   2014-02-14   &	18.50   $\pm$    0.16   &   6.68 &   13.94 \\ 
 & & &  \ha2  &   2014-02-14   &	18.59   $\pm$    0.17   &   6.88 &   14.36 \\ 
 & & &  \ha3  &   2015-02-03   &	18.33   $\pm$    0.13   &   7.22 &   15.02 \\ 
 & & &  \ha4  &   2015-02-03   &	18.65   $\pm$    0.16   &   7.25 &   15.07 \\ 
\hline 
\hline 
\label{tab:prelim} 
\end{longtable} 
}
\twocolumn
 
\end{document}